\documentclass[useAMS,usenatbib]{mn2e}

\usepackage{tabularx}
\usepackage{graphicx}
\usepackage[section]{placeins}
\usepackage{epstopdf}
\usepackage{array}
\usepackage{amssymb}
\usepackage{wrapfig}
\usepackage{upgreek}
\usepackage{color}
\usepackage{longtable}
\usepackage{supertabular}
\usepackage{units}
\usepackage{booktabs}
\usepackage{dcolumn}
\usepackage[fleqn]{amsmath}
\usepackage{multirow}

\def \arcsec {$^{\prime\prime}$}
\def \arcmin {$^\prime$}
\def \um {$\upmu$m}

\def \th {$^{{\rm th}}$}

\def \nh {N$_{2}$H$^{+}$}
\def \sun {$_{\odot}$}

\def \mnras {MNRAS}
\def \aap {A\&A}
\def \aj {AJ}
\def \pasp {PASP}
\def \apj {ApJ}
\def \apjs {ApJS}
\def \pasj {PASJ}
\def \nature {Nature}
\def \procspie {Proceedings of the SPIE}

\DeclareMathOperator\erf{erf}

\title[First SCUBA-2 observations of Cepheus]{The JCMT Gould Belt Survey: First results from SCUBA-2 observations of the Cepheus Flare Region}
\author[K. Pattle et al.]{K. Pattle$^{1}$, D. Ward-Thompson$^{1}$, J.M. Kirk$^{1}$, J. Di Francesco$^{3, 2}$, H. Kirk$^{3}$, \newauthor J.C. Mottram$^{4,5}$, J. Keown$^{2}$, J. Buckle$^{6, 7}$, S.F. Beaulieu$^{8}$, D.S. Berry$^{9}$, \newauthor H. Broekhoven-Fiene$^{2}$, M.J. Currie$^{9}$, M. Fich$^{8}$, J. Hatchell$^{10}$, T. Jenness$^{9, 11}$, \newauthor D. Johnstone$^{9, 3, 2}$, D. Nutter$^{12}$, J.E. Pineda$^{13, 14, 15}$, C. Quinn$^{12}$, C. Salji$^{6, 7}$, S. Tisi$^{8}$, \newauthor S. Walker-Smith$^{6, 7}$, M.R. Hogerheijde$^{4}$, P. Bastien$^{16}$, D. Bresnahan$^{1}$, H. Butner$^{17}$, \newauthor M. Chen$^{2}$, A. Chrysostomou$^{18}$, S. Coud\'{e}$^{16}$, C.J. Davis$^{19}$, E. Drabek-Maunder$^{20}$, \newauthor A. Duarte-Cabral$^{10}$, J. Fiege$^{21}$, P. Friberg$^{9}$, R. Friesen$^{22}$, G.A. Fuller$^{14}$, \newauthor S. Graves$^{9}$, J. Greaves$^{12}$, J. Gregson$^{23, 24}$, W. Holland$^{25, 26}$, G. Joncas$^{27}$, \newauthor L.B.G. Knee$^{3}$, S. Mairs$^{2}$, K. Marsh$^{12}$, B.C. Matthews$^{3, 2}$, G. Moriarty-Schieven$^{3}$, \newauthor C. Mowat$^{10}$, J. Rawlings$^{28}$, J. Richer$^{6, 7}$, D. Robertson$^{29}$, E. Rosolowsky$^{30}$, \newauthor D. Rumble$^{10}$, S. Sadavoy$^{5}$, H. Thomas$^{9}$, N. Tothill$^{31}$, S. Viti$^{28}$, G.J. White$^{23, 24}$, \newauthor  J. Wouterloot$^{9}$, J. Yates$^{28}$, M. Zhu$^{32}$\\
\\
Affiliations can be found after the references.}

\begin{document}

\date{}

\pagerange{\pageref{firstpage}--\pageref{lastpage}} \pubyear{2016}

\maketitle

\label{firstpage}

\begin{abstract}

We present observations of the Cepheus Flare obtained as part of the James Clerk Maxwell Telescope (JCMT) Gould Belt Legacy Survey (GBLS) with the SCUBA-2 instrument.  We produce a catalogue of sources found by SCUBA-2, and separate these into starless cores and protostars.  We determine masses and densities for each of our sources, using source temperatures determined by the \emph{Herschel} Gould Belt Survey.  We compare the properties of starless cores in four different molecular clouds: L1147/58, L1172/74, L1251 and L1228.  We find that the core mass functions for each region typically show shallower-than-Salpeter behaviour.  We find that L1147/58 and L1228 have a high ratio of starless cores to Class II protostars, while L1251 and L1174 have a low ratio, consistent with the latter regions being more active sites of current star formation, while the former are forming stars less actively.  We determine that, if modelled as thermally-supported Bonnor-Ebert spheres,  most of our cores have stable configurations accessible to them.  We estimate the external pressures on our cores using archival $^{13}$CO velocity dispersion measurements and find that our cores are typically pressure-confined, rather than gravitationally bound.  We perform a virial analysis on our cores, and find that they typically cannot be supported against collapse by internal thermal energy alone, due primarily to the measured external pressures.  This suggests that the dominant mode of internal support in starless cores in the Cepheus Flare is either non-thermal motions or internal magnetic fields.

\end{abstract}


\section{Introduction}

\begin{figure*} 
\centering
\includegraphics[width=0.7\textwidth]{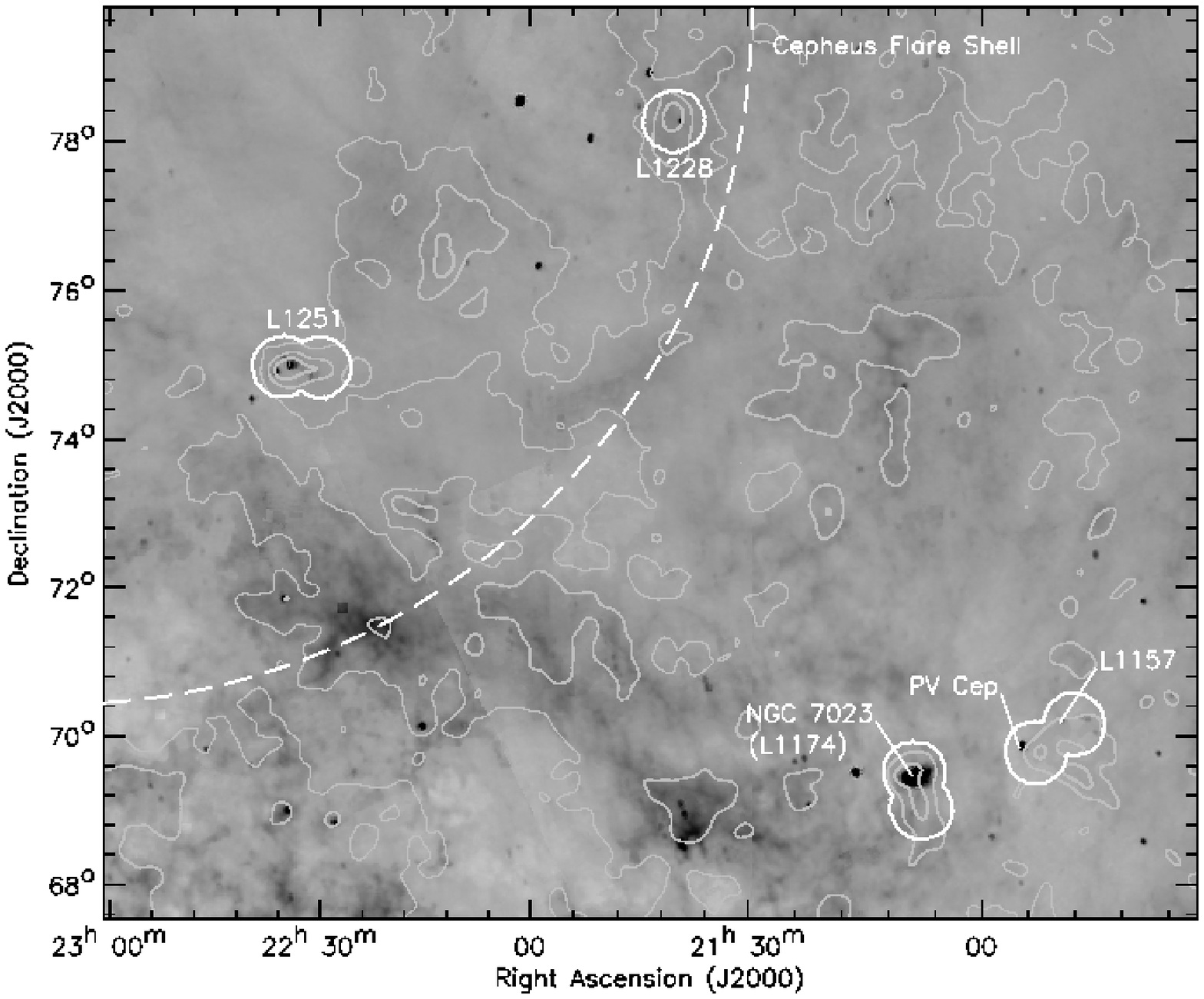}
\caption{A finding chart of the Cepheus region.  The greyscale image shows IRAS 100-\um\ emission \citep{mivilledeschenes2005}.  The grey contours show $A_{V}$ extinctions of 0.1, 0.5 and 1.0, smoothed with an 8-pixel Gaussian for clarity \citep{dobashi2005}.  The regions enclosed in solid white lines were observed as part of the JCMT GBS \citep{wardthompson2007}.  The reflection nebula L1174/NGC\,7023 is marked.  The L1172 region is immediately to the south of L1174.  The variable star PV Cep and the protostar L1157-mm, both in the L1147/58 region, are also marked.  The dashed white line shows the approximate position of the Cepheus Flare Shell (K09) -- the CFS.}
\label{fig:ceph_finding_chart}
\end{figure*}

The Cepheus Flare region is a collection of star-forming molecular clouds extending to $\sim 10-20$ degrees above the Galactic Plane at a Galactic Longitude of $\sim 110$ degrees \citep{hubble1934}.  Star formation is occurring at several different distances along the line of sight toward the Cepheus Flare: at $\sim 160$\,pc, where star formation is associated with the edge of the Local Bubble; at $\sim 300$\,pc, associated with the Gould Belt; and at $\sim 800$\,pc, associated with the Perseus arm of the Galaxy (\citealt{kun2008}, and references therein; \citealt{kirk2009}, hereafter K09).

The Gould Belt is a ring of molecular clouds and OB associations $\sim1\,$kpc in diameter and inclined $\sim 20^{\circ}$ to the Galactic Plane \citep{herschel1847,gould1879}.  The Gould Belt is considered a `laboratory' for the study of low-mass star formation, as most of the low-mass star forming regions within 500\,pc of the Earth are associated with it.  As a result, surveys aimed at mapping substantial fractions of the Gould Belt have been undertaken using the JCMT \citep{wardthompson2007}, the \emph{Herschel} Space Observatory \citep{andre2010}, and the \emph{Spitzer} Space Telecope \citep{evans2009}.

In this paper, we present SCUBA-2 observations of the intermediate-distance material in Cepheus associated with the Gould Belt.  These data were taken as part of the James Clerk Maxwell Telescope (JCMT) Gould Belt Legacy Survey (GBLS; \citealt{wardthompson2007}).  There are five dark cloud complexes in the Cepheus Flare which are associated with the Gould Belt:  L1147/48/52/55/57/58, L1172/74, L1247/51, L1228 and L1241 \citep{lynds1962}.  We present SCUBA-2 data for all or part of each of these regions, with the exception of L1241.

The Cepheus Flare is a sparsely-filled region in which star formation appears to be proceeding in a variety of different environments.  IRAS 100\um\ observations of the Cepheus Flare \citep{mivilledeschenes2005} are shown in Figure~\ref{fig:ceph_finding_chart}, with contours of $A_{V}$ extinction overlaid \citep{dobashi2005}.  The regions of highest visual extinction are not distributed evenly across the Cepheus Flare, but instead are principally located on its north-eastern and south-western sides.  In addition, Cepheus has a central region of relatively low extinction ($A_{V}<3$; \citealt{dobashi2005}) in which little star formation is occurring, although there is not a complete lack of molecular gas or young stars here \citep{tachihara2005}.  K09 found that YSOs in the Cepheus Flare are typically found in small, isolated groups, with a much higher fraction of distributed YSOs (the fraction of YSOs not associated with a group) than is typical: 41\% of YSOs in Cepheus are distributed, compared to an average of $\sim10$\% across clouds observed as part of the Spitzer c2d survey \citep{evans2009}.

\begin{table*} 
\centering
\caption{Cepheus regions observed as part of the JCMT GBLS, with approximate central positions in equatorial and galactic coordinates listed.}
\label{tab:cepheus_regions}
\begin{tabular}{c cc cc c c}
\toprule
 & R.A. (J2000) & Dec. (J2000) & l & b & Distance & Distance\\
Region & (hours:min) & (deg:arcmin) & (deg) & (deg) & (pc) & Reference\\
\midrule
L1147/58 & 21:02 & +68:00 & 104.0 & 14.1 & $325\pm13$ & \citet{straizys1992}\\
L1172/74 & 20:41 & +67:52 & 102.6 & 15.6 & $288\pm25$ & \citet{straizys1992}\\
L1251 & 22:34 & +75:14 & 114.4 & 14.7 & $300^{+50}_{-10}$ & \citet{kun2008}\\
L1228 & 20:58 & +77:38 & 111.7 & 20.2 & $200^{+100}_{-10}$ & \citet{kun2008}\\
\bottomrule
\end{tabular}
\end{table*}

\begin{figure} 
\centering
\includegraphics[width=0.49\textwidth]{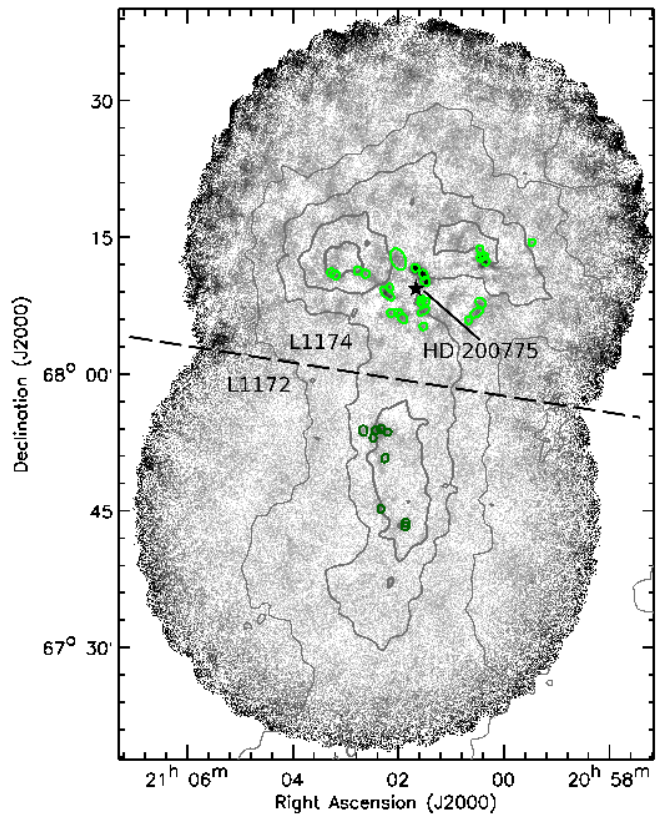}
\caption{SCUBA-2 850-\um\ observations of the L1172 (south) and L1174 (north) regions, with sources marked.  The ellipses show twice the FWHM size of each source.  Contours of $A_{\textsc{v}}$ of 0.5, 1.0, 1.5 and 2.0 magnitudes are shown for reference \citep{dobashi2005}.  The dashed line marks approximately the boundary between the L1172 and L1174 regions.  The position of the Herbig Ae/Be star HD\,200775 is marked.}
\label{fig:l1174_finding_chart} 
\end{figure}

The Cepheus Flare is defined by the interaction of a collection of shells with the local ISM, of which the most significant to the evolution of the region appears to be the Cepheus Flare Shell (CFS -- \citealt{grenier1989}; \citealt{olano2006}), an expanding supernova bubble with a radius $\sim 9.5^{\circ}$, whose centre is located to the east of the Cepheus Flare at Galactic coordinates $l\sim 120^{\circ}$, $b\sim 17^{\circ}$.  The approximate position of the CFS is marked on Figure~\ref{fig:ceph_finding_chart}.  The shell divides the north-eastern and south-western star-forming regions.  \citet{olano2006} suggest that star formation in the eastern regions of the Cepheus Flare has been triggered by the passage of the CFS.  K09 note that the current position of the CFS is consistent with that of L1228, and suggest that star formation in this region is being enhanced by the interaction with the shell.  A possible geometry of the clouds associated with the CFS is proposed by \citet{kun2008}.  In this geometry, the various intermediate-distance dark clouds are located approximately on the current surface of the CFS.  As the CFS has an approximate radius of $\sim 50$\,pc and is located at a distance of $\sim 300$\,pc from the Earth \citep{olano2006}, there are significant differences, both fractional and absolute, between the distances of the various dark clouds associated with the CFS, despite those dark clouds appearing along very similar lines of sight.  (See Table~\ref{tab:cepheus_regions} for distances.)

In this study we identify, and investigate the properties of, starless cores in the Cepheus Flare.  We investigate the cores' stability against collapse and the relative importance of gravity and external surface pressure in their confinement.  Previous analysis of GBS data of the Ophiuchus molecular cloud (an intermediate-mass star-forming region forming stars in a clustered manner; e.g. \citealt{wilking2008}) has suggested that dense starless cores in that region are typically confined by external surface pressure rather than self-gravity \citep{pattle2015}.  We here investigate whether starless cores in the various different environments found in the Cepheus Flare behave in a similar manner.

This paper is laid out as follows.  In Section 2, we discuss the observations and data reduction.  In Section 3, we discuss source extraction and characterisation, and present our catalogue of sources.  In Section 4, we discuss the properties of the starless cores in our catalogue.  In Section 5, we discuss the counting statistics of starless and protostellar sources in Cepheus.  In Section 6, we assess the stability of our cores using the Bonnor-Ebert criterion.  In Section 7, we discuss the energy balance in the starless cores in our catalogue, and put an upper limit on the degree to which the cores are virially bound.  In Section 8, we summarise our conclusions.

\section{Observations}

\begin{figure*} 
\centering
\includegraphics[width=0.9\textwidth]{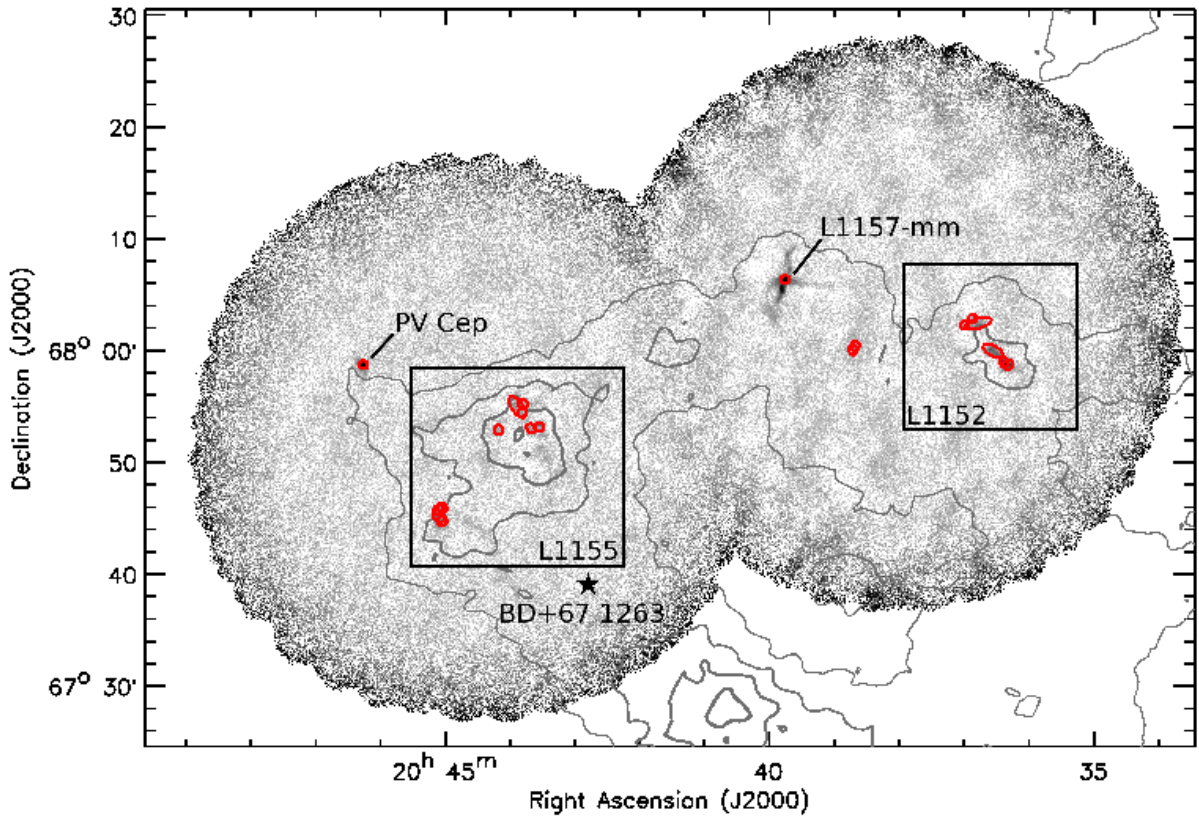}
\caption{SCUBA-2 850-\um\ observations of the L1147/L1158 region, with sources marked.  The ellipses show twice the FWHM size of each source.  Contours of $A_{\textsc{v}}$ of 0.5, 1.0, 1.5 and 2.0 magnitudes are shown for reference \citep{dobashi2005}.  The protostars PV Cep and L1157-mm are labelled, and the position of the A6V star BD$+67$ 1263 is marked.  The boxes mark approximately the extent of the L1152 and L1155 regions.}
\label{fig:l1147_finding_chart} 
\end{figure*}

The SCUBA-2 \citep{holland2013} observations used here form part of the JCMT GBLS \citep{wardthompson2007}.  Continuum observations at 850~\um\ and 450~\um\ were made using fully sampled 30\arcmin\ diameter circular regions \citep[PONG1800 mapping mode,][]{kackley2010} at resolutions of 14.1\arcsec\ and 9.6\arcsec\ respectively.  The Cepheus Flare was observed with SCUBA-2 in 41 observations taken between 2012 March 30 and 2014 October 24.  The L1174 field was observed four times in very dry (Grade 1; $\tau_{225\, {\rm GHz}}\!<\! 0.05$) weather.  The remainder of the fields were each observed six times in dry (Grade 2; $0.05\!<\!\tau_{225\,{\rm GHz} }\!<\! 0.08$) weather, except for one field, L1147/58 East (containing the star PV Cep, discussed below), which was observed seven times.  Larger regions were mosaicked with overlapping scans. Four final output maps were produced, the central co-ordinates of which are listed in Table~\ref{tab:cepheus_regions}.

The data were reduced using an iterative map-making technique \citep[\emph{makemap} in {\sc smurf},][]{chapin2013}, and gridded to 3\arcsec\ pixels at 850~\um\ and 2\arcsec\ pixels at 450~\um, as part of the Legacy Release 1 (LR1) GBLS data set (see \citealt{mairs2015}).  The iterations were halted when the map pixels, on average, changed by $<$0.1\% of the estimated map rms. The initial reductions of each individual scan were coadded to form a mosaic from which a mask based on signal-to-noise ratio was produced for each region.  The final mosaic was produced from a second reduction using this mask to define areas of emission. Detection of emission structure and calibration accuracy are robust within the masked regions, and are uncertain outside of the masked region.

A spatial filter of 10\arcmin\ is used in the reduction, which means that flux recovery is robust for sources with a Gaussian FWHM less than 2.5\arcmin. Sources between 2.5\arcmin\ and 7.5\arcmin\ in size will be detected, but both the flux and the size are underestimated because Fourier components with scales greater than 5\arcmin\ are removed by the filtering process. Detection of sources larger than 7.5\arcmin\ is dependent on the mask used for reduction.  The mask introduces further spatial filtering, as after all but the final iteration of the map-maker, all emission outside the region enclosed by the mask is suppressed.  Therefore, the recovery of extended structure outside of the masked regions is limited.

The data are calibrated in mJy/arcsec$^{2}$, using aperture Flux Conversion Factors (FCFs) of 2.34 and 4.71 Jy/pW/arcsec$^{2}$ at 850~\um\ and 450~\um, respectively, derived from average values of JCMT calibrators \citep{dempsey2013}.  The estimated $1-\sigma$ errors on the FCFs are 0.08 Jy/pW/arcsec$^{2}$ and 0.50 Jy/pW/arcsec$^{2}$ at 850~\um\ and 450~\um\ respectively.  The PONG scan pattern leads to lower noise levels in the map centre and overlap regions, while data reduction and emission artifacts can lead to small variations in the noise level over the whole map.

The SCUBA-2 850-\um\ data of Cepheus are shown in Figures~\ref{fig:l1174_finding_chart} (L1172/74), \ref{fig:l1147_finding_chart} (L1147/58), \ref{fig:l1251_finding_chart} (L1251) and \ref{fig:l1228_finding_chart} (L1228).  The sources we extract from the data are marked as coloured ellipses: light green in L1174, dark green in L1172, red in L1147/58, blue in L1251, and purple in L1228.  This colour coding is continued throughout this paper.

The emission measured in the 850-\um\ filter on SCUBA-2 can be contaminated by the CO J\,$=3\to 2$ transition \citep{drabek2012} which, with a rest wavelength of 867.6\,\um, is covered by the SCUBA-2 850-\um\ filter, which has a half-power bandwith of 85 \um\ \citep{holland2013}.  The only regions in the map which are expected to be substantially CO-contaminated are local to the PV Cep and L1157-mm protostars (discussed in Section~\ref{sec:ceph_regions}), with which there are strong outflows associated (the CO contribution from the outflow associated with L1157-mm is clearly visible as extensions north and south of the source in Figures~\ref{fig:l1147_finding_chart} and \ref{fig:zoom_figure}, below).  However, as can be seen in Figure~\ref{fig:l1147_finding_chart}, both PV Cep and L1157-mm are isolated objects, and CO emission from their outflows is unlikely to affect the fluxes measured for any of the other sources in the field.

\begin{table} 
\centering
\caption{The mean 1-$\sigma$ RMS noise levels in each of the regions observed, measured on the default LR1 pixel sizes of 2 arcsec at 450\um\ and 3 arcsec at 850\um.}
\label{tab:rms_noise}
\begin{tabular}{c r@{\,$\pm$\,}l r@{\,$\pm$\,}l}
\toprule
  & \multicolumn{2}{c}{450-\um\ RMS} & \multicolumn{2}{c}{850-\um\ RMS} \\
Region & \multicolumn{2}{c}{mJy/sqa} & \multicolumn{2}{c}{mJy/sqa} \\
\midrule
L1174 & 1.03 & 0.08 & 0.069 & 0.006 \\
\midrule
L1172 & 2.16 & 0.16 & 0.056 & 0.004 \\
L1155 & 2.44 & 0.11 & 0.056 & 0.004 \\
L1157 & 2.20 & 0.09 & 0.055 & 0.005 \\
\midrule
L1251 E & 1.77 & 0.09 & 0.059 & 0.003 \\
L1251 W & 0.93 & 0.04 & 0.054 & 0.005 \\
\midrule
L1228 & 0.87 & 0.05 & 0.059 & 0.007 \\
\bottomrule
\end{tabular}
\end{table}

Table~\ref{tab:rms_noise} lists the 1-$\sigma$ RMS noise levels in each of the regions observed, measured on the default LR1 pixel widths of 2 arcsec at 450~\um\ and 3 arcsec at 850~\um.  The 450-\um\ RMS noise levels vary somewhat between different regions observed in the same weather band.  This is due to the differing 450-\um\ sensitivity across Band 2 weather conditions. The 850-\um\ RMS noise is highest in L1174, despite this region having been observed in the best weather, due to the presence of the NGC\,7023 reflection nebula (see Section~\ref{sec:ceph_regions}).  The bright, extended emission from NGC\,7023 makes it more difficult for the data reduction process to converge on a solution.

The 450-\um\ and 850-\um\ SCUBA-2 data presented in this paper are available at: http://dx.doi.org/xx.xxxxx/yy.yyyy.

\section{Results}

\subsection{Cepheus Flare Region}
\label{sec:ceph_regions}

The Cepheus Flare consists of several distinct areas of high column density, each of which is at a different distance and likely to have a different star formation history.  Thus, we consider each separately in the following analysis, and summarise their properties here.

L1172/L1174 is a site of clustered star formation.  The dark cloud L1174, shown in the northern part of Figure~\ref{fig:l1174_finding_chart}, is coincident with the NGC\,7023 reflection nebula, also known as the Iris Nebula \citep{herschel1802}.  The nebula is illuminated by the Herbig Ae/Be star HD\,200775 (R.A.\,(J2000)\,$=\allowbreak 21^{h}\,01^{m}\,39.920^{s}$, Dec.\,(J2000)\,$=\allowbreak +68^{\circ}\,09^{\prime}\,47.76^{\prime\prime}$; \citealt{vanleeuwen2007}), of spectral classification B2Ve \citep{guetter1968}.  The position of HD\,200775 is marked on Figure~\ref{fig:l1174_finding_chart}, although HD\,200775 itself is not visible in the SCUBA-2 data.

L1172 lies to the south of L1174, and is also shown in Figure~\ref{fig:l1174_finding_chart}. It is forming stars much less actively than the neighbouring L1174.

L1147/L1158 contains the Lynds dark nebulae L1147, L1148, L1152, L1155, L1157, and L1158 \citep{lynds1962}.  This region is considered to be the least affected by the CFS, and to be forming stars with a low efficiency (K09).  Only L1147, L1152, and L1155 were observed with SCUBA-2.  All of the emission seen in the western area shown in Figure~\ref{fig:l1147_finding_chart} is associated with L1152, except for the bright protostar L1157-mm and its associated outflow \citep{kun2008}, which are discussed below.  All of the emission in the eastern region of Figure~\ref{fig:l1147_finding_chart} is associated with L1155, with the exception of the bright point source in the north-east, the star PV Cep (\citealt{li1994}; discussed below).

Both L1152 and L1155 appear relatively quiescent (K09). There is some evidence that L1155 may be undergoing external heating: \citet{nutter2009} found evidence for a $\sim2\,$K temperature gradient across one of the cores in the region, L1155C, which they ascribed to the effects of the nearby A6V star BD$+67$ 1263 (marked on Figure~\ref{fig:l1147_finding_chart}).

The SCUBA-2 field contains two bright PMS stars: PV Cep (R.A.\,(J2000)\,$=\allowbreak 20^{h}\,45^{m}\,53.943^{s}$, Dec.\,(J2000)\,$=\allowbreak +67^{\circ}\,57^{\prime}\,38.66^{\prime\prime}$; \citealt{cutri2003}) and L1157-mm (R.A.\,(J2000)\,$=\allowbreak 20^{h}\,39^{m}\,06.2^{s}$, Dec.\,(J2000)\,$=\allowbreak \,+68^{\circ}02^{\prime}15^{\prime\prime}$; K09).  PV Cep is a highly variable \citep{kun2009} A5 Herbig Ae/Be star \citep{li1994}, with which an extended ouflow is associated \citep{reipurth1997}.  PV Cep has a high westerly proper motion of $\sim20$\,km\,s$^{-1}$, and is likely to have escaped from the NGC\,7023 cluster, which is discussed below \citep{goodman2004}.  L1157-mm is a Class 0 protostar with an extremely strong molecular outflow \citep{chini2001}.  The outflow is highly visible in the 850-\um\ SCUBA-2 observations, and can be seen in Figures~\ref{fig:l1147_finding_chart} and \ref{fig:zoom_figure}, below.

\begin{figure*} 
\centering
\includegraphics[width=0.9\textwidth]{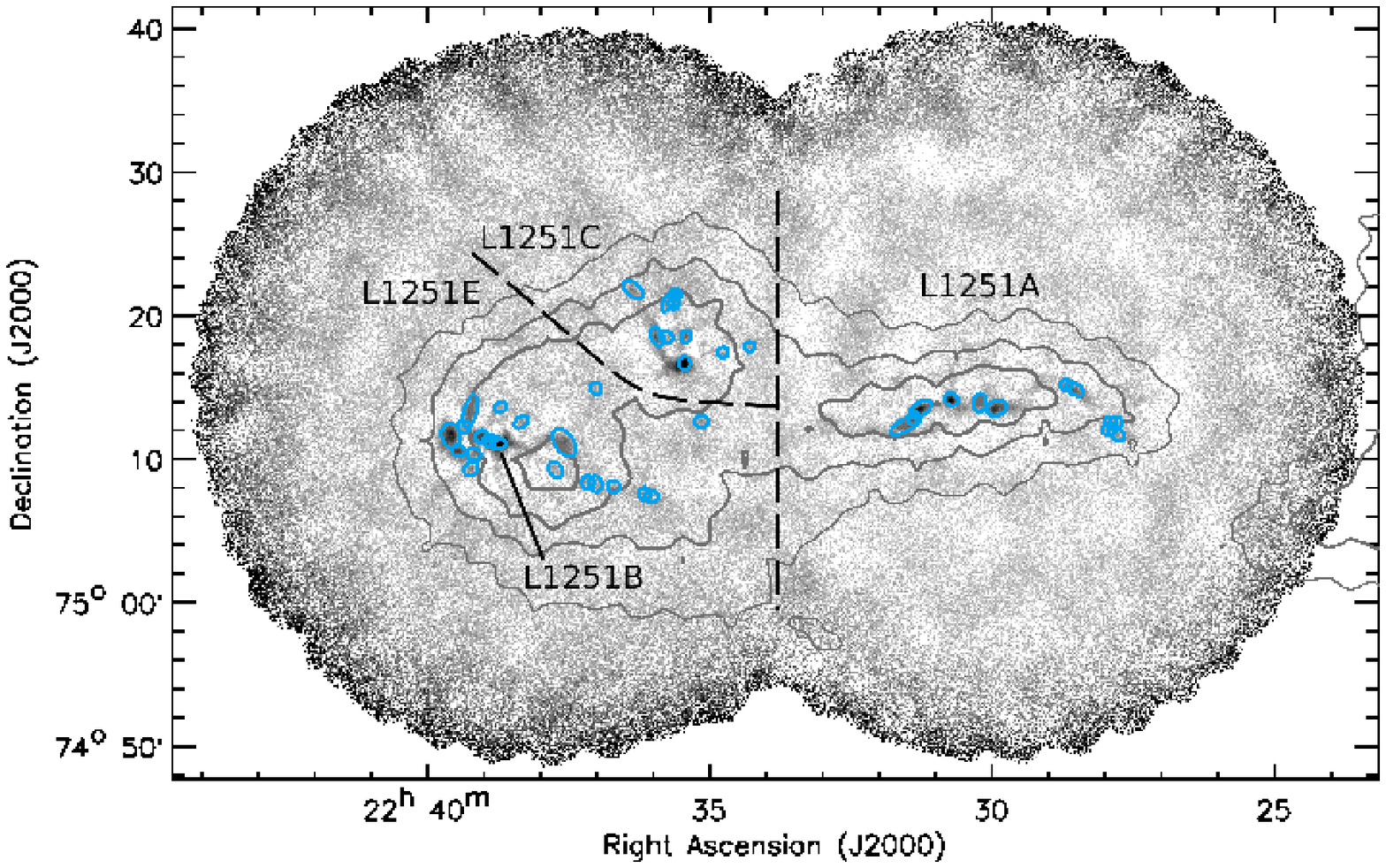}
\caption{SCUBA-2 850-\um\ observations of the L1251 region, with sources marked.  The ellipses show twice the FWHM size of each source.  Contours of $A_{\textsc{v}}$ of 0.5, 1.0, 1.5 and 2.0 magnitudes are shown for reference \citep{dobashi2005}.  The dashed lines mark approximately the boundaries between the L1251A, L1251C and L1251E regions.  The protostellar cluster L1251B is labelled.}
\label{fig:l1251_finding_chart} 
\end{figure*}

L1251, shown in Figure~\ref{fig:l1251_finding_chart}, consists of three submillimetre-bright regions, the western L1251A, the central L1251C, and the eastern L1251E \citep{sato1994}, surrounded by a network of filaments.  L1251 appears to be actively forming stars; in particular, there is a small group of young stars, L1251B, embedded within the L1251E region (\citealt{sato1994}; \citealt{lee2007}).  K09 suggest that star formation in L1251 may have been triggered or enhanced by the passage of the CFS $\sim 4$\,Myr ago.

L1228, shown in Figure~\ref{fig:l1228_finding_chart}, is a small cloud which is likely to be located on the near side of the CFS, unlike the other clouds discussed here \citep{kun2008}.  L1228 runs $\sim3^{\circ}$ along an approximately North-South axis.  As can be seen from the extinction contours in Figure~\ref{fig:ceph_finding_chart}, only the central part of L1228 was observed by the JCMT GBLS. K09 note that L1228 is at a location consistent with the current position of the CFS, and suggest that star formation here may be in the process of being enhanced by interaction with the shell.

Enlargements of the regions of significant 850-\um\ emission within each of the areas observed with SCUBA-2 are shown in Figure~\ref{fig:zoom_figure}.

\subsection{Source Extraction}
\label{sec:source_extraction}

\begin{figure} 
\centering
\includegraphics[width=0.49\textwidth]{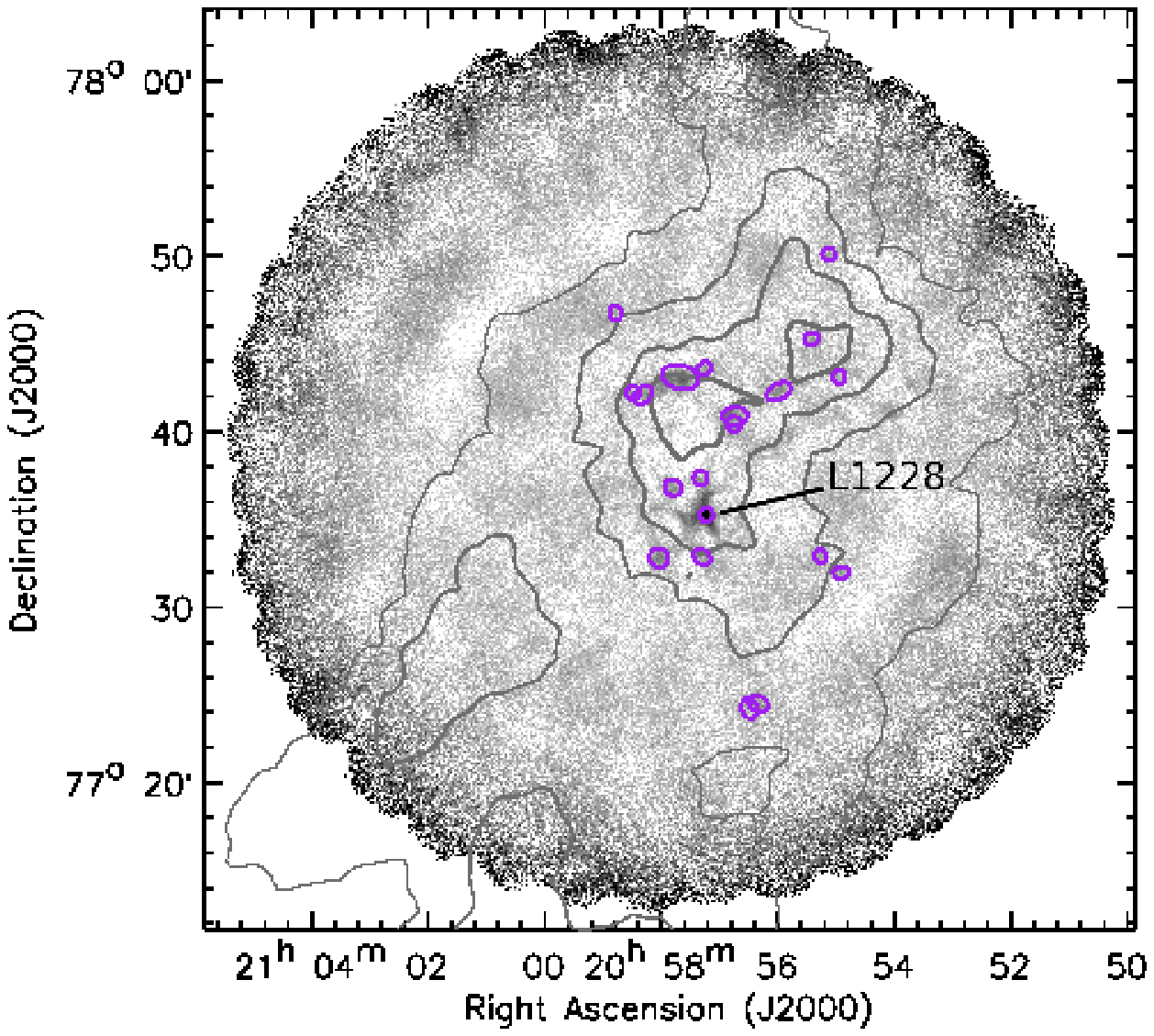}
\caption{SCUBA-2 850-\um\ observations of the L1228 region, with sources marked.  The ellipses show twice the FWHM size of each source.  Contours of $A_{\textsc{v}}$ of 0.5, 1.0, 1.5 and 2.0 magnitudes are shown for reference \citep{dobashi2005}.  The protostar L1228 is labelled.}
\label{fig:l1228_finding_chart} 
\end{figure}

We identified sources in the SCUBA-2 850-\um\ data using CSAR (Cardiff Source-extraction AlgoRithm; \citealt{kirk2013}). CSAR is a dendrogram-based source-finding algorithm, which was run in its non-hierarchical mode. CSAR identifies a source based on a peak in the emission map and assigns neighbouring pixels to that source if those pixels are above an assigned signal-to-noise criterion, and continues to do so until the contour level at which the source becomes confused with its neighbours is reached.

\begin{figure*} 
\centering
\includegraphics[width=\textwidth]{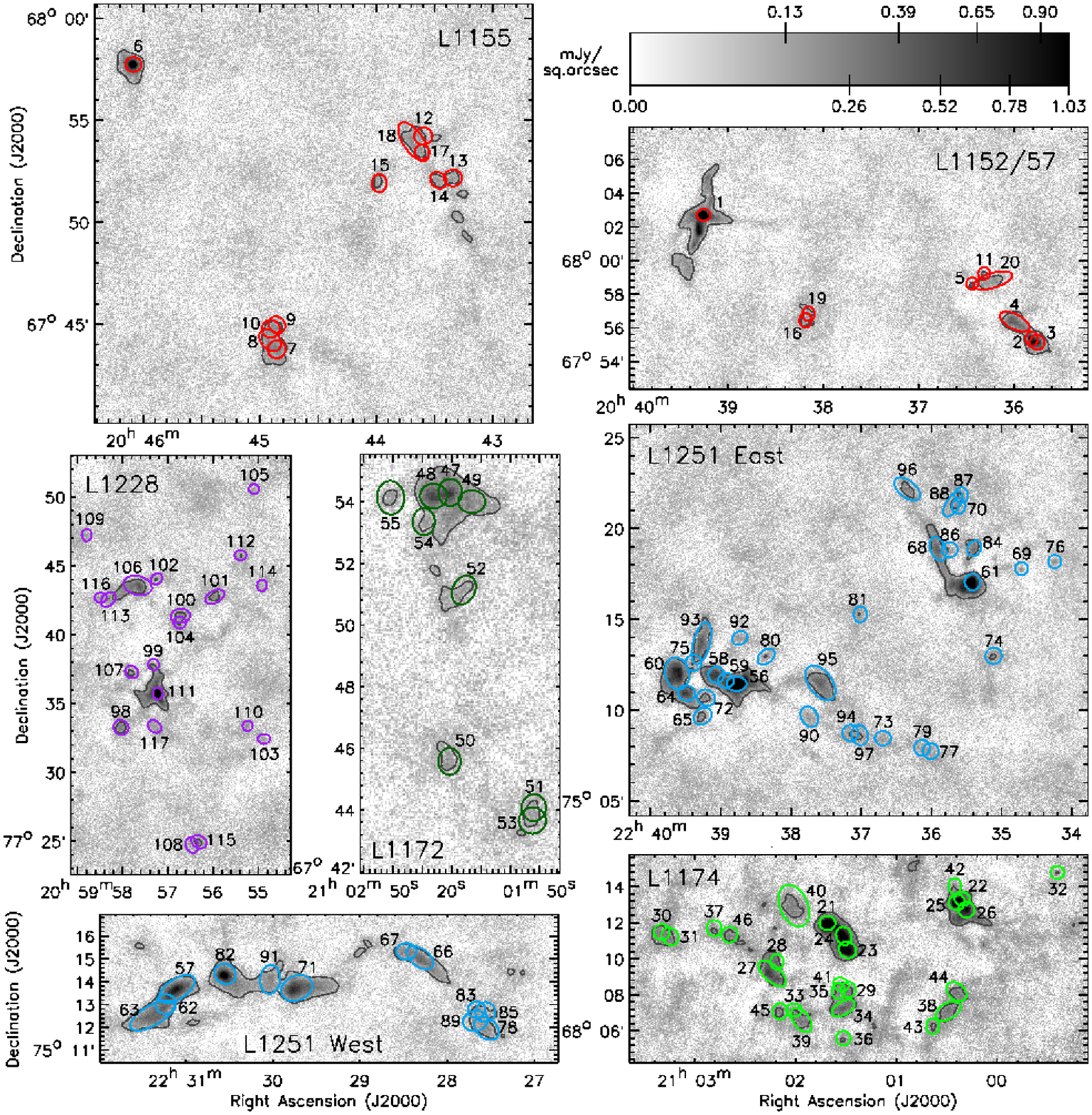}
\caption{SCUBA-2 850-\um\ observations of regions of significant emission.  Sources extracted in this work are numbered as in Table~\ref{tab:cepheus_sources} and colour-coded by region: red -- L1147/58; light green -- L1172; dark green -- L1174; blue -- L1251; purple -- L1228.  The data are shown in square-root scaling.}
\label{fig:zoom_figure} 
\end{figure*}

\setlength{\tabcolsep}{3pt} 
\begin{table*}
\centering
\caption{Sources identified in SCUBA-2 850-\um\ emission by CSAR and characterised using multiple-Gaussian fitting in the Cepheus Flare region.  FWHMs are as measured, without deconvolution.  Position angles are measured east of north, and listed for elliptical sources only.  `Model' peak and total flux density values are the results of the multiple-Gaussian fitting process, and are used in subsequent mass calculations.  `Photometry' peak and total flux density values are determined from aperture photometry, using the source sizes shown in Figure~\ref{fig:zoom_figure}, and hence flux density will be double-counted in some of these measurements.  `Photometry' measurements are used in the subsequent calculation of flux-ratio-determined temperature.  450-\um\ signal-to-noise is measured on peak value.  Sources marked with `*' overlap significantly with at least one other source, listed in the final column.  See text for details.}
\label{tab:cepheus_sources}
\begin{tabular}{c c c r@{\,$\times$\,}l c c c c c c c c c c c}
\toprule
 & & & \multicolumn{2}{c}{ } & & \multicolumn{2}{c}{Model} & \multicolumn{5}{c}{Photometry} & & & \\ \cmidrule(r){7-8} \cmidrule(l){9-13}
Source & RA & Dec & \multicolumn{2}{c}{FWHM} & Angle & F$_{\nu(850)}^{peak}$ & F$_{\nu(850)}^{total}$ & F$_{\nu(850)}^{peak}$ & F$_{\nu(450)}^{peak}$ & F$_{\nu(850)}^{total}$ & F$_{\nu(450)}^{total}$ & 450\um & Type & Region & Overlaps\\
Index & (J2000) & (J2000) & \multicolumn{2}{c}{(arcsec)} & ($^{\circ}$) & (mJy/sqa) & (Jy) & \multicolumn{2}{c}{(mJy/sqa)} & \multicolumn{2}{c}{(Jy)} & S/N & & & \\
\midrule
1 & 20:39:05.28 & 68:02:20.40 & 21.6 & 24.0 & 100.1 & 3.07 & 1.80 & 5.28 & 45.4 & 2.31 & 11.42 & 20.6 & P & 47/58 & -- \\
2* & 20:35:45.11 & 67:53:02.40 & 21.6 & 26.4 & 5.4 & 0.83 & 0.54 & 1.23 & 10.8 & 0.98 & 5.72 & 4.9 & P & 47/58 & 3 \\
3* & 20:35:41.76 & 67:52:48.00 & 26.4 & 26.4 & -- & 0.78 & 0.62 & 1.17 & 10.8 & 1.14 & 5.88 & 4.9 & C & 47/58 & 2 \\
4 & 20:35:54.72 & 67:54:10.80 & 57.8 & 26.4 & 152.0 & 0.39 & 0.68 & 0.49 & 8.7 & 0.91 & 4.49 & 4.0 & C & 47/58 & -- \\
5* & 20:36:18.96 & 67:56:42.00 & 21.6 & 21.6 & -- & 0.10 & 0.05 & 0.40 & 7.7 & 0.08 & 0.87 & 3.5 & P & 47/58 & 20 \\
6 & 20:45:53.28 & 67:57:39.60 & 23.4 & 21.6 & 170.2 & 1.66 & 0.95 & 2.87 & 24.4 & 1.27 & 7.87 & 11.1 & P & 47/58 & -- \\
7* & 20:44:48.48 & 67:43:12.00 & 26.4 & 26.4 & -- & 0.16 & 0.13 & 0.35 & 8.0 & 0.42 & 2.49 & 3.6 & C & 47/58 & 8 \\
8* & 20:44:51.60 & 67:43:40.80 & 37.3 & 26.4 & 125.0 & 0.14 & 0.15 & 0.35 & 8.0 & 0.51 & 2.73 & 3.6 & C & 47/58 & 7,10 \\
9* & 20:44:47.52 & 67:44:24.00 & 23.1 & 26.4 & 43.0 & 0.12 & 0.08 & 0.31 & 7.4 & 0.26 & 0.79 & 3.4 & C & 47/58 & 10 \\
10* & 20:44:50.88 & 67:44:13.20 & 26.4 & 26.4 & -- & 0.16 & 0.12 & 0.32 & 7.4 & 0.35 & 1.40 & 3.4 & C & 47/58 & 8,9 \\
11* & 20:36:10.80 & 67:57:14.40 & 21.6 & 21.6 & -- & 0.16 & 0.08 & 0.30 & 6.2 & 0.11 & 0.52 & 2.8 & P & 47/58 & 20 \\
12* & 20:43:24.48 & 67:53:09.60 & 26.4 & 25.7 & 170.0 & 0.07 & 0.06 & 0.32 & 7.2 & 0.24 & 0.77 & 3.3 & C & 47/58 & 18 \\
13* & 20:43:10.56 & 67:51:00.00 & 26.4 & 24.3 & 10.0 & 0.10 & 0.07 & 0.27 & 7.3 & 0.23 & 0.39 & 3.3 & C & 47/58 & 14 \\
14* & 20:43:18.24 & 67:50:56.40 & 21.6 & 26.4 & 37.5 & 0.10 & 0.06 & 0.27 & 7.9 & 0.2 & 1.25 & 3.6 & C & 47/58 & 13 \\
15 & 20:43:49.20 & 67:51:00.00 & 21.6 & 26.4 & 173.0 & 0.09 & 0.06 & 0.27 & 7.6 & 0.16 & 0.75 & 3.4 & C & 47/58 &  -- \\
16* & 20:38:06.96 & 67:55:30.00 & 26.4 & 21.6 & 80.0 & 0.06 & 0.04 & 0.26 & 6.1 & 0.19 & 0.28 & 2.8 & C & 47/58 & 19 \\
17* & 20:43:25.68 & 67:52:22.80 & 21.6 & 21.6 & 177.0 & 0.08 & 0.04 & 0.30 & 8.2 & 0.17 & 0.94 & 3.7 & C & 47/58 & 18 \\
18* & 20:43:29.76 & 67:52:55.20 & 66.4 & 29.8 & 121.7 & 0.13 & 0.30 & 0.32 & 8.2 & 0.68 & 3.36 & 3.7 & C & 47/58 & 12,17 \\
19* & 20:38:04.57 & 67:55:51.60 & 21.6 & 26.4 & 0.0 & 0.04 & 0.02 & 0.26 & 6.3 & 0.19 & 0.44 & 2.9 & C & 47/58 & 16 \\
20* & 20:36:05.76 & 67:56:45.60 & 72.2 & 26.4 & 19.6 & 0.20 & 0.43 & 0.29 & 7.6 & 0.48 & 2.35 & 3.4 & C & 47/58 & 5,11 \\
21 & 21:01:40.81 & 68:12:03.60 & 26.4 & 23.7 & 10.0 & 1.33 & 0.94 & 2.03 & 18.4 & 1.94 & 16.91 & 18.4 & C & L1174 & -- \\
22* & 21:00:19.68 & 68:13:22.80 & 22.8 & 26.4 & 100.0 & 0.76 & 0.52 & 1.98 & 14.6 & 1.47 & 8.07 & 14.6 & P & L1174 & 25,26 \\
23* & 21:01:28.80 & 68:10:33.60 & 29.8 & 26.4 & 147.2 & 1.61 & 1.43 & 1.77 & 19.0 & 2.14 & 18.08 & 19.0 & P & L1174 & 24 \\
24* & 21:01:30.96 & 68:11:20.40 & 31.2 & 25 & 112.8 & 1.47 & 1.30 & 1.61 & 17.3 & 2.20 & 19.26 & 17.3 & P & L1174 & 23 \\
25* & 21:00:23.04 & 68:13:12.00 & 26.4 & 26.4 & -- & 1.17 & 0.93 & 1.98 & 14.6 & 1.72 & 10.83 & 14.6 & P & L1174 & 22,26 \\
26* & 21:00:17.28 & 68:12:46.80 & 26.4 & 26.4 & -- & 0.99 & 0.78 & 1.12 & 7.2 & 1.36 & 7.33 & 7.2 & C & L1174 & 22,25 \\
27* & 21:02:13.92 & 68:09:14.40 & 57.2 & 26.4 & 127.9 & 0.62 & 1.05 & 0.70 & 7.3 & 1.48 & 11.64 & 7.3 & C & L1174 & 28 \\
28* & 21:02:11.04 & 68:09:54.00 & 21.6 & 26.4 & 10.0 & 0.20 & 0.13 & 0.47 & 5.6 & 0.46 & 3.58 & 5.6 & C & L1174 & 27 \\
29* & 21:01:28.32 & 68:08:20.40 & 26.4 & 24.6 & 71.8 & 0.24 & 0.18 & 0.41 & 4.8 & 0.44 & 2.71 & 4.8 & C & L1174 & 35,41 \\
30* & 21:03:20.16 & 68:11:31.20 & 26.4 & 26.4 & -- & 0.12 & 0.10 & 0.52 & 3.7 & 0.50 & 1.96 & 3.7 & C & L1174 & 31 \\
31* & 21:03:15.12 & 68:11:16.80 & 30.8 & 26.4 & 115.8 & 0.17 & 0.15 & 0.52 & 3.2 & 0.55 & 1.49 & 3.2 & C & L1174 & 30 \\
32 & 20:59:22.56 & 68:14:49.20 & 22.3 & 21.6 & 10.0 & 0.18 & 0.10 & 0.39 & 3.2 & 0.15 & 0.22 & 3.2 & P & L1174 & -- \\
33* & 21:02:00.72 & 68:07:12.00 & 26.4 & 21.6 & 172.8 & 0.10 & 0.06 & 0.36 & 3.4 & 0.29 & 0.68 & 3.4 & C & L1174 & 39 \\
34 & 21:01:31.20 & 68:07:19.20 & 42.3 & 21.9 & 24.4 & 0.26 & 0.27 & 0.48 & 5.2 & 0.73 & 4.69 & 5.2 & C & L1174 & -- \\
35* & 21:01:34.32 & 68:08:16.80 & 21.6 & 26.4 & 0.0 & 0.07 & 0.04 & 0.40 & 4.2 & 0.34 & 1.69 & 4.2 & P & L1174 & 29,41 \\
36 & 21:01:31.20 & 68:05:38.40 & 24 & 24 & -- & 0.22 & 0.15 & 0.35 & 3.8 & 0.25 & 1.44 & 3.8 & C & L1174 & -- \\
37 & 21:02:48.72 & 68:11:45.60 & 24.9 & 26.4 & 10.0 & 0.10 & 0.08 & 0.34 & 2.7 & 0.30 & 0.81 & 2.7 & C & L1174 & -- \\
38 & 21:00:28.56 & 68:07:08.40 & 45.5 & 26.4 & 40.3 & 0.25 & 0.34 & 0.43 & 4.7 & 0.78 & 3.55 & 4.7 & C & L1174 & -- \\
39* & 21:01:56.39 & 68:06:39.60 & 46.1 & 26.4 & 136.3 & 0.22 & 0.30 & 0.40 & 3.1 & 0.74 & 2.37 & 3.1 & C & L1174 & 33 \\
40 & 21:02:00.96 & 68:13:01.20 & 73.6 & 46 & 122.7 & 0.22 & 0.86 & 0.41 & 4.3 & 1.61 & 6.08 & 4.3 & C & L1174 & -- \\
41* & 21:01:32.64 & 68:08:38.40 & 26.4 & 21.6 & 161.8 & 0.23 & 0.15 & 0.41 & 4.2 & 0.33 & 1.96 & 4.2 & P & L1174 & 29,35 \\
42 & 21:00:24.25 & 68:14:06.00 & 26.4 & 21.6 & 95.6 & 0.10 & 0.07 & 0.37 & 3.9 & 0.26 & 0.38 & 3.9 & C & L1174 & -- \\
43 & 21:00:37.92 & 68:06:18.00 & 21.6 & 26.4 & 1.4 & 0.14 & 0.09 & 0.31 & 4.1 & 0.27 & 0.74 & 4.1 & C & L1174 & -- \\
44 & 21:00:23.52 & 68:08:13.20 & 38 & 26.4 & 148.7 & 0.18 & 0.20 & 0.40 & 4.1 & 0.57 & 2.61 & 4.1 & C & L1174 & -- \\
45 & 21:02:09.12 & 68:07:08.40 & 25.6 & 26.4 & 170.0 & 0.15 & 0.11 & 0.39 & 3.6 & 0.36 & 1.46 & 3.6 & C & L1174 & -- \\
46 & 21:02:39.60 & 68:11:24.00 & 27 & 26.4 & 175.2 & 0.17 & 0.14 & 0.38 & 3.5 & 0.41 & 0.70 & 3.5 & C & L1174 & -- \\
47* & 21:02:20.64 & 67:54:21.60 & 23.5 & 26.4 & 177.6 & 0.41 & 0.29 & 0.72 & 8.4 & 0.70 & 3.72 & 3.8 & P & L1172 & 48 \\
48* & 21:02:26.40 & 67:54:14.40 & 26.4 & 24 & 170.0 & 0.38 & 0.28 & 0.65 & 7.1 & 0.65 & 3.27 & 3.2 & P & L1172 & 47 \\
49 & 21:02:13.20 & 67:54:03.60 & 22.3 & 26.4 & 80.0 & 0.10 & 0.07 & 0.36 & 5.9 & 0.35 & 1.49 & 2.7 & C & L1172 & -- \\
50 & 21:02:20.64 & 67:45:36.00 & 21.6 & 26.4 & 170.0 & 0.09 & 0.06 & 0.27 & 8.0 & 0.18 & 1.14 & 3.7 & C & L1172 & -- \\
51* & 21:01:51.60 & 67:44:06.00 & 23.8 & 26.4 & 170.0 & 0.08 & 0.05 & 0.24 & 7.5 & 0.20 & 0.72 & 3.4 & C & L1172 & 53 \\
52 & 21:02:15.84 & 67:51:10.80 & 29.5 & 21.6 & 53.4 & 0.10 & 0.07 & 0.26 & 6.1 & 0.21 & 1.00 & 2.8 & C & L1172 & -- \\
53* & 21:01:52.08 & 67:43:40.80 & 26.4 & 25.7 & 10.0 & 0.08 & 0.06 & 0.24 & 7.5 & 0.21 & 0.61 & 3.4 & C & L1172 & 51 \\
54 & 21:02:29.76 & 67:53:24.00 & 21.6 & 26.4 & 170.0 & 0.07 & 0.05 & 0.24 & 5.6 & 0.16 & 0.31 & 2.5 & C & L1172 & -- \\
55 & 21:02:41.28 & 67:54:10.80 & 33 & 26.4 & 84.7 & 0.17 & 0.17 & 0.26 & 5.3 & 0.19 & 0.55 & 2.4 & C & L1172 & -- \\
56* & 22:38:47.04 & 75:11:31.20 & 33.4 & 23.8 & 9.9 & 3.54 & 3.19 & 4.22 & 34.9 & 3.67 & 23.33 & 19.7 & P & L1251 & 58,59 \\
\end{tabular}
\end{table*}
\addtocounter{table}{-1}
\begin{table*}
\centering
\caption{-- continued.}
\begin{tabular}{c c c r@{\,$\times$\,}l c c c c c c c c c c c}
\toprule
 & & & \multicolumn{2}{c}{ } & & \multicolumn{2}{c}{Model} & \multicolumn{5}{c}{Photometry} & & & \\ \cmidrule(r){7-8} \cmidrule(l){9-13}
Source & RA & Dec & \multicolumn{2}{c}{FWHM} & Angle & F$_{\nu(850)}^{peak}$ & F$_{\nu(850)}^{total}$ & F$_{\nu(850)}^{peak}$ & F$_{\nu(450)}^{peak}$ & F$_{\nu(850)}^{total}$ & F$_{\nu(450)}^{total}$ & 450\um & Type & Region & Overlaps\\ \cmidrule(r){9-10} \cmidrule(l){11-12}
Index & (J2000) & (J2000) & \multicolumn{2}{c}{(arcsec)} & ($^{\circ}$) & (mJy/sqa) & (Jy) & \multicolumn{2}{c}{(mJy/sqa)} & \multicolumn{2}{c}{(Jy)} & S/N & & & \\
\midrule
57* & 22:31:04.32 & 75:13:37.20 & 53.7 & 26.4 & 27.2 & 0.92 & 1.48 & 1.02 & 7.2 & 1.57 & 7.71 & 7.7 & P & L1251 & 62,63 \\
58* & 22:39:04.56 & 75:12:00.00 & 26.4 & 26.4 & -- & 0.76 & 0.60 & 0.70 & 8.4 & 0.98 & 3.47 & 4.8 & C & L1251 & 59 \\
59* & 22:38:56.16 & 75:11:42.00 & 26.4 & 24.3 & 177.6 & 0.65 & 0.47 & 3.50 & 28.9 & 1.08 & 5.37 & 16.3 & C & L1251 & 56,58 \\
60* & 22:39:38.40 & 75:12:03.60 & 53.3 & 39.0 & 93.2 & 0.80 & 1.89 & 0.92 & 6.6 & 2.51 & 7.00 & 3.7 & C & L1251 & 64 \\
61 & 22:35:22.56 & 75:17:06.00 & 27.1 & 25.3 & 80.0 & 1.89 & 1.47 & 2.81 & 23.0 & 2.18 & 14.62 & 13.0 & P & L1251 & -- \\
62* & 22:31:12.48 & 75:12:57.60 & 26.4 & 26.4 & -- & 0.34 & 0.27 & 0.69 & 4.5 & 0.66 & 3.50 & 4.8 & C & L1251 & 57,63 \\
63* & 22:31:22.08 & 75:12:28.80 & 65.7 & 26.4 & 19.9 & 0.41 & 0.81 & 0.45 & 4.5 & 1.03 & 6.41 & 4.8 & C & L1251 & 57,62 \\
64* & 22:39:30.00 & 75:10:58.80 & 28.7 & 24.4 & 158.9 & 0.58 & 0.46 & 0.73 & 6.4 & 0.80 & 2.61 & 3.6 & C & L1251 & 60 \\
65 & 22:39:16.08 & 75:09:43.20 & 32.3 & 25.6 & 49.5 & 0.19 & 0.18 & 0.35 & 5.1 & 0.31 & 0.51 & 2.9 & C & L1251 & -- \\
66* & 22:28:15.36 & 75:14:38.40 & 40.6 & 23.4 & 146.6 & 0.42 & 0.45 & 0.45 & 3.8 & 0.58 & 2.40 & 4.1 & C & L1251 & 67 \\
67* & 22:28:24.72 & 75:14:56.40 & 26.4 & 21.6 & 170.0 & 0.15 & 0.10 & 0.39 & 3.5 & 0.30 & 1.36 & 3.8 & C & L1251 & 66 \\
68* & 22:35:52.32 & 75:18:57.60 & 41.0 & 24.5 & 109.8 & 0.39 & 0.45 & 0.51 & 8.3 & 0.66 & 3.78 & 4.7 & C & L1251 & 86 \\
69 & 22:34:39.84 & 75:17:49.20 & 21.6 & 21.6 & -- & 0.14 & 0.08 & 0.27 & 4.2 & 0.08 & 0.74 & 2.4 & P & L1251 & -- \\
70* & 22:35:34.08 & 75:21:18.00 & 21.6 & 26.4 & 172.3 & 0.11 & 0.07 & 0.31 & 5.0 & 0.24 & 0.78 & 2.8 & C & L1251 & 87,88 \\
71 & 22:29:41.52 & 75:13:30.00 & 42.6 & 32.1 & 15.0 & 0.66 & 1.02 & 0.75 & 5.7 & 1.32 & 7.10 & 6.1 & C & L1251 & -- \\
72 & 22:39:13.20 & 75:10:44.40 & 30.4 & 26.4 & 163.5 & 0.18 & 0.17 & 0.36 & 6.4 & 0.33 & 1.08 & 3.6 & C & L1251 & -- \\
73 & 22:36:40.80 & 75:08:31.20 & 26.4 & 23.8 & 10.0 & 0.08 & 0.06 & 0.26 & 7.0 & 0.21 & 1.07 & 4.0 & C & L1251 & -- \\
74 & 22:35:04.80 & 75:13:01.20 & 27.7 & 26.4 & 29.7 & 0.14 & 0.12 & 0.27 & 5.2 & 0.29 & 2.19 & 2.9 & C & L1251 & -- \\
75* & 22:39:24.24 & 75:12:39.60 & 26.4 & 26.4 & -- & 0.15 & 0.12 & 0.32 & 6.1 & 0.28 & -0.17 & 3.4 & C & L1251 & -- \\
76 & 22:34:10.80 & 75:18:10.80 & 21.6 & 21.6 & -- & 0.14 & 0.07 & 0.26 & 4.7 & 0.12 & 0.31 & 2.7 & P & L1251 & 93 \\
77* & 22:35:59.76 & 75:07:48.00 & 26.4 & 26.4 & -- & 0.09 & 0.07 & 0.25 & 4.9 & 0.27 & 0.30 & 2.8 & C & L1251 & -- \\
78* & 22:27:31.44 & 75:11:24.00 & 36.1 & 25.8 & 143.7 & 0.16 & 0.17 & 0.31 & 2.9 & 0.36 & 1.62 & 3.1 & C & L1251 & 79 \\
79* & 22:36:07.20 & 75:07:58.80 & 25.7 & 26.4 & 0.0 & 0.08 & 0.06 & 0.24 & 5.5 & 0.23 & 0.06 & 3.1 & C & L1251 & 83,85,89 \\
80 & 22:38:21.60 & 75:13:01.20 & 31.8 & 21.6 & 39.1 & 0.16 & 0.13 & 0.26 & 6.0 & 0.18 & 1.32 & 3.4 & C & L1251 & 77 \\
81 & 22:37:00.00 & 75:15:21.60 & 26.4 & 21.6 & 96.8 & 0.09 & 0.06 & 0.28 & 4.9 & 0.21 & 0.24 & 2.8 & C & L1251 & -- \\
82 & 22:30:30.72 & 75:14:13.20 & 29.2 & 25.8 & 146.8 & 1.04 & 0.88 & 1.41 & 12.8 & 1.13 & 5.79 & 13.7 & P & L1251 & -- \\
83* & 22:27:37.69 & 75:12:14.40 & 23.1 & 21.6 & 10.0 & 0.10 & 0.06 & 0.31 & 2.9 & 0.18 & 0.90 & 3.1 & C & L1251 & -- \\
84 & 22:35:20.64 & 75:18:57.60 & 27.3 & 21.6 & 65.4 & 0.32 & 0.21 & 0.42 & 6.7 & 0.33 & 2.29 & 3.8 & P & L1251 & 78,85,89 \\
85* & 22:27:31.68 & 75:12:07.20 & 26.4 & 26.4 & 8.4 & 0.06 & 0.05 & 0.31 & 2.9 & 0.25 & 1.07 & 3.1 & C & L1251 & -- \\
86* & 22:35:42.00 & 75:18:54.00 & 26.4 & 24.9 & 10.0 & 0.14 & 0.10 & 0.35 & 6.6 & 0.25 & 1.69 & 3.8 & C & L1251 & 78,83,89 \\
87* & 22:35:31.44 & 75:21:54.00 & 22.1 & 26.4 & 10.0 & 0.09 & 0.06 & 0.28 & 4.7 & 0.25 & 0.92 & 2.7 & C & L1251 & 68 \\
88* & 22:35:38.88 & 75:21:25.20 & 47.6 & 21.6 & 64.1 & 0.08 & 0.09 & 0.31 & 5.3 & 0.42 & 1.63 & 3.0 & C & L1251 & 70,88 \\
89* & 22:27:38.87 & 75:11:45.60 & 33.3 & 26.4 & 3.3 & 0.10 & 0.10 & 0.31 & 3.3 & 0.31 & 1.69 & 3.5 & C & L1251 & 70,87 \\
90 & 22:37:44.16 & 75:09:43.20 & 35.8 & 26.4 & 129.3 & 0.16 & 0.17 & 0.27 & 5.5 & 0.28 & 0.20 & 3.1 & C & L1251 & 78,83,85 \\
91 & 22:29:59.76 & 75:13:55.20 & 38.1 & 26.4 & 73.9 & 0.20 & 0.23 & 0.35 & 4.3 & 0.49 & 3.50 & 4.6 & P & L1251 & -- \\
92 & 22:38:44.40 & 75:14:02.40 & 26.4 & 21.6 & 18.2 & 0.13 & 0.08 & 0.26 & 6.7 & 0.18 & 0.73 & 3.8 & C & L1251 & -- \\
93* & 22:39:17.52 & 75:13:44.40 & 71.2 & 27.5 & 77.1 & 0.42 & 0.94 & 0.50 & 7.5 & 0.85 & 0.66 & 4.2 & C & L1251 & 75 \\
94* & 22:37:08.88 & 75:08:49.20 & 26.4 & 26.4 & -- & 0.10 & 0.08 & 0.28 & 5.5 & 0.30 & 1.13 & 3.1 & C & L1251 & 97 \\
95 & 22:37:34.57 & 75:11:34.80 & 65.6 & 38.3 & 134.0 & 0.38 & 1.07 & 0.46 & 6.4 & 1.50 & 5.33 & 3.6 & C & L1251 & -- \\
96 & 22:36:18.72 & 75:22:15.60 & 50.3 & 27.2 & 130.8 & 0.22 & 0.34 & 0.35 & 5.3 & 0.56 & 0.30 & 3.0 & C & L1251 & -- \\
97* & 22:37:00.71 & 75:08:42.00 & 33.7 & 26.4 & 114.4 & 0.11 & 0.11 & 0.27 & 5.7 & 0.37 & 1.76 & 3.2 & C & L1251 & 94 \\
98 & 20:58:02.16 & 77:33:18.00 & 33.7 & 31.2 & 126.0 & 0.24 & 0.28 & 0.34 & 3.1 & 0.52 & 2.00 & 3.5 & C & L1228 & -- \\
99 & 20:57:18.24 & 77:37:51.60 & 24.0 & 24.0 & -- & 0.20 & 0.13 & 0.27 & 3.9 & 0.23 & 1.32 & 4.4 & C & L1228 & -- \\
100* & 20:56:41.28 & 77:41:24.00 & 42.7 & 30.0 & 15.8 & 0.18 & 0.26 & 0.30 & 3.5 & 0.60 & 2.16 & 3.9 & C & L1228 & 104 \\
101 & 20:55:54.24 & 77:42:46.80 & 44.5 & 26.4 & 18.2 & 0.19 & 0.25 & 0.31 & 3.2 & 0.48 & 1.86 & 3.6 & C & L1228 & -- \\
102 & 20:57:13.68 & 77:44:06.00 & 26.4 & 21.6 & 35.0 & 0.11 & 0.07 & 0.26 & 3.5 & 0.20 & 1.00 & 3.9 & C & L1228 & -- \\
103 & 20:54:49.44 & 77:32:24.00 & 26.4 & 21.6 & 170.0 & 0.12 & 0.07 & 0.23 & 2.3 & 0.15 & 0.36 & 2.6 & C & L1228 & -- \\
104* & 20:56:42.24 & 77:40:55.20 & 26.4 & 26.4 & -- & 0.12 & 0.09 & 0.29 & 3.5 & 0.32 & 1.11 & 3.9 & C & L1228 & 100 \\
105 & 20:54:59.04 & 77:50:34.80 & 21.6 & 21.6 & -- & 0.04 & 0.02 & 0.24 & 3.1 & 0.14 & 0.52 & 3.5 & C & L1228 & -- \\
106 & 20:57:39.60 & 77:43:37.20 & 62.8 & 40.7 & 175.2 & 0.36 & 1.03 & 0.53 & 4.4 & 1.33 & 5.85 & 4.9 & C & L1228 & -- \\
107 & 20:57:47.76 & 77:37:19.20 & 30.0 & 26.5 & 147.4 & 0.17 & 0.15 & 0.30 & 2.5 & 0.31 & 0.64 & 2.8 & C & L1228 & -- \\
108* & 20:56:27.37 & 77:24:43.20 & 35.2 & 26.4 & 118.4 & 0.07 & 0.07 & 0.26 & 2.8 & 0.29 & 0.29 & 3.1 & C & L1228 & 115 \\
109 & 20:58:49.68 & 77:47:16.80 & 26.2 & 21.6 & 100.0 & 0.07 & 0.04 & 0.21 & 2.7 & 0.19 & 0.30 & 3.0 & C & L1228 & -- \\
110 & 20:55:11.28 & 77:33:21.60 & 21.8 & 21.6 & 55.0 & 0.09 & 0.05 & 0.24 & 2.7 & 0.16 & 0.28 & 3.0 & C & L1228 & -- \\
111 & 20:57:12.24 & 77:35:45.60 & 26.4 & 21.8 & 100.5 & 2.20 & 1.43 & 3.65 & 29.0 & 2.03 & 9.00 & 32.2 & P & L1228 & -- \\
112 & 20:55:18.48 & 77:45:46.80 & 25.1 & 21.6 & 0.5 & 0.06 & 0.04 & 0.31 & 3.3 & 0.21 & 1.04 & 3.7 & C & L1228 & -- \\
113* & 20:58:19.92 & 77:42:36.00 & 37.4 & 25.9 & 35.0 & 0.11 & 0.12 & 0.25 & 3.2 & 0.32 & 1.42 & 3.5 & C & L1228 & 116 \\
114 & 20:54:49.44 & 77:43:33.60 & 21.6 & 24.6 & 173.7 & 0.08 & 0.05 & 0.22 & 2.7 & 0.13 & 0.33 & 3.0 & C & L1228 & -- \\
115* & 20:56:18.00 & 77:24:57.60 & 34.6 & 26.4 & 142.4 & 0.16 & 0.17 & 0.28 & 2.6 & 0.38 & 0.51 & 2.9 & C & L1228 & 108 \\
116* & 20:58:30.24 & 77:42:43.20 & 26.4 & 21.6 & 3.0 & 0.11 & 0.07 & 0.25 & 3.2 & 0.19 & 1.01 & 3.5 & C & L1228 & 113 \\
117 & 20:57:17.05 & 77:33:21.60 & 31.9 & 24.1 & 129.7 & 0.13 & 0.11 & 0.26 & 2.6 & 0.28 & 0.75 & 2.9 & C & L1228 & -- \\

\bottomrule
\end{tabular}
\end{table*}
\setlength{\tabcolsep}{6pt}

We gridded each of the SCUBA-2 850-\um\ maps onto 6-arcsec pixels before performing the source extraction.  The LR1 default pixel size is 3~arcsec at 850~\um.  However, the beam noise resulting from this oversampling of the data prevented CSAR from finding closed contours around extended low-surface-brightness sources.  Source extraction was performed on the low-variance regions of the maps, where the variance, as measured in the variance array, was very low, $<0.005$\,(mJy/arcsec$^{2}$)$^{2}$.  The criteria chosen for a robustly-detected source were a peak flux density $F_{\nu}^{peak}\geq 5\,\sigma$ and a minimum of a $1\,\sigma$ drop in flux density between adjacent sources (i.e. a local minimum in flux density at least $1\,\sigma$ less than peak value of the fainter of the two sources), where $\sigma$ is the RMS noise level of the data. We adopted $1\,\sigma$ values of 0.041 mJy/arcsec$^{2}$ in L1174, and 0.028 mJy/arcsec$^{2}$ elsewhere on 6-arcsec pixels at 850~\um.

\begin{table*} 
\centering
\caption{The protostellar sources in our catalogue, with their identification and evolutionary class from K09, and alternative identifications.  With the exception of L1157-mm and stars with an identification of the form $X\! X$ Cep, alternative identifications are given in the following order of preference: IRAS Point or Faint Source Catalogs (IRAS -- \citealt{beichman1988}; \citealt{moshir1992}), 2MASS All-Sky Catalog of Point Sources (2MASS -- \citealt{cutri2003}), \emph{Spitzer} Gould Belt Survey (SSTgbs -- K09).  For the L1251B cluster, designations from \citet{lee2006} are also given.  K09 identifications given in brackets indicate an offset between our source central coordinates and the coordinates of the K09 source greater than the JCMT 850-\um\ beam size, but less than the radius of the source as listed in Table~\ref{tab:cepheus_sources}.}
\label{tab:cepheus_protostars}
\begin{tabular}{c c c c }
\hline
Source ID & K09 ID & K09 Class & Alternative ID \\
\hline
1 & 134 & I & L1157-mm \\
2 & 1 & I & IRAS 20353+6742 \\
5 & 3 & II & IRAS 20359+6475 \\
6 & 135 & I & PV Cep \\
11 & 2 & II & 2MASS J2036+1165+6757093 \\
22 & 17 & I & SSTgbs J2100207+6813169 \\
22 & (100) & F & SSTgbs J2100224+6813042 \\
23 & 27 & II & 2MASS J21012637+6810385 \\
23 & 137 & II & SSTgbs J2101271+6810380 \\
24 & 34 & I & 2MASS J21013280+6811204 \\
25 & 18 & I & SSTgbs J2100221+6812585 \\
25 & (100) & F & SSTgbs J2100224+6813042 \\
32 & 15 & II & FT Cep \\
35 & (104) & F & PW Cep \\
41 & (104) & F & PW Cep \\
47 & 49 & I & IRAS 21017+6742 \\
48 & 50 & F & SSTgbs J2102273+6754186 \\
48 & (53) & II & 2MASS J21022993+6754083 \\
56 & 89 & I & 2MASS J22384282+7511369; L1251B IRS 4 \\
56 & 90 & I & SSTgbs J2238469+7511337; L1251B IRS 1 \\
56 & 92 & I & 2MASS J22385287+7511235; L1251B IRS 2 \\
56 & 107 & III & IRAS 22376+7455; L1251B IRS 3 \\
56 & 108 & III & SSTgbs J2238440+7511266; L1251B IRS 5 \\
56 & 109 & II & 2MASS J22384807+7511488; L1251B IRS 6 \\
57 & 68 & II & SSTgbs J2231056+7513372 \\
61 & 143 & I & IRAS 22343+7501 \\
69 & 69 & F & 2MASS J22344051+7517444 \\
76 & 142 & F & IRAS 22331+7502 \\
82 & 67 & I & SSTgbs J2230318+7514094 \\
84 & (76) & II & 2MASS J22351668+7518471 \\
91 & 66 & I & IRAS 22290+7458 \\
111 & 9 & F & IRAS 20582+7724; L1228 \\
\hline
\end{tabular}
\end{table*}

We identified 27 sources in L1147/58, 26 sources in L1174, 9 sources in L1172, 42 sources in L1251 and 20 sources in L1228.  Of the 27 sources in L1147/58, 7 were rejected due to their being associated with the L1157-mm outflow and hence likely to be artefacts resulting from CO contamination in the SCUBA-2 850-\um\ data.  Rejecting these left us with 20 reliable sources in L1147/58.  There were no sources in other regions which we considered likely to be CO artefacts.

The sources we identified in each cloud are shown in detail on Figure~\ref{fig:zoom_figure}, and on Figures~\ref{fig:l1147_finding_chart}--\ref{fig:l1228_finding_chart} for reference.  Due to the significant overlap between some of the sources, we fitted each source using a multiple-Gaussian fitting routine.  This model, which utilises the fitting routine \emph{mpfit} \citep{mpfit}, is described in detail by \citet{pattle2015}.  The fitting routine models the flux density of sources in crowded regions by fitting a two-dimensional Gaussian and an inclined-plane background to each of a set of associated sources simultaneously.  Sources are considered to be neighbours if they are separated by less than twice the FWHM of the larger source.  Groups to be fitted simultaneously are defined such that each source in a group is a neighbour to at least one other source in the group, and no source has any neighbours outside of the group.  The source positions and sizes determined using CSAR were supplied as initial input to the fitting routine.  The Gaussian fitting routine was constrained such that for each source, the $x$ and $y$ coordinates of the source could vary no more than 6 arcsec from their initial position, the source semi-major and semi-minor axes could not vary by more than 10 per\,cent of their initial values, and the source position angle could vary by no more than 5$^{\circ}$.  The total flux of the source was constrained to be positive.

It should be noted that while the Gaussian model is a popular and widely-used choice of model for characterising the properties of starless cores (e.g. \citealt{wardthompson1994}; \citealt{hirota2002}; \citealt{enoch2008}; \citealt{gomez2014}; \citealt{pattle2015}), the underlying geometry of a starless core is unlikely to obey a Gaussian distribution, instead typically showing a flat central plateau and power-law wings (e.g. \citealt{alves2001}), which may be characterised using a Bonnor-Ebert geometry (\citealt{ebert1955}; \citealt{bonnor1956}) or a Plummer-like geometry (\citealt{plummer1911}; \citealt{whitworth1996}).  However, the Gaussian model remains a very useful tool for characterising the properties of ensembles of starless cores, due to its analytic tractability.  Gaussian fits may underestimate core size \citep{terebey1993}, typically fitting the central plateau of the core and underestimating the extent of the wings.  However, two arguments mitigate against the effect of this on our core sample.  Firstly, if we were significantly underestimating the size of our cores, then we would expect to see positive annuli of unfitted flux in the residuals of our Gaussian fits, which is not the case.  Secondly, it can be shown that for Gaussian and Plummer-like distributions with the same total mass and central density, the characteristic sizes of the two distributions are very similar, $R_{\rm Plummer}=1.17\,R_{\rm Gaussian}$, where $R_{\rm Plummer}$ is the characteristic size of the Plummer-like distribution and $R_{\rm Gaussian}$ is the Gaussian width (assuming a power-law index for the Plummer-like distribution of 4; see \citealt{pattle2016} for derivations of the masses of the two distributions).  This suggests that we are unlikely to be significantly underestimating the size of our cores by using a Gaussian distribution.

In this analysis we are concerned with the ensemble properties of starless cores in the Cepheus molecular cloud, and so require an approximate size and mass estimate for each core, which can be usefully provided by a Gaussian fit to the data.  Future detailed analyses of the interior structure of starless cores using SCUBA-2 data will require more sophisticated modelling of core geometries.

For each of our sources, Table~\ref{tab:cepheus_sources} lists the position, angular size, orientation, peak and total flux densities, signal-to-noise ratio at 450~\um, classification as starless or protostellar, and the region in which the source is located.  For the 850-\um\ flux densities, both the modelled values and the values determined from aperture photometry are listed.  For the 450-\um\ flux densities, only values determined from aperture photometry are listed.  The aperture photometry measurements were made using elliptical apertures with major and minor axis diameters of twice the FWHM values listed in Table~\ref{tab:cepheus_sources}, and as shown in Figures~\ref{fig:l1147_finding_chart}--\ref{fig:zoom_figure}.

Prior to aperture photometry measurements being made, the 450-\um\ data were convolved to match the resolution of the 850-\um\ data using a convolution kernel constructed as described by \citet{pattle2015}, following the method proposed by \citet{aniano2011}.  The convolution kernel used was constructed using the SCUBA-2 450-\um\ and 850-\um\ beam models given by \citet{dempsey2013}.  However, the peak 450-\um\ flux densities, and the 450-\um\ signal-to-noise ratios, were determined from the original, non-convolved map.

We emphasise that due to the significant overlap between many of the sources (see Figures~\ref{fig:l1147_finding_chart}--\ref{fig:l1228_finding_chart}), there will be double-counting of pixels in many of the flux densities determined from aperture photometry, and the flux density values determined from aperture photometry are likely to be overestimates of the amount of emission associated with a source.  The aperture-photometry-determined peak flux densities are those of the brightest pixel in the source aperture, and so may be identical for overlapping sources.  The modelling-determined peak flux densities are the best-fit peak flux densities assuming the sources obey Gaussian distributions.

It can be seen in Table~\ref{tab:cepheus_sources} that the aperture-photometry-determined 850-\um\ flux densities are typically $\sim 30$\,per\,cent higher than the model 850-\um\ flux densities in isolated (non-overlapping) sources.  This is due to the inclined-plane background which is fitted to the measured emission along with the Gaussian source model.

Note that the 450-\um\ and 850-\um\ aperture-photometry-determined flux densities do not have the SCUBA-2 aperture photometry corrections discussed by \citet{dempsey2013} applied to them.  The SCUBA-2 aperture photometry corrections are determined for point sources, and account for flux in the secondary beam of the JCMT not enclosed by a small aperture (the JCMT's secondary beam has a FWHM of 25\,arcsec at 450\um\ and 48\,arcsec at 850~\um; see \citealt{dempsey2013}) .  We do not use these aperture photometry corrections in this work, as their applicability to either extended sources or non-circular apertures is not certain. Furthermore, for aperture diameters from 25 to 50 arcsec (i.e. the vast majority of our sources), the 450-\um\ and 850-\um\ aperture photometry corrections are identical, while for sources larger than 50 arcsec, the difference between the 450-\um\ and 850-\um\ corrections is very small, typically $\sim 1$\,per\,cent \citep{dempsey2013}.  As we are using the aperture-photometry-determined flux densities only as a ratio quantity (see \ref{sec:scuba2_temps}, below), use of the aperture photometry corrections (or otherwise) should not affect our results.  However, as aperture-photometry-corrected flux densities may be useful for other purposes, we direct the reader to \citet{dempsey2013} for further information.

In the analysis that follows, we use the best-fit model 850-\um\ total flux densities in order to determine source masses.  The ratio of the 450-\um\ and 850-\um\ aperture-photometry-determined total flux densities are used to determine source temperatures, for those sources with a peak 450-\um\ signal-to-noise ratio $\geq 3$ -- see Section~\ref{sec:scuba2_temps} below.

\subsection{Source Characterisation}

Of the 117 sources in our Cepheus Flare catalogue, 23 were associated with at least one protostar in the K09 Spitzer catalogue.  (The K09 catalogue lists 143 protostellar sources and covers all of the regions observed with SCUBA-2.)  Protostar associations are listed in Table~\ref{tab:cepheus_protostars}, along with the K09 source with which they are associated, the evolutionary class of that source (as determined from the infrared spectral index, $\alpha_{\textsc{ir}}$ by K09), and alternative identifications.  It should be noted that due to the $\sim 300$\,pc distances to the Cepheus Flare clouds, a single SCUBA-2 source in Cepheus may be associated with more than one protostellar object.  In particular, source 56 contains six embedded sources, the L1251B group.

The K09 Spitzer catalogue is the only systematic protostar catalogue produced from \emph{Spitzer} observations of Cepheus to date.  We compared the K09 results to a more limited recent study by \citet{dunham2013}, who revise the classification of a number of protostars detected by the \emph{Spitzer} c2d \citep{evans2009} and Gould Belt (P.I. L. Allen; see, e.g., K09) surveys.  \citet{dunham2013} extend the methods developed by \citet{evans2009} for correcting protostellar fluxes and luminosities for extinction, providing corrected classifications for \emph{Spitzer}-detected protostars associated with at least one submillimetre detection at wavelengths $\geq 350$\um.  \citet{dunham2013} include 20 protostars in Cepheus in their sample, all of which are included in the K09 catalogue.  The \citet{dunham2013} extinction corrections alter the $\alpha_{\textsc{ir}}$ classification of two of the 20 stars which they consider in Cepheus, both of which we detect with SCUBA-2: Source 48 (K09 Source 50), which is reclassified from Flat to Class II, and Source 91 (K09 Source 66), which is reclassified from Class I to Flat.  Source 111 (K09 source 9) also moves from Class I to the Class I/Flat boundary.  As these extinction-corrected classifications are available for only a subset of the \emph{Spitzer} sources in Cepheus, and as only a small minority of the source classifications are changed by the correction for extinction, we continue to use the classifications given in K09 throughout this work.  This is in order to use a self-consistent set of source classifications.

\begin{table*} 
\centering
\caption{Properties of the sources.  See text for discussion.}
\label{tab:cepheus_properties}
\begin{tabular}{c D{,}{\,\pm\,}{-1} D{,}{\,\pm\,}{-1} D{,}{\,\pm\,}{-1} D{,}{\,\pm\,}{-1} D{,}{\,\pm\,}{-1} D{,}{\,\pm\,}{-1} c}
\toprule
Source & \multicolumn{1}{c}{$T_{\rm Herschel}$} & \multicolumn{1}{c}{$T_{\text{SCUBA-2}}$} & \multicolumn{1}{c}{M($T_{\rm Herschel}$)} & \multicolumn{1}{c}{M($T_{\text{SCUBA-2}}$)} & \multicolumn{1}{c}{$N$(H$_{2}$)} & \multicolumn{1}{c}{$n$(H$_{2}$)} & Deconv. \\
Index & \multicolumn{1}{c}{(K)} & \multicolumn{1}{c}{(K)} & \multicolumn{1}{c}{(M$_{\odot}$)} & \multicolumn{1}{c}{(M$_{\odot}$)} &\multicolumn{1}{c}{($\times10^{21}$ cm$^{-2}$)} & \multicolumn{1}{c}{($\times10^{4}$ cm$^{-3}$)} & FWHM (pc) \\
\midrule
1 & 14.8,0.2 & 10.7,1.7 & 2.43,0.07 & 4.36,1.35 & 42.66,1.30 & 36.81,1.12 & 0.028 \\
2 & 12.4,0.3 & \multicolumn{1}{c}{--} & 0.99,0.05 & \multicolumn{1}{c}{--} & 14.91,0.80 & 11.93,0.64 & 0.030 \\
3 & 11.5,0.2 & \multicolumn{1}{c}{--} & 1.31,0.06 & \multicolumn{1}{c}{--} & 14.77,0.66 & 10.21,0.45 & 0.035 \\
4 & 10.6,0.1 & \multicolumn{1}{c}{--} & 1.70,0.07 & \multicolumn{1}{c}{--} & 7.17,0.31 & 3.03,0.13 & 0.057 \\
5 & 13.1,0.2 & \multicolumn{1}{c}{--} & 0.08,0.01 & \multicolumn{1}{c}{--} & 1.78,0.30 & 1.68,0.28 & 0.026 \\
6 & 16.2,0.2 & 13.5,2.8 & 1.11,0.03 & 1.51,0.55 & 20.19,0.58 & 17.75,0.51 & 0.028 \\
7 & 11.7,0.0 & \multicolumn{1}{c}{--} & 0.26,0.02 & \multicolumn{1}{c}{--} & 2.93,0.19 & 2.02,0.13 & 0.035 \\
8 & 11.5,0.1 & \multicolumn{1}{c}{--} & 0.32,0.02 & \multicolumn{1}{c}{--} & 2.29,0.16 & 1.26,0.09 & 0.044 \\
9 & 11.7,0.1 & \multicolumn{1}{c}{--} & 0.17,0.02 & \multicolumn{1}{c}{--} & 2.31,0.25 & 1.76,0.19 & 0.032 \\
10 & 11.5,0.1 & \multicolumn{1}{c}{--} & 0.26,0.02 & \multicolumn{1}{c}{--} & 2.93,0.21 & 2.02,0.14 & 0.035 \\
11 & 12.3,0.0 & \multicolumn{1}{c}{--} & 0.15,0.01 & \multicolumn{1}{c}{--} & 3.21,0.30 & 3.03,0.28 & 0.026 \\
12 & 11.5,0.1 & \multicolumn{1}{c}{--} & 0.12,0.02 & \multicolumn{1}{c}{--} & 1.36,0.19 & 0.96,0.14 & 0.035 \\
13 & 11.5,0.0 & \multicolumn{1}{c}{--} & 0.15,0.02 & \multicolumn{1}{c}{--} & 1.94,0.21 & 1.42,0.15 & 0.033 \\
14 & 11.6,0.0 & \multicolumn{1}{c}{--} & 0.13,0.02 & \multicolumn{1}{c}{--} & 1.95,0.24 & 1.56,0.19 & 0.030 \\
15 & 12.4,0.0 & \multicolumn{1}{c}{--} & 0.11,0.01 & \multicolumn{1}{c}{--} & 1.64,0.20 & 1.32,0.16 & 0.030 \\
16 & 13.3,0.1 & \multicolumn{1}{c}{--} & 0.06,0.01 & \multicolumn{1}{c}{--} & 0.94,0.18 & 0.75,0.15 & 0.030 \\
17 & 11.5,0.1 & \multicolumn{1}{c}{--} & 0.09,0.02 & \multicolumn{1}{c}{--} & 1.79,0.33 & 1.68,0.31 & 0.026 \\
18 & 11.3,0.1 & \multicolumn{1}{c}{--} & 0.65,0.03 & \multicolumn{1}{c}{--} & 2.05,0.11 & 0.75,0.04 & 0.066 \\
19 & 13.2,0.0 & \multicolumn{1}{c}{--} & 0.04,0.01 & \multicolumn{1}{c}{--} & 0.57,0.18 & 0.45,0.14 & 0.030 \\
20 & 11.5,0.1 & \multicolumn{1}{c}{--} & 0.91,0.04 & \multicolumn{1}{c}{--} & 2.99,0.14 & 1.12,0.05 & 0.065 \\
21 & 16.7,0.5 & 23.3,9.6 & 0.82,0.05 & 0.50,0.29 & 13.68,0.86 & 11.52,0.72 & 0.029 \\
22 & 13.1,0.1 & 11.9,2.1 & 0.68,0.02 & 0.81,0.27 & 12.00,0.39 & 10.40,0.34 & 0.028 \\
23 & 27.6,1.5 & 21.8,8.2 & 0.60,0.05 & 0.83,0.45 & 7.31,0.58 & 5.25,0.42 & 0.034 \\
24 & 19.2,1.1 & 23.5,9.8 & 0.91,0.08 & 0.68,0.40 & 11.17,1.04 & 8.07,0.75 & 0.034 \\
25 & 13.2,0.1 & 13.7,2.9 & 1.19,0.03 & 1.12,0.41 & 17.06,0.41 & 13.30,0.32 & 0.031 \\
26 & 11.9,0.3 & 11.6,2.0 & 1.21,0.07 & 1.27,0.42 & 17.40,0.94 & 13.57,0.73 & 0.031 \\
27 & 20.3,1.4 & 18.9,6.0 & 0.68,0.08 & 0.75,0.36 & 3.70,0.42 & 1.78,0.20 & 0.051 \\
28 & 21.8,0.6 & 18.7,6.0 & 0.08,0.01 & 0.09,0.05 & 1.45,0.17 & 1.31,0.15 & 0.027 \\
29 & 26.9,1.1 & \multicolumn{1}{c}{--} & 0.08,0.01 & \multicolumn{1}{c}{--} & 1.22,0.14 & 1.00,0.11 & 0.030 \\
30 & 15.5,0.2 & \multicolumn{1}{c}{--} & 0.09,0.01 & \multicolumn{1}{c}{--} & 1.36,0.17 & 1.06,0.13 & 0.031 \\
31 & 15.1,0.2 & \multicolumn{1}{c}{--} & 0.16,0.02 & \multicolumn{1}{c}{--} & 1.81,0.18 & 1.27,0.13 & 0.035 \\
32 & 14.1,0.1 & \multicolumn{1}{c}{--} & 0.11,0.01 & \multicolumn{1}{c}{--} & 2.87,0.33 & 2.97,0.34 & 0.024 \\
33 & 25.3,0.3 & \multicolumn{1}{c}{--} & 0.03,0.01 & \multicolumn{1}{c}{--} & 0.57,0.10 & 0.51,0.09 & 0.027 \\
34 & 19.5,0.4 & 14.0,3.1 & 0.18,0.02 & 0.32,0.12 & 1.81,0.15 & 1.17,0.10 & 0.038 \\
35 & 28.4,1.2 & \multicolumn{1}{c}{--} & 0.02,0.01 & \multicolumn{1}{c}{--} & 0.34,0.10 & 0.31,0.09 & 0.027 \\
36 & 18.2,0.4 & \multicolumn{1}{c}{--} & 0.11,0.01 & \multicolumn{1}{c}{--} & 2.09,0.22 & 1.87,0.20 & 0.027 \\
37 & 15.5,0.0 & \multicolumn{1}{c}{--} & 0.07,0.01 & \multicolumn{1}{c}{--} & 1.16,0.16 & 0.94,0.13 & 0.030 \\
38 & 19.5,0.7 & \multicolumn{1}{c}{--} & 0.23,0.02 & \multicolumn{1}{c}{--} & 1.67,0.16 & 0.92,0.09 & 0.044 \\
39 & 24.2,0.6 & \multicolumn{1}{c}{--} & 0.15,0.01 & \multicolumn{1}{c}{--} & 1.07,0.09 & 0.58,0.05 & 0.045 \\
40 & 18.8,1.3 & \multicolumn{1}{c}{--} & 0.62,0.08 & \multicolumn{1}{c}{--} & 1.38,0.18 & 0.43,0.06 & 0.079 \\
41 & 31.0,1.1 & \multicolumn{1}{c}{--} & 0.05,0.01 & \multicolumn{1}{c}{--} & 1.01,0.12 & 0.91,0.11 & 0.027 \\
42 & 13.2,0.1 & \multicolumn{1}{c}{--} & 0.08,0.01 & \multicolumn{1}{c}{--} & 1.61,0.27 & 1.45,0.25 & 0.027 \\
43 & 21.2,0.3 & \multicolumn{1}{c}{--} & 0.05,0.01 & \multicolumn{1}{c}{--} & 1.02,0.14 & 0.92,0.13 & 0.027 \\
44 & 18.9,0.5 & \multicolumn{1}{c}{--} & 0.15,0.01 & \multicolumn{1}{c}{--} & 1.30,0.13 & 0.80,0.08 & 0.040 \\
45 & 23.6,0.5 & \multicolumn{1}{c}{--} & 0.06,0.01 & \multicolumn{1}{c}{--} & 0.88,0.11 & 0.70,0.09 & 0.031 \\
46 & 16.3,0.3 & \multicolumn{1}{c}{--} & 0.12,0.01 & \multicolumn{1}{c}{--} & 1.69,0.17 & 1.29,0.13 & 0.032 \\
47 & 12.0,0.0 & \multicolumn{1}{c}{--} & 0.45,0.01 & \multicolumn{1}{c}{--} & 7.56,0.21 & 6.40,0.18 & 0.029 \\
48 & 12.1,0.1 & \multicolumn{1}{c}{--} & 0.42,0.02 & \multicolumn{1}{c}{--} & 6.81,0.28 & 5.68,0.23 & 0.029 \\
49 & 11.8,0.1 & \multicolumn{1}{c}{--} & 0.10,0.01 & \multicolumn{1}{c}{--} & 1.91,0.23 & 1.69,0.20 & 0.028 \\
50 & 12.9,0.0 & \multicolumn{1}{c}{--} & 0.07,0.01 & \multicolumn{1}{c}{--} & 1.42,0.19 & 1.29,0.17 & 0.027 \\
51 & 12.7,0.1 & \multicolumn{1}{c}{--} & 0.08,0.01 & \multicolumn{1}{c}{--} & 1.27,0.18 & 1.06,0.15 & 0.029 \\
52 & 12.5,0.0 & \multicolumn{1}{c}{--} & 0.10,0.01 & \multicolumn{1}{c}{--} & 1.61,0.19 & 1.34,0.15 & 0.029 \\
53 & 12.6,0.1 & \multicolumn{1}{c}{--} & 0.08,0.01 & \multicolumn{1}{c}{--} & 1.20,0.16 & 0.95,0.13 & 0.031 \\
54 & 12.3,0.1 & \multicolumn{1}{c}{--} & 0.07,0.01 & \multicolumn{1}{c}{--} & 1.33,0.22 & 1.20,0.20 & 0.027 \\
55 & 12.5,0.1 & \multicolumn{1}{c}{--} & 0.24,0.01 & \multicolumn{1}{c}{--} & 2.58,0.15 & 1.73,0.10 & 0.036 \\
56 & 14.9,0.3 & 13.9,3.0 & 3.61,0.14 & 4.07,1.51 & 39.85,1.52 & 27.26,1.04 & 0.036 \\
57 & 11.0,0.1 & 10.7,1.6 & 2.89,0.09 & 3.09,0.95 & 15.59,0.48 & 7.46,0.23 & 0.051 \\
58 & 11.1,0.1 & \multicolumn{1}{c}{--} & 1.16,0.03 & \multicolumn{1}{c}{--} & 15.37,0.42 & 11.51,0.31 & 0.032 \\
59 & 13.1,0.5 & 10.8,1.7 & 0.66,0.06 & 0.96,0.30 & 9.83,0.84 & 7.81,0.67 & 0.031 \\
60 & 10.3,0.2 & \multicolumn{1}{c}{--} & 4.22,0.21 & \multicolumn{1}{c}{--} & 14.75,0.72 & 5.68,0.28 & 0.063 \\
61 & 18.4,0.1 & 14.9,3.5 & 1.19,0.02 & 1.67,0.66 & 16.04,0.27 & 12.15,0.21 & 0.032 \\
62 & 10.8,0.1 & \multicolumn{1}{c}{--} & 0.55,0.03 & \multicolumn{1}{c}{--} & 7.30,0.33 & 5.46,0.25 & 0.032 \\
63 & 11.0,0.2 & \multicolumn{1}{c}{--} & 1.57,0.07 & \multicolumn{1}{c}{--} & 6.72,0.29 & 2.86,0.12 & 0.057 \\
\end{tabular}
\end{table*}
\addtocounter{table}{-1}
\begin{table*}
\centering
\caption{-- continued.}
\begin{tabular}{c D{,}{\,\pm\,}{-1} D{,}{\,\pm\,}{-1} D{,}{\,\pm\,}{-1} D{,}{\,\pm\,}{-1} D{,}{\,\pm\,}{-1} D{,}{\,\pm\,}{-1} c}
\toprule
Source & \multicolumn{1}{c}{$T_{\rm Herschel}$} & \multicolumn{1}{c}{$T_{\text{SCUBA-2}}$} & \multicolumn{1}{c}{M($T_{\rm Herschel}$)} & \multicolumn{1}{c}{M($T_{\text{SCUBA-2}}$)} & \multicolumn{1}{c}{$N$(H$_{2}$)} & \multicolumn{1}{c}{$n$(H$_{2}$)} & Deconv. \\
Index & \multicolumn{1}{c}{(K)} & \multicolumn{1}{c}{(K)} & \multicolumn{1}{c}{(M$_{\odot}$)} & \multicolumn{1}{c}{(M$_{\odot}$)} &\multicolumn{1}{c}{($\times10^{21}$ cm$^{-2}$)} & \multicolumn{1}{c}{($\times10^{4}$ cm$^{-3}$)} & FWHM (pc) \\
\midrule
64 & 10.2,0.1 & \multicolumn{1}{c}{--} & 1.05,0.04 & \multicolumn{1}{c}{--} & 13.82,0.46 & 10.31,0.35 & 0.033 \\
65 & 11.4,0.1 & \multicolumn{1}{c}{--} & 0.33,0.02 & \multicolumn{1}{c}{--} & 3.45,0.25 & 2.30,0.16 & 0.036 \\
66 & 11.1,0.1 & \multicolumn{1}{c}{--} & 0.87,0.03 & \multicolumn{1}{c}{--} & 7.58,0.26 & 4.62,0.16 & 0.040 \\
67 & 11.4,0.1 & \multicolumn{1}{c}{--} & 0.18,0.01 & \multicolumn{1}{c}{--} & 3.18,0.26 & 2.76,0.23 & 0.028 \\
68 & 11.4,0.1 & \multicolumn{1}{c}{--} & 0.81,0.03 & \multicolumn{1}{c}{--} & 6.63,0.23 & 3.91,0.14 & 0.041 \\
69 & 13.0,0.1 & \multicolumn{1}{c}{--} & 0.11,0.01 & \multicolumn{1}{c}{--} & 2.64,0.27 & 2.70,0.28 & 0.024 \\
70 & 12.4,0.1 & \multicolumn{1}{c}{--} & 0.11,0.01 & \multicolumn{1}{c}{--} & 2.03,0.22 & 1.76,0.19 & 0.028 \\
71 & 10.7,0.1 & 11.6,2.0 & 2.09,0.07 & 1.80,0.59 & 11.78,0.39 & 5.75,0.19 & 0.050 \\
72 & 11.1,0.1 & \multicolumn{1}{c}{--} & 0.32,0.02 & \multicolumn{1}{c}{--} & 3.46,0.24 & 2.36,0.16 & 0.036 \\
73 & 12.5,0.0 & \multicolumn{1}{c}{--} & 0.09,0.01 & \multicolumn{1}{c}{--} & 1.35,0.17 & 1.09,0.14 & 0.030 \\
74 & 12.0,0.0 & \multicolumn{1}{c}{--} & 0.19,0.01 & \multicolumn{1}{c}{--} & 2.36,0.16 & 1.71,0.12 & 0.034 \\
75 & 11.1,0.0 & \multicolumn{1}{c}{--} & 0.23,0.02 & \multicolumn{1}{c}{--} & 3.09,0.20 & 2.31,0.15 & 0.032 \\
76 & 14.4,0.1 & \multicolumn{1}{c}{--} & 0.09,0.01 & \multicolumn{1}{c}{--} & 2.12,0.23 & 2.17,0.24 & 0.024 \\
77 & 13.0,0.0 & \multicolumn{1}{c}{--} & 0.10,0.01 & \multicolumn{1}{c}{--} & 1.30,0.14 & 0.97,0.10 & 0.032 \\
78 & 12.1,0.1 & \multicolumn{1}{c}{--} & 0.28,0.02 & \multicolumn{1}{c}{--} & 2.50,0.18 & 1.54,0.11 & 0.039 \\
79 & 12.9,0.0 & \multicolumn{1}{c}{--} & 0.09,0.01 & \multicolumn{1}{c}{--} & 1.18,0.15 & 0.90,0.11 & 0.032 \\
80 & 12.8,0.0 & \multicolumn{1}{c}{--} & 0.18,0.01 & \multicolumn{1}{c}{--} & 2.47,0.17 & 1.87,0.13 & 0.032 \\
81 & 13.5,0.0 & \multicolumn{1}{c}{--} & 0.08,0.01 & \multicolumn{1}{c}{--} & 1.38,0.18 & 1.20,0.16 & 0.028 \\
82 & 11.9,0.1 & 11.1,1.8 & 1.49,0.05 & 1.70,0.54 & 17.66,0.56 & 12.53,0.39 & 0.034 \\
83 & 11.7,0.0 & \multicolumn{1}{c}{--} & 0.10,0.01 & \multicolumn{1}{c}{--} & 2.15,0.28 & 2.07,0.27 & 0.025 \\
84 & 12.2,0.1 & \multicolumn{1}{c}{--} & 0.34,0.02 & \multicolumn{1}{c}{--} & 5.74,0.26 & 4.85,0.22 & 0.029 \\
85 & 11.8,0.1 & \multicolumn{1}{c}{--} & 0.08,0.01 & \multicolumn{1}{c}{--} & 1.06,0.18 & 0.79,0.13 & 0.032 \\
86 & 12.0,0.1 & \multicolumn{1}{c}{--} & 0.17,0.01 & \multicolumn{1}{c}{--} & 2.41,0.20 & 1.88,0.16 & 0.031 \\
87 & 12.6,0.1 & \multicolumn{1}{c}{--} & 0.09,0.01 & \multicolumn{1}{c}{--} & 1.48,0.20 & 1.26,0.17 & 0.029 \\
88 & 12.3,0.1 & \multicolumn{1}{c}{--} & 0.15,0.02 & \multicolumn{1}{c}{--} & 1.16,0.13 & 0.67,0.08 & 0.042 \\
89 & 11.9,0.1 & \multicolumn{1}{c}{--} & 0.17,0.02 & \multicolumn{1}{c}{--} & 1.66,0.18 & 1.07,0.12 & 0.038 \\
90 & 11.8,0.1 & \multicolumn{1}{c}{--} & 0.29,0.02 & \multicolumn{1}{c}{--} & 2.57,0.15 & 1.57,0.09 & 0.040 \\
91 & 11.9,0.2 & \multicolumn{1}{c}{--} & 0.38,0.03 & \multicolumn{1}{c}{--} & 3.10,0.22 & 1.82,0.13 & 0.041 \\
92 & 12.5,0.0 & \multicolumn{1}{c}{--} & 0.13,0.01 & \multicolumn{1}{c}{--} & 2.24,0.21 & 1.94,0.18 & 0.028 \\
93 & 10.8,0.1 & \multicolumn{1}{c}{--} & 1.90,0.07 & \multicolumn{1}{c}{--} & 7.13,0.27 & 2.84,0.11 & 0.061 \\
94 & 12.1,0.0 & \multicolumn{1}{c}{--} & 0.13,0.01 & \multicolumn{1}{c}{--} & 1.66,0.16 & 1.24,0.12 & 0.032 \\
95 & 11.4,0.1 & \multicolumn{1}{c}{--} & 1.96,0.06 & \multicolumn{1}{c}{--} & 5.57,0.17 & 1.93,0.06 & 0.070 \\
96 & 12.4,0.1 & \multicolumn{1}{c}{--} & 0.54,0.02 & \multicolumn{1}{c}{--} & 3.02,0.11 & 1.48,0.05 & 0.050 \\
97 & 12.3,0.1 & \multicolumn{1}{c}{--} & 0.17,0.01 & \multicolumn{1}{c}{--} & 1.65,0.14 & 1.05,0.09 & 0.038 \\
98 & 11.7,0.1 & \multicolumn{1}{c}{--} & 0.22,0.01 & \multicolumn{1}{c}{--} & 3.79,0.16 & 3.25,0.14 & 0.028 \\
99 & 11.8,0.1 & \multicolumn{1}{c}{--} & 0.10,0.01 & \multicolumn{1}{c}{--} & 3.85,0.25 & 4.97,0.32 & 0.019 \\
100 & 11.6,0.1 & \multicolumn{1}{c}{--} & 0.20,0.01 & \multicolumn{1}{c}{--} & 2.76,0.14 & 2.11,0.11 & 0.032 \\
101 & 12.4,0.0 & \multicolumn{1}{c}{--} & 0.17,0.01 & \multicolumn{1}{c}{--} & 2.58,0.12 & 2.07,0.10 & 0.030 \\
102 & 12.2,0.1 & \multicolumn{1}{c}{--} & 0.05,0.01 & \multicolumn{1}{c}{--} & 2.09,0.25 & 2.71,0.32 & 0.019 \\
103 & 13.1,0.1 & \multicolumn{1}{c}{--} & 0.05,0.01 & \multicolumn{1}{c}{--} & 1.88,0.20 & 2.44,0.26 & 0.019 \\
104 & 11.8,0.1 & \multicolumn{1}{c}{--} & 0.07,0.01 & \multicolumn{1}{c}{--} & 2.12,0.19 & 2.37,0.21 & 0.022 \\
105 & 13.2,0.0 & \multicolumn{1}{c}{--} & 0.01,0.00 & \multicolumn{1}{c}{--} & 0.66,0.25 & 1.01,0.38 & 0.016 \\
106 & 11.3,0.2 & \multicolumn{1}{c}{--} & 0.86,0.04 & \multicolumn{1}{c}{--} & 5.39,0.23 & 2.78,0.12 & 0.047 \\
107 & 11.7,0.1 & \multicolumn{1}{c}{--} & 0.12,0.01 & \multicolumn{1}{c}{--} & 2.91,0.19 & 2.99,0.19 & 0.024 \\
108 & 13.3,0.1 & \multicolumn{1}{c}{--} & 0.04,0.01 & \multicolumn{1}{c}{--} & 0.89,0.13 & 0.82,0.12 & 0.026 \\
109 & 13.7,0.1 & \multicolumn{1}{c}{--} & 0.02,0.00 & \multicolumn{1}{c}{--} & 1.00,0.18 & 1.31,0.23 & 0.019 \\
110 & 13.3,0.1 & \multicolumn{1}{c}{--} & 0.03,0.00 & \multicolumn{1}{c}{--} & 1.70,0.26 & 2.58,0.39 & 0.016 \\
111 & 12.5,0.1 & 9.8,1.4 & 0.98,0.02 & 1.57,0.46 & 38.47,0.64 & 49.75,0.83 & 0.019 \\
112 & 12.5,0.0 & \multicolumn{1}{c}{--} & 0.03,0.00 & \multicolumn{1}{c}{--} & 1.17,0.21 & 1.58,0.29 & 0.018 \\
113 & 12.3,0.1 & \multicolumn{1}{c}{--} & 0.08,0.01 & \multicolumn{1}{c}{--} & 1.57,0.13 & 1.42,0.12 & 0.027 \\
114 & 13.0,0.0 & \multicolumn{1}{c}{--} & 0.03,0.00 & \multicolumn{1}{c}{--} & 1.32,0.21 & 1.81,0.28 & 0.018 \\
115 & 13.0,0.0 & \multicolumn{1}{c}{--} & 0.11,0.01 & \multicolumn{1}{c}{--} & 2.20,0.11 & 2.06,0.11 & 0.026 \\
116 & 12.5,0.1 & \multicolumn{1}{c}{--} & 0.05,0.01 & \multicolumn{1}{c}{--} & 1.90,0.22 & 2.47,0.28 & 0.019 \\
117 & 11.8,0.0 & \multicolumn{1}{c}{--} & 0.09,0.01 & \multicolumn{1}{c}{--} & 2.22,0.17 & 2.33,0.17 & 0.023 \\
\bottomrule
\end{tabular}
\end{table*}

Temperatures for each of our sources were supplied by Di Francesco et al. (2016, in prep.).  These temperatures were determined from SED fitting to the 160--500-\um\ \emph{Herschel} observations taken toward the Cepheus Flare as part of the \emph{Herschel} Gould Belt Survey (GBS) \citep{andre2007}.  The \emph{Herschel} data were fitted by Di Francesco et al. (2016, in prep.) using the model
\begin{equation}
F_{\nu}=\frac{MB_{\nu}(T)\kappa_{\nu}}{D^{2}},
\label{eq:herschel_sed}
\end{equation}
where $F_{\nu}$ is the measured flux density, $B_{\nu}(T)$ is the Planck function, $M$ is the source mass, $D$ is the distance to the source, and the \citet{beckwith1990} parameterisation of dust opacity,  $\kappa_{\nu}\,=\,0.1(\nu/10^{12}{\rm Hz})^{\beta}$\,cm$^{2}$g$^{-1}$, is used, assuming a dust emissivity index $\beta=2.0$.  We use this model for dust opacity and this value of dust emissivity index throughout the rest of this work, in order to combine the \emph{Herschel} data with our own in a self-consistent manner.  This model of dust properties, adopted by the \emph{Herschel} GBS, is described in detail by, e.g., \citet{konyves2015}.

We note that combined SCUBA-2 and \emph{Herschel} observations have demonstrated variations in $\beta$ toward star-forming regions in the range $\beta=1.6-2.0$ \citep{sadavoy2013} and $\beta=1.0-2.7$ \citep{chen2016}, with lower values of $\beta$ typically observed toward protostellar cores.  \citet{sadavoy2013} found $\beta\approx 2.0$ toward filaments and moderately dense material, suggesting that $\beta=2.0$ is a representative value for the starless cores in our sample, but may be less appropriate for the protostellar sources which we observe.

The SED fitting process is described in detail by \citet{konyves2015}.  It must be emphasised that the only quantity derived from the \emph{Herschel} data which we use is the source temperature.  We discuss our own determinations of source masses -- from their SCUBA-2 850\um\ flux densities -- below.  All of our sources were observed as part of the \emph{Herschel} GBS. However, the sources on the western edge of L1152 are on the very edge of the \emph{Herschel} field, and hence their temperatures may be less reliable than those in other parts of the region.  Temperatures of cores without embedded sources are typically in the range 9--15\,K, except in the NGC\,7023 (L1174) region, where temperatures of up to $\sim50$\,K are measured.

Source masses were determined using the \citet{hildebrand1983} formulation
\begin{equation}
M=\frac{F_{\nu}^{total}(850\upmu{\rm m})D^{2}}{\kappa_{\nu(850\upmu{\rm m})}B_{\nu(850\upmu{\rm m})}(T)},
\label{eq:mass}
\end{equation}
where $F_{\nu}(850\upmu{\rm m})$ is the best-fit model flux density at 850~\um, $D$ is the source distance as listed in Table~\ref{tab:cepheus_regions}, $B_{\nu(850\upmu{\rm m})}(T)$ is the Planck function, and $\kappa_{\nu(850\upmu{\rm m})}$ is the dust mass opacity as parameterised by \citet{beckwith1990}, where $\beta$ is again taken to be 2.0.  Note that equation~\ref{eq:mass} is functionally identical to equation~\ref{eq:herschel_sed}.  However, we determine the masses of our sources using our model SCUBA-2 850-\um\ flux densities and \emph{Herschel}-determined temperatures, whereas equation~\ref{eq:herschel_sed} was used to determine best-fit source temperatures by fitting the flux densities measured in four \emph{Herschel} wavebands to each pixel in the \emph{Herschel} observations.  We use the mean fitted temperature in the pixels enclosed by the source apertures shown on Figure~\ref{fig:zoom_figure}.  Detection of a SCUBA-2 source does not necessarily mean that there is a \emph{Herschel} source at the same position.

Mean source volume densities were determined using the equation
\begin{equation}
  n({\rm H}_{2})=\frac{M}{\mu m_{\textsc{h}}}\frac{1}{\frac{4}{3}\pi R^{3}},
\label{eq:density}
\end{equation}
where $R$ is the equivalent deconvolved mean FWHM of the source.  The equivalent deconvolved mean FWHM was taken to be the geometric mean of the best-fit major and minor FWHMs, with the JCMT 850\um\ effective beam FWHM (14.1\arcsec) subtracted in quadrature. The mean molecular weight $\mu$ was taken to be 2.86, assuming that the gas is $\sim 70$\% H$_{2}$ by mass \citep{kirk2013}.  We give densities in terms of H$_{2}$ rather than of total density of particles as the density ranges traced by different molecular species are typically expressed in terms of H$_{2}$ number density (see, e.g., \citealt{difrancesco2007}).  The range of densities traced by isotopologues of CO is relevant to our determination of core stability in Section~\ref{sec:virial}, below, and so we express particle number density in terms of $n({\rm H}_{2})$ throughout this work.  Assuming a typical mean particle mass of 2.3 amu, our H$_{2}$ number densities can be converted to total gas particle number densities by multiplication by a factor of 1.24.

Mean source column densities were determined using the equation
\begin{equation}
N({\rm H}_{2})=\frac{M}{\mu m_{\textsc{h}}}\frac{1}{\pi R^{2}},
\label{eq:col_density}
\end{equation}
with symbols defined as above.

\begin{figure} 
\centering
\includegraphics[width=0.47\textwidth]{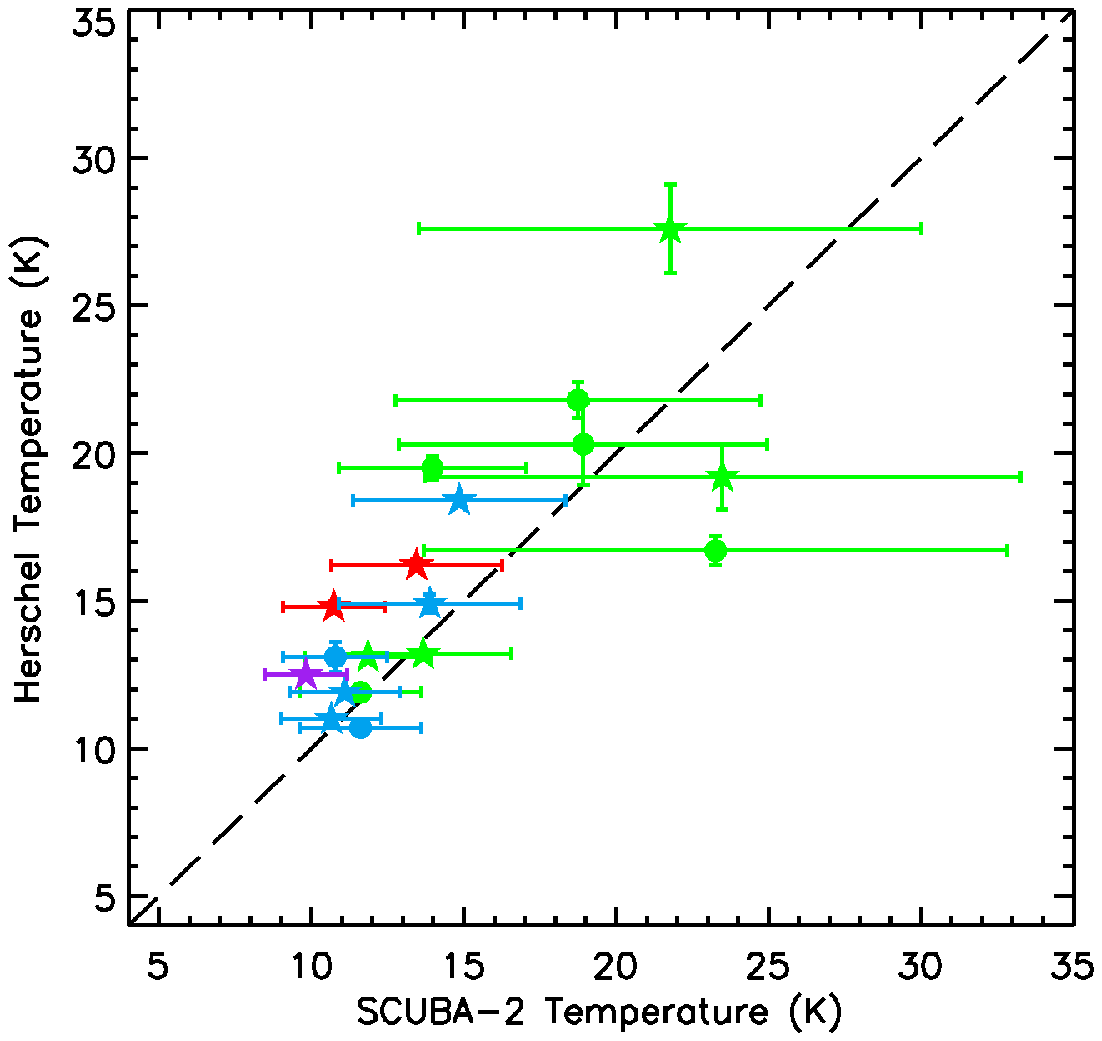}
\caption{Comparison of SCUBA-2- and \emph{Herschel}-derived (Di Francesco et al., 2016, in prep.) temperatures.  Red sources lie in L1147/58, light green sources in L1174, blue sources in L1251, and purple sources in L1228.  Error bars show 1-$\sigma$ uncertainties as listed in Table~\ref{tab:cepheus_properties}.  The dashed line is the 1:1 line.}
\label{fig:scuba2_temps}
\includegraphics[width=0.47\textwidth]{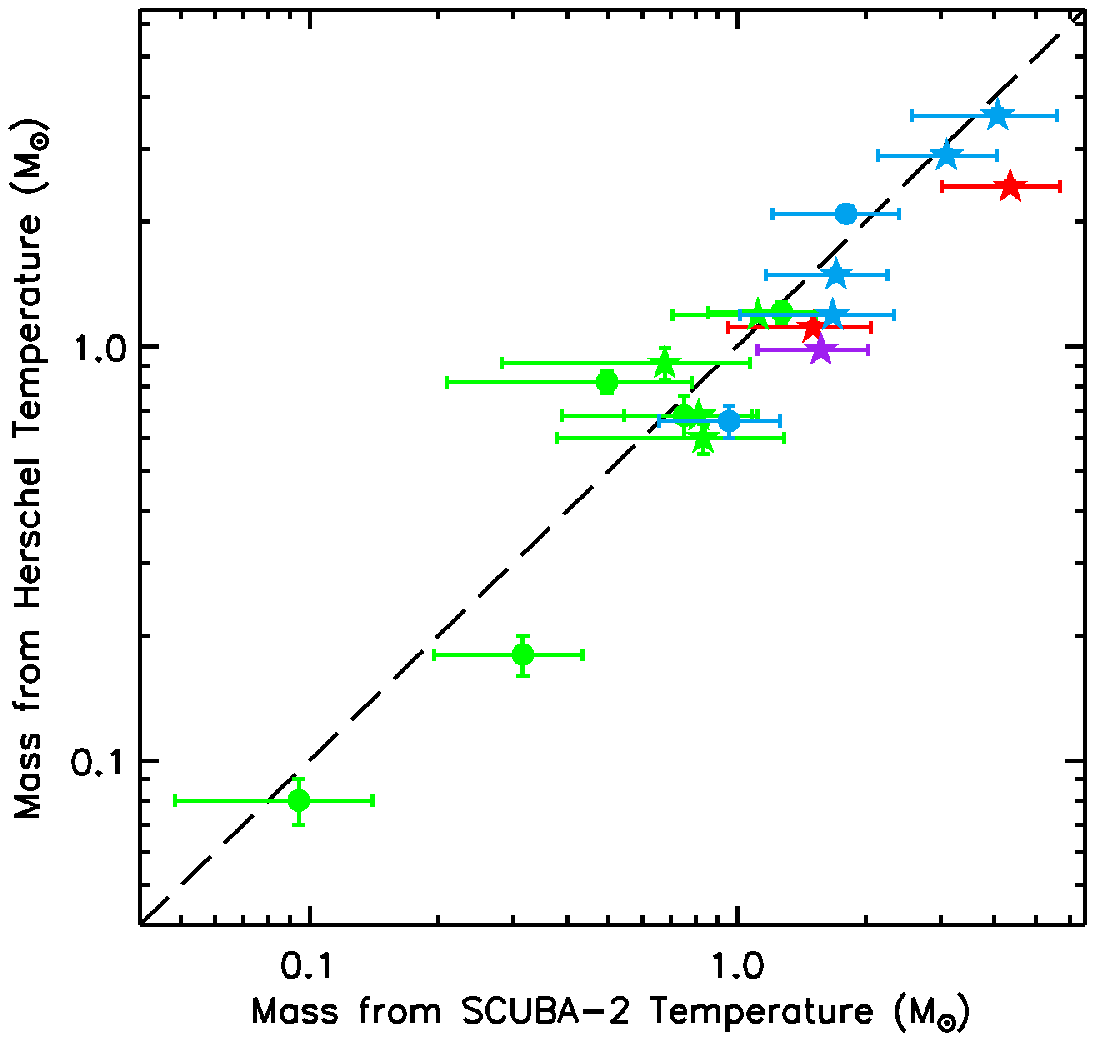}
\caption{Comparison of source masses determined using SCUBA-2- and \emph{Herschel}-derived (Di Francesco et al., 2016, in prep.) temperatures.  Red sources lie in L1147/58, light green sources in L1174, blue sources in L1251, and purple sources in L1228.  Error bars show 1-$\sigma$ uncertainties as listed in Table~\ref{tab:cepheus_properties}.  The dashed line is the 1:1 line.}
\label{fig:scuba2_masses}
\end{figure}

The derived properties of our sources: temperature, mass, column density, volume density, and deconvolved FWHM, are listed in Table~\ref{tab:cepheus_properties}.  For the protostellar sources in our catalogue, the temperatures, and hence the masses, determined from the dust emission are those of the protostellar envelopes, and not of the protostars themselves (see, e.g., \citealt{pattle2015}).  The modified blackbody model used to fit temperatures is applicable only to envelope-dominated sources; the temperatures and masses determined for the Class II and III protostars in our catalogue (11 sources, listed in Table~\ref{tab:cepheus_protostars}) may not be representative.

\subsection{SCUBA-2-derived temperatures and masses}
\label{sec:scuba2_temps}

\begin{figure*} 
\centering
\includegraphics[width=0.8\textwidth]{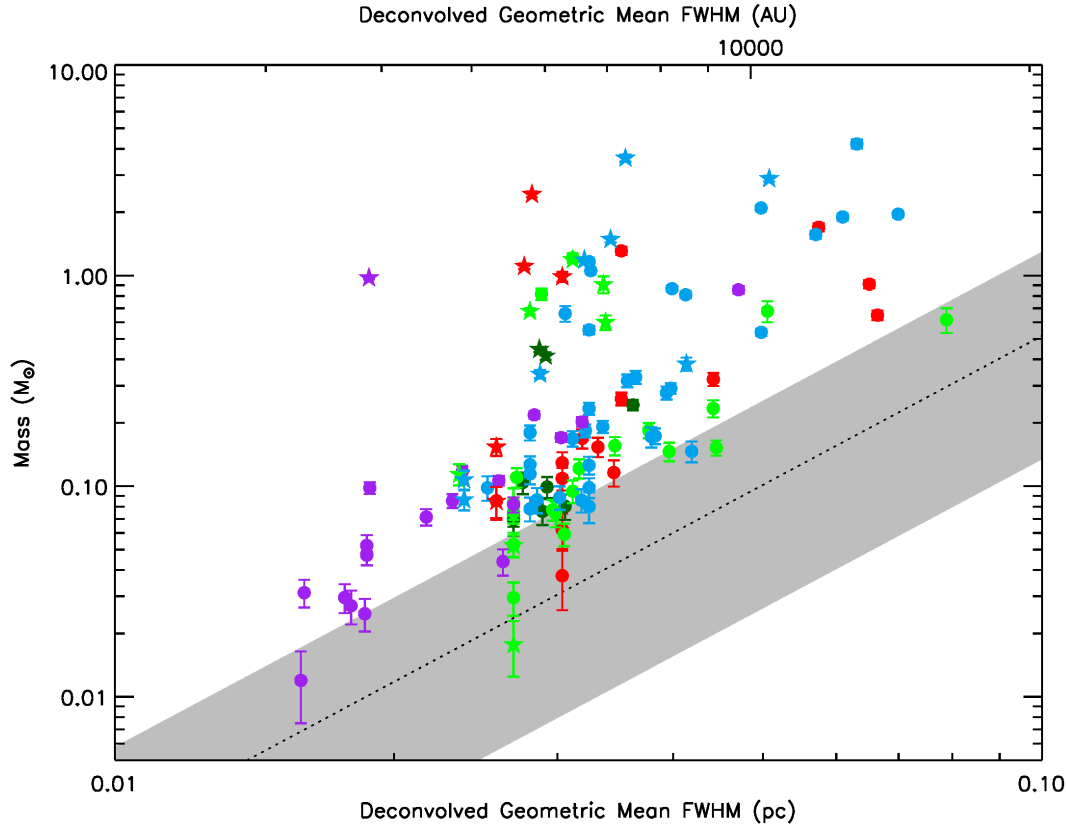}
\caption{Comparison of source mass (determined using \emph{Herschel}-derived temperatures) and source size (geometric mean of modelled major and minor FWHMs) for the sources in Cepheus. Circles represent starless cores; stars represent cores with embedded protostars.  Red sources lie in L1147/58, light green sources in L1174, dark green sources in L1172, blue sources in L1251, and purple sources in L1228.}
\label{fig:ceph_mass_size}
\end{figure*}

We derived temperatures for our sources from the ratio of the total SCUBA-2 450-\um\ and 850-\um\ flux densities measured along the line of sight for each source.  The temperature of a source can be determined from measurements of flux densities $F_{\nu_1}$ and $F_{\nu_2}$ at frequencies respectively $\nu_{1}$ and $\nu_{2}$ using the relation
\begin{equation}
\frac{F_{\nu_1}}{F_{\nu_2}}=\left(\frac{\nu_1}{\nu_2}\right)^{3+\beta}\frac{e^{\frac{h\nu_2}{k_{\textsc{b}}T}}-1}{e^{\frac{h\nu_1}{k_{\textsc{b}}T}}-1},
\label{eq:fd_ratio}
\end{equation}
where $T$ is the source temperature and $\beta$ is the dust opacity index, as before (see, e.g., \citealt{buckle2015}).  Assuming $\beta=2.0$, then in the case of the ratio of SCUBA-2 450-\um\ and 850-\um\ fluxes this relation becomes
\begin{equation}
\frac{F_{\nu(450\upmu{\rm m})}}{F_{\nu(850\upmu{\rm m})}}=24.05\times\frac{e^{\nicefrac{16.96}{T}}-1}{e^{\nicefrac{32.02}{T}}-1}.
\label{eq:fd_ratio_s2}
\end{equation}
This equation can be solved numerically for source temperature $T$.  This analysis presumes that the 450-\um\ and 850-\um\ emission traces the same dust population, and that the line-of-sight temperature variation is minimal (see, e.g., \citealt{shetty2009}).  A detailed analysis of dust temperatures determined from SCUBA-2 450-\um\ and 850-\um\ observations of the W40 region was performed by \citet{rumble2016}.  We choose $\beta=2.0$ for consistency with the \emph{Herschel}-derived temperature measurements, as described above.

As discussed in Section~\ref{sec:source_extraction}, 450-\um\ aperture photometry measurements were made using data which were convolved to match the resolution of the 850-\um\ data using a convolution kernel constructed as described by \citet{pattle2015}, using the SCUBA-2 450-\um\ and 850-\um\ beam models given by \citet{dempsey2013}.

\begin{figure*} 
\centering
\includegraphics[width=0.8\textwidth]{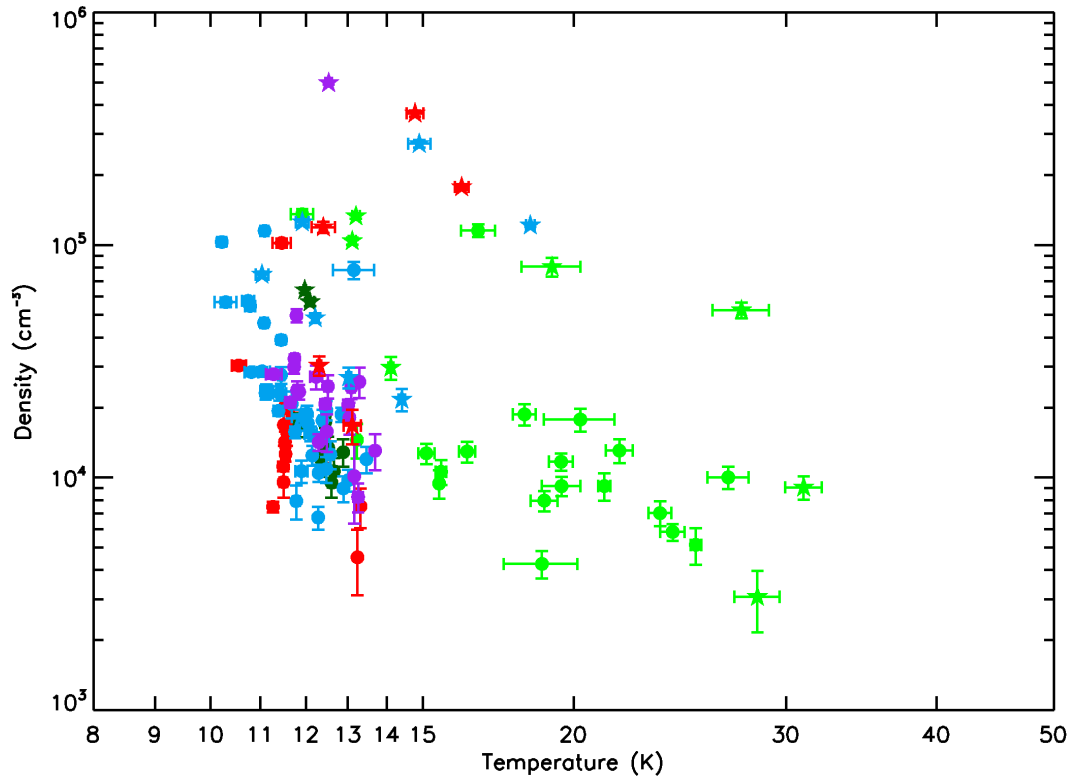}
\caption{Comparison of temperature and density for the sources in Cepheus.  Colour and symbol coding is as in Figure~\ref{fig:ceph_mass_size} (red -- L1147/58; light green -- L1172; dark green -- L1174; blue -- L1251; purple -- L1228).}
\label{fig:ceph_temp_density}
\end{figure*}

We derived temperatures for those of our sources with a detection $\geq 5\sigma$ at SCUBA-2 450\um.  These temperatures are listed in Table~\ref{tab:cepheus_properties}.  Figure~\ref{fig:scuba2_temps} compares the temperatures derived using SCUBA-2 and \emph{Herschel} data, showing that SCUBA-2-derived and \emph{Herschel}-derived source temperatures (Di Francesco et al., 2016, in prep.) are typically in agreement, albeit with large uncertainties on the SCUBA-2-derived temperatures.

This analysis suggests that determining source temperatures using only the ratio of SCUBA-2 450-\um\ and 850-\um\ data will produce reliable results in low-temperature cores.  This is as expected, as equation~\ref{eq:fd_ratio} is insensitive to temperature in the Rayleigh-Jeans (RJ) limit ($h\nu/k_{\textsc{b}}T\ll 1$).  The 450-\um\ data point will fall on the RJ tail of the spectral energy distribution if $T\gg 32$\,K, while the 850-\um\ data point will fall on the RJ tail if $T\gg 17$\,K.  It can be seen in Figure~\ref{fig:scuba2_temps} that the uncertainties on our SCUBA-2-derived temperatures increase substantially when $T>20$\,K, due to the decreasing sensitivity of the flux density ratio to temperature as source temperature increases.

Figure~\ref{fig:scuba2_temps} shows that there is a slight tendency for source temperatures determined from \emph{Herschel} measurements to be higher than those determined from SCUBA-2 measurements.  While this behaviour is not statistically significant, this is consistent with the shorter-wavelength \emph{Herschel} observations being sensitive to emission from warmer material than the longer-wavelength SCUBA-2 observations.

We calculated the masses of each of our sources using equation~\ref{eq:mass}, our SCUBA-2-derived temperatures, and our best-fit model 850-\um\ flux densities.  Source masses determined from SCUBA-2 temperatures are listed in Table~\ref{tab:cepheus_properties}.  Figure~\ref{fig:scuba2_masses} compares the masses derived using SCUBA-2-based and \emph{Herschel}-based source temperatures (Di Francesco et al., 2016, in prep.), and shows that the two measures of mass are generally in agreement.  There is a slight tendency to SCUBA-2-temperature masses to be higher than \emph{Herschel}-temperature masses.  This is a result of the tendency for \emph{Herschel}-derived temperatures to be higher than those derived from SCUBA-2 data.

Thus, we conclude that there is reasonable agreement between SCUBA-2-derived and \emph{Herschel}-derived source temperatures for our sources, and that SCUBA-2-derived source temperatures will be accurate when there is a good ($>5\sigma$) source detection at 450-\um, and when neither the 450-\um\ nor the 850-\um\ data point falls on the Rayleigh-Jeans tail of the spectral energy distribution.  For a detailed comparison of core properties determined from SCUBA-2 and \emph{Herschel} data, see \citet{wardthompson2016}.

\section{Discussion of Derived Properties}
\label{sec:ceph_discussion}

The masses and sizes of our sources are shown in Figure~\ref{fig:ceph_mass_size}.  Our sources typically occupy the upper part of the mass/size plane, in which prestellar cores are expected to lie \citep{simpson2008}, being overdense relative to transient, unbound structure (c.f. \citealt{andre2010}). The grey band on Figure~\ref{fig:ceph_mass_size} shows the region in which transient, gravitationally-unbound CO clumps are expected to lie \citep{elmegreen1996}.

\begin{figure*} 
\centering
\includegraphics[width=\textwidth]{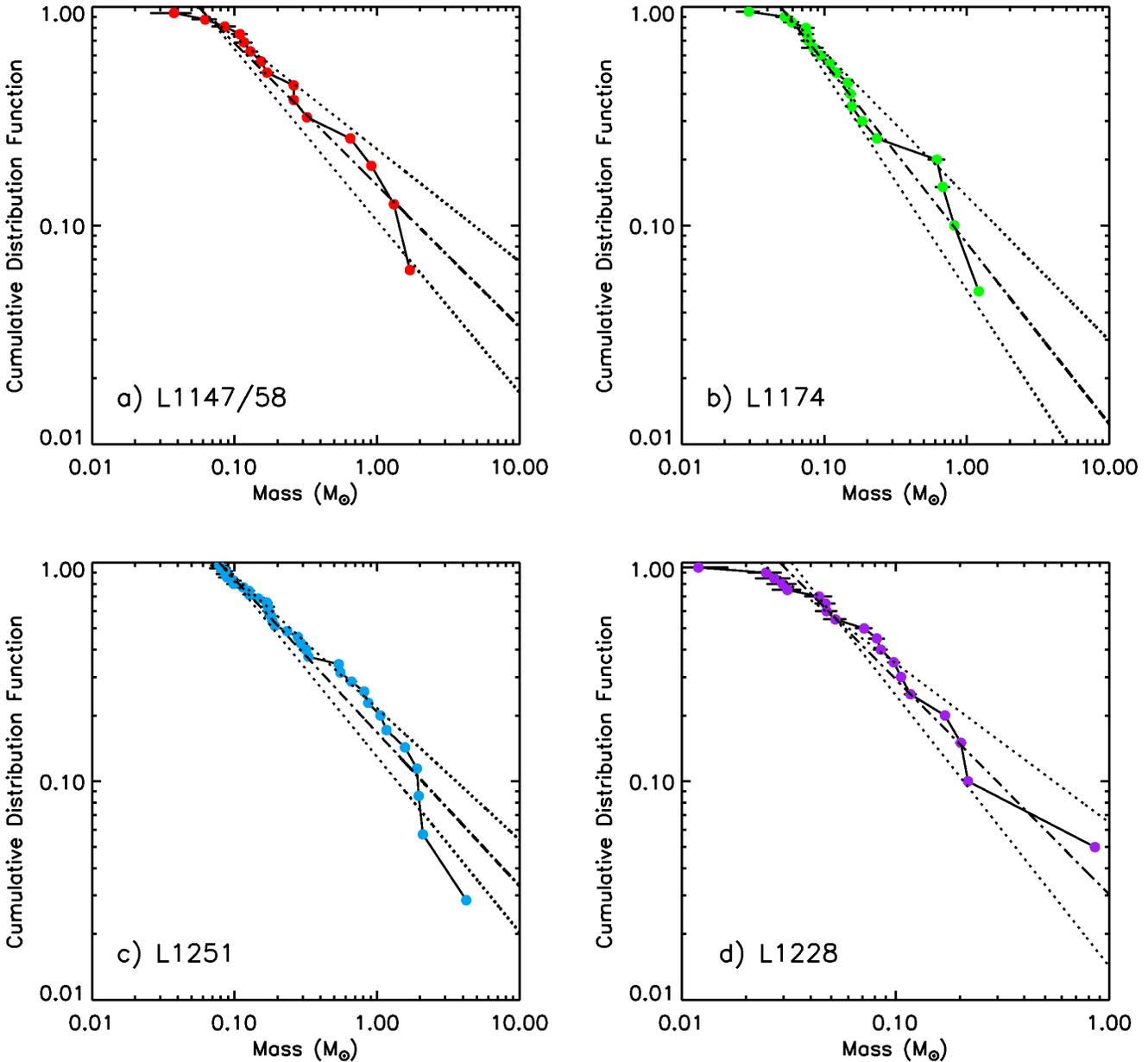}
\caption{Cumulative probability plots by region.  Colour coding is as in Figure~\ref{fig:ceph_mass_size} (red -- L1147/58; light green -- L1172; blue -- L1251; purple -- L1228).}
\label{fig:ceph_cumulative_prob_regions}
\end{figure*}

\begin{figure*} 
\centering
\includegraphics[width=\textwidth]{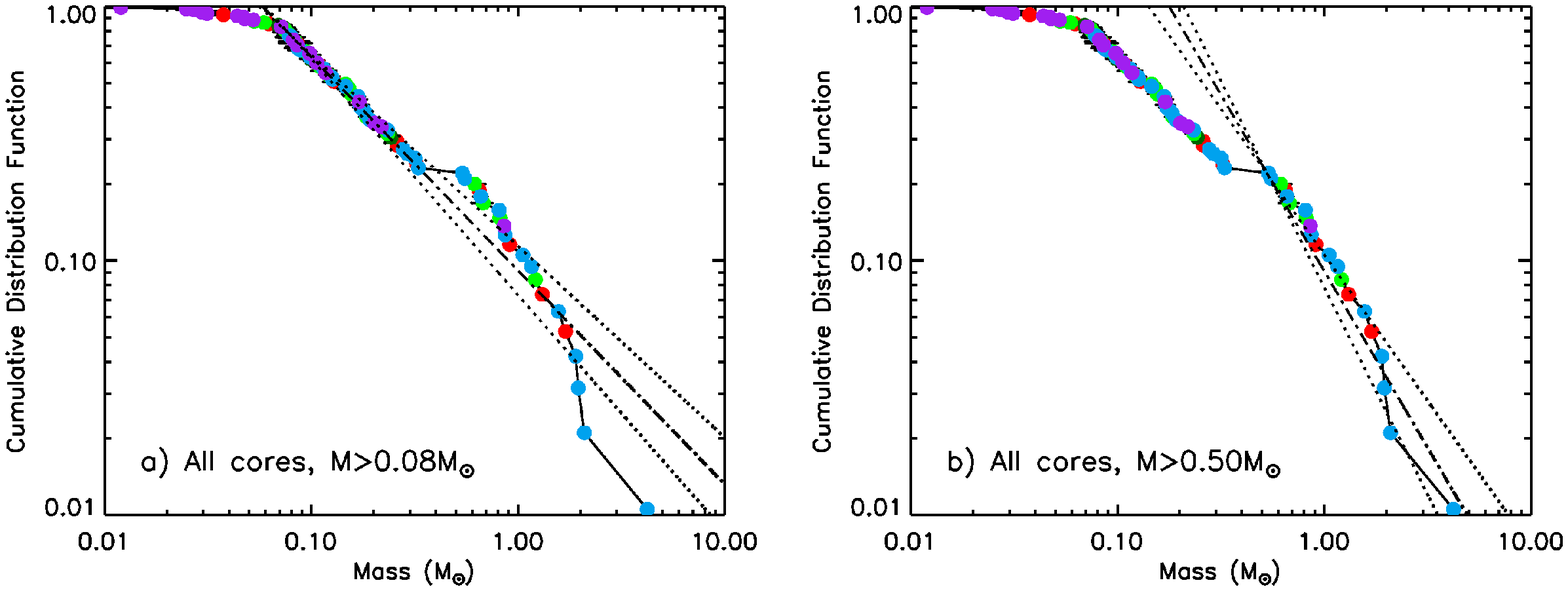}
\caption{Cumulative probability plots for the starless cores in Cepheus.  Left panel: power-law distribution for cores with masses $>0.08$\,M\sun.  Right panel: power-law distribution for cores with masses $>0.5$\,M\sun.  Colour coding is as in Figure~\ref{fig:ceph_mass_size} (red -- L1147/58; light green -- L1172; dark green -- L1174; blue -- L1251; purple -- L1228).}
\label{fig:ceph_cumulative_prob_all}
\end{figure*}

The temperatures and volume densities of our sources are shown in Figure~\ref{fig:ceph_temp_density}.  It can be seen that with the exception of sources in L1174 -- the reflection nebula -- the cores in our sample have a narrow range of temperatures ($\sim 9-15$\,K).

In order to determine a mass function for each set of starless cores in our sample, we analysed the cumulative distribution functions of starless core masses for each region in Cepheus, using the maximum likelihood estimator for an infinite power-law distribution \citep{koen2006,maschberger2009}.  Throughout the following discussion we assume that the masses of cores can be modelled by a power-law function,
\begin{equation}
\xi(M)dM\propto M^{-\alpha}dM,
\end{equation}
where $\xi(M)dM$ is the number of cores in the mass range $M$ to $M+dM$.

The empirical cumulative distribution function $\hat{F}$ is given, for the $i$\th\ data point in our sample, by
\begin{equation}
 \hat{F}(X_{i})\equiv \frac{i}{n+1},
\label{eq:cdf}
\end{equation}
where $n$ is the number of data points $X$. The maximum likelihood (ML) estimator for the exponent $\alpha$ of an infinite power-law distribution is
\begin{equation}
\alpha_{ml}=1+\frac{n}{\left(\sum_{i=1}^{n}{\rm ln}\left(X_{i}\right)\right)-n{\rm ln}\left({\rm min}(X)\right)}.
\label{eq:ml_estimator}
\end{equation}
The unbiased maximum likelihood (UML) estimator, $\alpha_{uml}$ is then
\begin{equation}
\alpha_{uml}=1+\frac{n-1}{n}(\alpha_{ml}-1).
\label{eq:uml_estimator}
\end{equation}
Uncertainties were estimated by performing a set of Monte Carlo experiments, drawing a set of data points randomly from our distribution of masses, from which $\alpha_{\rm ml}$ was recalculated.  The error quoted is the standard deviation of the distribution of $\alpha_{uml}$ which results from this procedure.

In this analysis we consider only starless cores, excluding all sources with embedded objects.  This is in order to construct cumulative mass distributions comparable to the core mass function (CMF).

The cumulative mass distribution functions for each region of Cepheus except L1172 are shown in Figure~\ref{fig:ceph_cumulative_prob_regions}, while the cumulative mass distribution function for all of the starless cores in our sample combined is shown in Figure~\ref{fig:ceph_cumulative_prob_all}.  The maximum-likelihood-estimator mass functions for each region are listed in Table~\ref{tab:ceph_cmfs}.  L1172 is excluded from this analysis as the region contains only 7 starless cores, too few to accurately constrain the power-law index of the region's core mass function.

As can be seen from Figure~\ref{fig:ceph_cumulative_prob_regions} and Table~\ref{tab:ceph_cmfs}, the core mass function in each region in the Cepheus Flare other than L1172 can be characterised by a power law above a mass of 0.05\,M$_{\odot}$.  The L1147/58, L1174 and L1251 $\alpha_{uml}$ values are similar, and show a high-mass CMF slope of $\alpha_{uml}=1.8\pm0.2$, $2.0\pm0.2$ and $1.8\pm0.1$ respectively.  The L1228 region, however, has a high-mass CMF slope of $\alpha_{uml,{\rm L1228}}=2.3\pm0.3$.  Whether this difference in CMF slope is indicative of a difference in behaviour between L1228 and the remainder of the sample, or merely of the small sample sizes in each region, is difficult to determine.

We combined the cores from each individual region in order to overcome the problem of small number statistics.  The cumulative mass distribution for all of the starless cores we detect in Cepheus (including those in L1172) is shown in Figure~\ref{fig:ceph_cumulative_prob_all}.  There appears to be a break in starless core masses between 0.3 and 0.5 M$_{\odot}$, with no starless cores being detected in this mass range.  Determining $\alpha_{uml}$ over the mass range $M>0.08\,$M$_{\odot}$ gives a power-law index of $1.9\pm 0.1$, with starless cores in the mass range $0.08-0.3$\,M$_{\odot}$ conforming well to a power-law distribution (see left panel of Figure~\ref{fig:ceph_cumulative_prob_all}).  Determining $\alpha_{uml}$ for the high-mass cores only ($M>0.5\,$M$_{\odot}$) gives a steeper power-law index, of $2.6\pm0.3$. Whether these high-mass starless cores represent a different population is not clear.

Both \citet{chabrier2003} and \citet{kroupa2001} predict a power-law index of $2.3$ for the high-mass end of the stellar Initial Mass Function (IMF), consistent with the \citet{salpeter1955} high-mass IMF.  Previously, a number of authors have suggested a link between the stellar IMF and the CMF (e.g. \citealt{motte1998}; \citealt{nutter2007}).  In Cepheus we see a high-mass CMF slope of $2.6\pm0.3$ (when $M>0.5\,$M$_{\odot}$), consistent with the Kroupa-Chabrier-Salpeter value.

The break in core masses can be seen in Figure~\ref{fig:ceph_mass_size}, for both starless cores (as discussed in this section) and for cores with embedded protostars.  Inspection of Figure~\ref{fig:ceph_mass_size} further shows that these most massive cores have a higher average radius than the rest of the population.  This might suggest that the more massive cores without embedded sources are a separate population of starless `clumps'; objects which might be expected to fragment to form multiple starless cores.  The lower-mass population of starless clumps might, due to their large radii and low masses and temperatures, be below the detectability limit of SCUBA-2 \citep{wardthompson2016}.  However, whether these highest-mass objects are in fact a separate population is by no means certain.

\begin{table} 
\centering
\caption{Maximum-likelihood-estimator power law indices for cores in Cepheus}
\begin{tabular}{c r@{\,$\pm$\,}l c}
\bottomrule
Region & \multicolumn{2}{c}{$\alpha_{uml}$} & Mass Range \\
\midrule
L1147/L1158 & 1.8 & 0.2 & $>0.05\,$M$_{\odot}$ \\
L1174 & 2.0 & 0.2 & $>0.05\,$M$_{\odot}$ \\
L1251 & 1.8 & 0.1 & $>0.05\,$M$_{\odot}$ \\
L1228 & 2.3 & 0.3 & $>0.05\,$M$_{\odot}$ \\
All & 1.9 & 0.1 & $>0.08\,$M$_{\odot}$ \\
All & 2.6 & 0.3 & $>0.5\,$M$_{\odot}$ \\
\bottomrule
\end{tabular}
\label{tab:ceph_cmfs}
\end{table}

\section{Counting Statistics}

\begin{figure} 
\centering
\includegraphics[width=0.47\textwidth]{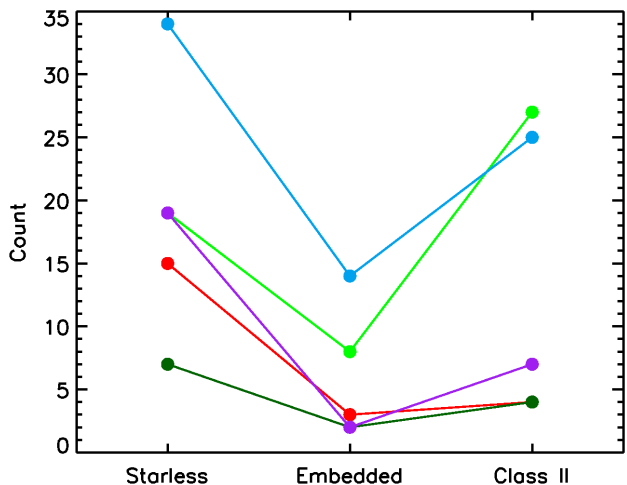}
\caption{Regional statistics of absolute number of starless, embedded and Class II sources in Cepheus.  Colour coding is as in Figure~\ref{fig:ceph_mass_size} (red -- L1147/58; light green -- L1172; dark green -- L1174; blue -- L1251; purple -- L1228).}
\label{fig:ceph_region_stats}
\end{figure}

\begin{figure} 
\centering
\includegraphics[width=0.47\textwidth]{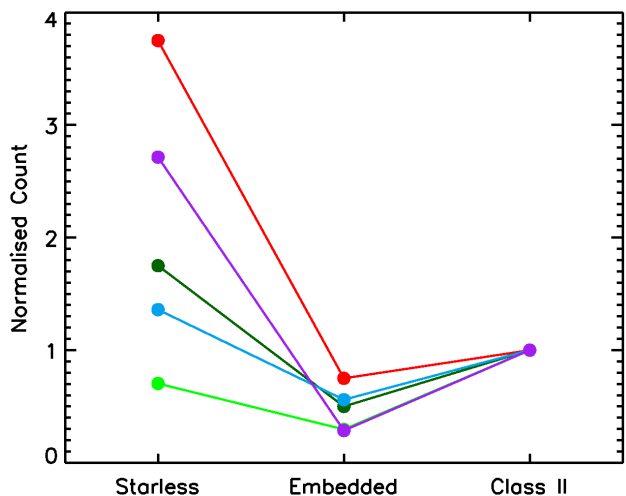}
\caption{Regional statistics of starless, embedded and Class II sources in Cepheus, normalised to the number of Class II sources.  Colour coding is as in Figure~\ref{fig:ceph_mass_size} (red -- L1147/58; light green -- L1172; dark green -- L1174; blue -- L1251; purple -- L1228).}
\label{fig:ceph_region_stats_normalised}
\end{figure}

We compared the number of starless cores in our sample with the number of embedded (Class I and Flat) and Class II sources detected by K09 in the same area, in order to make a crude estimate of the relative level of star formation activity in the different regions of the Cepheus Flare.  The absolute number counts are shown in Figure~\ref{fig:ceph_region_stats}, while the counts normalised to the number of Class II sources in the region are shown in Figure~\ref{fig:ceph_region_stats_normalised}.

Figure~\ref{fig:ceph_region_stats} shows that in absolute terms, L1251 contains the highest number of both starless cores and embedded sources, and the second highest number of Class II sources. L1174 contains the highest number of Class II sources; a natural result for a region in which clustered star formation has been ongoing for some time \citep{kun2008}.  L1174 has the second highest number of embedded sources after L1251, and the joint second-highest number of starless cores, along with L1228.  L1228, L1147/L1158 and L1172 have low number counts of both embedded and Class II sources.  This shows that the sites of ongoing active star formation, L1251 and L1174, have the highest absolute number of sources in almost all categories, while the regions of less active star formation generally have lower numbers of starless cores as well as of embedded sources.  However, in order to determine the evolutionary status of each region, the ratio of starless cores to embedded sources must be considered.

Figure~\ref{fig:ceph_region_stats_normalised} shows the number of sources of each type in each region, normalised to the number of Class II sources.  Again, a difference in behaviour can be seen between the less active regions, L1147/58 and L1228, and the active regions L1174 and L1251.  In the less active regions, there is a high ratio of starless cores to Class II sources: $\sim 3.8:1$ in L1148/L1157, and $\sim2.7:1$ in L1228. However, in the active star-forming regions, this ratio is much lower: $\sim 1.4:1$ in L1251, while in L1174 Class II sources outnumber starless cores, with a ratio $\sim0.7:1$.  L1172 shows an intermediate behaviour, with a ratio $\sim1.8:1$.  However, the low counting statistics in all classes in L1172 make any interpretation of this result difficult.

Since we do not know whether the star formation rate (SFR) in Cepheus has been constant over time, we consider two scenarios: a constant SFR over a very long time or a relatively short and finite burst of star formation.

In the first scenario, we can interpret those regions with a lower ratio of starless cores to Class II sources (L1251 and L1174) as having a high (but constant) SFR, i.e. converting gas into stars efficiently.  Those regions with a higher ratio of starless cores to Class II sources (L1147/58 and L1228), we interpret as having a lower (but still constant) SFR, i.e. these regions are forming stars less efficiently.

In the second scenario, we can interpret those regions with a higher ratio of starless cores to Class II sources as being at an earlier evolutionary stage than those with a lower starless core to Class II ratio, i.e. the regions with a high ratio have thus far converted only a small amount of their reservoir of available material into stars, while the regions with a low ratio have significantly depleted their local reservoir of dense gas.

The ratios which we observe are likely to result from a combination of these effects.  We can attempt to determine which effect is more likely to dominate for each region by considering what we know of their star formation histories and the current or historical influences on them.  However, all of the following intepretation must be used with care, as our absolute number counts of cores and Class II sources may not be large enough to put our conclusions on a strongly statistically-significant footing.

Star formation in L1251 may have been triggered or enhanced by passage of the Cepheus Flare Shell (CFS) through the region $\sim 4\,$Myr ago, while L1228 may currently be interacting with the CFS (see Section~\ref{sec:ceph_regions}, above).  This is consistent with the high starless-core-to-Class-II ratio in these regions resulting from the second scenario described above, with the low core-to-Class-II ratio in L1251 indicating that the region is significantly more evolved than L1228, in which star formation has only recently been triggered or enhanced.

\begin{figure*} 
\centering
\includegraphics[width=\textwidth]{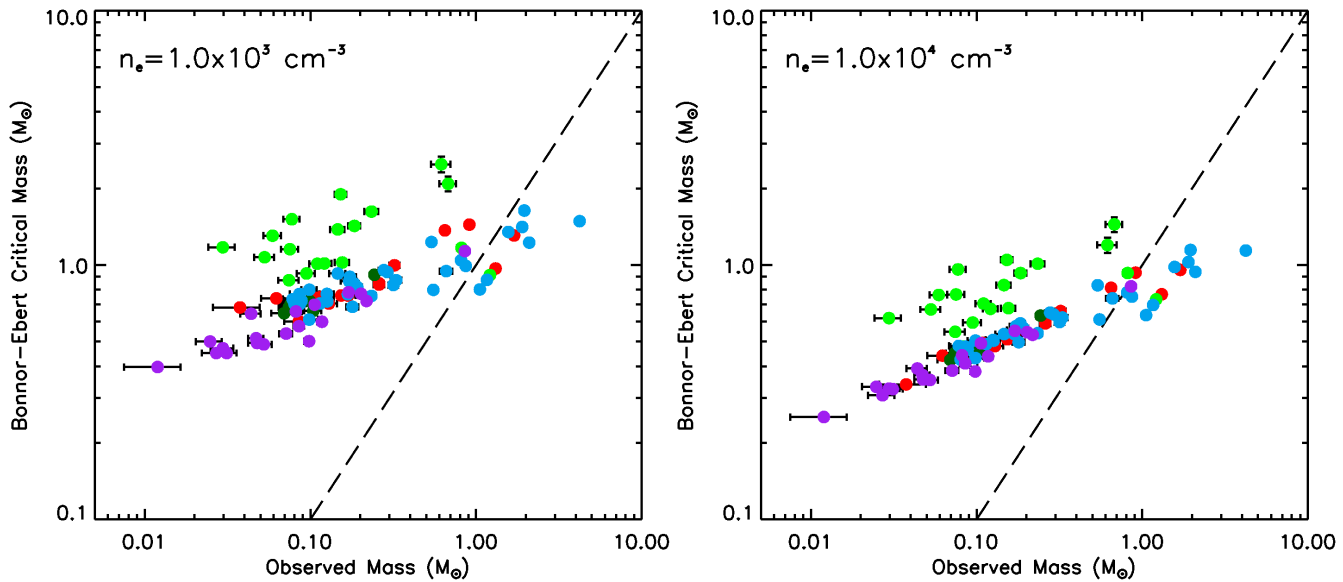}
\caption{BE stability plot for starless cores in Cepheus, with observed mass plotted against the Bonnor-Ebert critically-stable mass.  Left panel: bounding density $n_{e}=10^{3}\,{\rm cm}^{-3}$; right panel: bounding density $n_{e}=10^{4}\,{\rm cm}^{-3}$.  Cores to the right of the dashed line have no stable Bonnor-Ebert configuration.  Colour coding is as in Figure~\ref{fig:ceph_mass_size} (red -- L1147/58; light green -- L1172; dark green -- L1174; blue -- L1251; purple -- L1228).}
\label{fig:ceph_be}
\end{figure*}

L1147/58, with a high core-to-Class-II ratio, shows no significant signs of recent external influence (see Section~\ref{sec:ceph_regions}, above), suggesting that here the first scenario might be more likely, and star formation is an ongoing, inefficient process.  The clustered star formation in L1174 (low core-to-Class-II ratio) is more difficult to interpret; star formation has been ongoing in this region for some time, suggesting that L1174 might be running out of gas to convert into stars, thus favouring the first scenario.

\section{Bonnor-Ebert Stability Analysis}

The Bonnor-Ebert (BE) model of a starless core (\citealt{ebert1955}; \citealt{bonnor1956}) is frequently used as a measure of the stability of starless cores \citep[e.g.][]{alves2001}.  The BE model treats a core as an isothermal, self-gravitating, polytropic sphere bounded by external pressure.  For a given core temperature and external pressure, there is a maximum mass at which the core can be stable against gravitational collapse.  The critically-stable Bonnor-Ebert mass is frequently used as a proxy for virial mass (e.g. \citealt{konyves2015}).  In the following analysis, we consider the stability of the starless cores in Cepheus against gravitational collapse according to the Bonnor-Ebert model, under the assumption that our cores can be accurately characterised as Bonnor-Ebert spheres.  As discussed above, detailed modelling of core geometries is beyond the scope of this study, and so we cannot definitively state whether our cores have morphologies consistent with the Bonnor-Ebert model.

\subsection{Choice of bounding radius}
\label{sec:edge_radius}

As discussed above, the BE model treats cores as being bound by external pressure.  We have hitherto modelled our cores as having Gaussian density distributions, without a defined edge radius.

We define our edge radius $r_{e}$ as
\begin{equation}
r_{e}=\alpha\sqrt{2\,{\rm ln}\,\left(\frac{\rho_{0}}{\rho_{e}}\right)},
\label{eq:edge_radius}
\end{equation}
where $\rho_{0}$ is the central density of the core and $\rho_{e}$ is the density at the pressure-confined edge, assuming that the core obeys a Gaussian density distribution at all radii smaller than the edge radius.

The central density $\rho_{0}$ can be determined from the modelled mass $M$ and Gaussian width $\alpha$ of each core (listed in Table~\ref{tab:cepheus_properties}):
\begin{equation}
\rho_{0}=\frac{M}{(2\pi\alpha^{2})^{\frac{3}{2}}}.
\label{eq:peak_density}
\end{equation}

We here consider two different bounding densities,
\begin{equation}
\rho_{e}=\mu m_{\textsc{h}}\times10^{4}\,{\rm cm}^{-3},
\label{eq:edge_density}
\end{equation}
and
\begin{equation}
\rho_{e}=\mu m_{\textsc{h}}\times10^{3}\,{\rm cm}^{-3},
\label{eq:edge_density}
\end{equation}
 as representing a physically plausible range of densities at which our cores might be bound, and for consistency with our analysis of external pressure based on measurements of $^{13}$CO linewidths in Section~\ref{sec:virial}, below.  These choices of bounding density are consistent with our measurements: Figure~\ref{fig:ceph_temp_density} shows that the mean density of our cores is $\gtrsim 10^{4}$\,cm$^{-3}$ in almost all cases (the exceptions being warm, low-column-density cores in L1174), and in all of our cores, the central density inferred using equation~\ref{eq:peak_density} is $>10^{4}$\,cm$^{-3}$ (see Table~\ref{tab:cepheus_pressure}).

\subsection{Critically-stable Bonnor-Ebert sphere}

The mass at which a BE sphere at temperature $T$, with sound speed $c_{s}(T)$, and bounded by external pressure $P_{\textsc{ext}}$, is critically stable against gravitational collapse is given by
\begin{equation}
M_{\textsc{be},crit}=1.18\frac{c_{s}(T)^{4}}{P_{\textsc{ext}}^{1/2}G^{3/2}}.
\label{eq:BE}
\end{equation}
This can alternatively be expressed in terms of the critically-stable Bonnor-Ebert radius $R_{\textsc{be},crit}$,
\begin{equation}
M_{\textsc{be},crit}=2.4\frac{c_{s}^{2}}{G}R_{\textsc{be},crit}=2.4\frac{k_{\textsc{b}}T}{\mu m_{\textsc{h}}G}R_{\textsc{be},crit}.
\label{eq:ceph_be}
\end{equation}

In an attempt to determine whether our starless cores are likely to be virially bound, we determined their Bonnor-Ebert critically-stable masses (\citealt{ebert1955}; \citealt{bonnor1956}), under the assumption that $R_{\textsc{be},crit}=r_{e}$ .  The critically-stable Bonnor-Ebert masses of our cores are listed in Table~\ref{tab:cepheus_pressure}, and are plotted against our observed core masses in Figure~\ref{fig:ceph_be}.  Figure~\ref{fig:ceph_be} suggests that the majority of our cores have stable, pressure-confined (i.e. non-critical) Bonnor-Ebert solutions.  Our choice of bounding density does not significantly affect which of our cores have stable Bonnor-Ebert solutions.

\section{Energy Balance and Stability}
\label{sec:virial}

We attempted to assess the energy balance of the starless cores in the Cepheus molecular cloud and to determine the applicability of our Bonnor-Ebert analysis by estimating the external pressure on our cores using measurements presented by \citet{yonekura1997} (hereafter Y97).  Y97 conducted a large-scale $^{13}$CO $J=1\to 0$ survey of the Cepheus Flare region using two 4-m telescopes at Nagoya University.  Their observations had a resolution of 2.4 arcmin.  Each of the regions in our survey is entirely covered by a different, single, Y97 source: Y97 Source 8 for L1147/58, Y97 Source 14 for L1172 and L1174, Y97 Source 79 for L1251, and Y97 Source 66 for L1228.  Thus, we can estimate only a single value for external pressure in each region, which we then assume is representative for all of the cores within that region.

Previous studies of starless cores suggested that external gas pressure might be instrumental in confining dense cores in at least some cases (\citealt{maruta2010}; \citealt{pattle2015}).  We apply the method used by \citet{pattle2015} to estimate the external pressure on starless cores in the Ophiuchus molecular cloud to our sample of starless cores.

We consider the gas pressure in material traced by $^{13}$CO to be the external pressure acting on our starless cores, since CO is expected to trace the outer layers, or envelopes, of starless cores \citep{difrancesco2007}.  Higher-density tracers such as \nh\ are expected to trace the denser inner material of the cores themselves.

We estimate the external pressure $P_{\textsc{ext}}$ in each region from the linewidths measured by Y97 using the ideal gas law
\begin{equation}
P_{\textsc{ext}}\approx\rho_{^{13}\textsc{co}}\langle\sigma_{gas,^{13}\textsc{co}}^{2}\rangle.
\end{equation}
We consider two different models of the gas pressure in material traced by $^{13}$CO.  In the first instance, we assume that our cores are bounded at the maximum gas density traced by $^{13}$CO, $\rho_{^{13}\textsc{co}}\sim 10^{4}$\,cm$^{-3}$ \citep{difrancesco2007}.  We also consider the case in which our cores are bounded at the `typical' density of gas traced by $^{13}$CO, $\rho_{^{13}\textsc{co}}\sim 10^{3}$\,cm$^{-3}$ \citep{difrancesco2007}.  Hereafter, we refer to these models as `high-bounding-density' and `low-bounding-density' respectively.

The measured mean density of material within a molecular cloud depends strongly on the volume over which it is being assessed.  Using the values of total mass and surface area for Cepheus listed by \citet{dunham2015}, $2610\pm 170$\,M\sun\ and 38\,pc$^{2}$ respectively, we find, assuming a spherical geometry, $\langle n({\rm H}_{2})\rangle=210\,$cm$^{-3}$.  This value is determined over all areas mapped by \emph{Spitzer} with $A_{V}\gtrsim 3$.  However, K09 list a total cloud mass of 1003\,M\sun\ over a total area of 0.4\,pc$^{2}$, considering only areas with $A_{V}>5$.  Again assuming a spherical geometry, this implies a mean volume density $\langle n({\rm H}_{2})\rangle=7.5\times 10^{4}\,$cm$^{-3}$.  Moreover, the sizes and masses listed for individual clumps by K09 suggest densities $\sim 10^{5}$\,cm$^{-3}$ in at least some star-forming clumps.  Where in this range of densities our `bounding density' -- i.e. the \emph{minimum density of material associated with a potentially star-forming core} -- is likely to fall is not immediately clear.  However, the apparent threshold for star formation of $A_{V}\sim7$ (or higher) in local clouds (e.g. \citealt{molinari2014}, and references therein) suggests that star-forming cores are likely to exist within regions with densities significantly higher than the mean value in the cloud.

As discussed above, we typically find mean core densities $\sim 10^{4}$\,cm$^{-3}$ in our sample, and infer peak core densities $>10^{4}$\,cm$^{-3}$ but generally $<10^{5}$\,cm$^{-3}$ (with some exceptions in the densest cores).  Hence, we assume that the bounding densities of our cores cannot significantly exceed $10^{4}$\,cm$^{-3}$, and are likely to be lower.  Assuming that potentially star-forming cores exist within regions of density higher than the background cloud average, we consider $10^{3}-10^{4}$\,cm$^{-3}$ to be representative of the range of densities at which our cores are likely to be bound.

Y97 find the highest $^{13}$CO linewidth in L1251, the lowest in L1147/L1158, and the same, intermediate, value in L1172, L1174 and L1228.  It is possible that there are, locally, higher external pressures within these regions than are captured by the low-resolution Y97 measurements.

In the following analysis, we treat both the thermal and non-thermal components of the velocity dispersion in $^{13}$CO as representing a hydrostatic pressure on our cores -- i.e. we are treating the non-thermal component of the velocity dispersion as a modification to the sound speed in the gas (the microturbulent assumption; \citealt{chandrasekhar1951a},b\nocite{chandrasekhar1951b}).  Whilst on the majority of size scales in molecular clouds this has been demonstrated to be an invalid assumption, it has been shown that in both the compressible and incompressible cases, turbulence can provide support against cloud collapse (and hence, conversely, can provide an `inward' pressure promoting collapse) on scales smaller than the thermal Jeans wavelength in the cloud (\citealt{maclow2004}, and references therein).  For typical conditions in our cores, $T\sim 15$\,K and $n({\rm H}_{2})\sim 10^{4}$\,cm$^{-3}$, the thermal Jeans wavelength ($\lambda_{J}=c_{s}\sqrt{\pi/(G\rho)}$, where $c_{s}$ is sound speed and $\rho$ is gas density; \citealt{jeans1928}) in our cores is $\lambda_{J}\sim 0.2$\,pc, an order of magnitude larger than the size scale of our cores (see Tables~\ref{tab:cepheus_properties} and \ref{tab:cepheus_pressure}).  Hence, the assumption that the non-thermal component of the velocity dispersion can be treated as a hydrostatic pressure is justifiable in the case of our cores.

\setlength{\tabcolsep}{4pt}
\begin{table*} 
\centering
\caption{Data relating to starless cores' virial stability: (1) core ID, (2) modelled peak density, (3) gas velocity dispersion determined from low-resolution Y97 $^{13}$CO measurements, (4,5,6,7) modelled bounding radius, external pressure, Bonnor-Ebert critically-stable mass, and ratio of observed to BE-critical mass, for a bounding density of $n$(H$_{2}$)\,$=10^{3}$\,cm$^{-3}$, (8,9,10,11) as 4--7, for a bounding density of  $n$(H$_{2}$)\,$=10^{4}$\,cm$^{-3}$.}
\label{tab:cepheus_pressure}
\begin{tabular}{c r@{\,$\pm$\,}l c c c r@{\,$\pm$\,}l r@{\,$\pm$\,}l c c r@{\,$\pm$\,}l r@{\,$\pm$\,}l}
\toprule
\multicolumn{4}{c}{} & \multicolumn{6}{c}{$n$(H$_{2}$)\,$=10^{3}$\,cm$^{-3}$} & \multicolumn{6}{c}{$n$(H$_{2}$)\,$\,=10^{4}$\,cm$^{-3}$} \\ \cmidrule(rl){5-10} \cmidrule(rl){11-16}
Source & \multicolumn{2}{c}{$n_{0}$} & $\sigma_{^{13}\textsc{co},gas}$ & \multirow{2}{*}{$\displaystyle\frac{r_{^{13}\textsc{co}}}{FWHM}$} & $P_{\textsc{ext}}/k_{\textsc{b}}$ & \multicolumn{2}{c}{$M_{\textsc{be}}$} & \multicolumn{2}{c}{\multirow{2}{*}{$\displaystyle\frac{M}{M_{\textsc{be}}}$}} & \multirow{2}{*}{$\displaystyle\frac{r_{^{13}\textsc{co}}}{FWHM}$} & $P_{\textsc{ext}}/k_{\textsc{b}}$ & \multicolumn{2}{c}{$M_{\textsc{be}}$} & \multicolumn{2}{c}{\multirow{2}{*}{$\displaystyle\frac{M}{M_{\textsc{be}}}$}} \\
ID & \multicolumn{2}{c}{($\times 10^{4}$\,cm$^{-3}$)} & (kms$^{-1}$) & & (Kcm$^{-3}$) & \multicolumn{2}{c}{(M$_{\odot}$)} & \multicolumn{2}{c}{} & & (Kcm$^{-3}$) & \multicolumn{2}{c}{(M$_{\odot}$)} & \multicolumn{2}{c}{} \\
\midrule
3 & 35.45 & 1.58 & 0.4 & 1.46 & 0.6 & 0.969 & 0.017 & 1.35 & 0.08 & 1.13 & 6.2 & 0.768 & 0.013 & 1.71 & 0.10 \\
4 & 22.23 & 0.63 & 0.4 & 1.40 & 0.6 & 1.307 & 0.018 & 1.30 & 0.08 & 1.06 & 6.2 & 0.957 & 0.013 & 1.77 & 0.10  \\
7 & 5.82 & 0.98 & 0.4 & 1.21 & 0.6 & 0.847 & 0.003 & 0.31 & 0.02 & 0.80 & 6.2 & 0.596 & 0.002 & 0.44 & 0.03  \\
8 & 5.87 & 0.70 & 0.4 & 1.21 & 0.6 & 0.998 & 0.008 & 0.32 & 0.03 & 0.80 & 6.2 & 0.659 & 0.005 & 0.49 & 0.04  \\
9 & 61.65 & 1.77 & 0.4 & 1.52 & 0.6 & 0.759 & 0.008 & 0.22 & 0.03 & 1.22 & 6.2 & 0.525 & 0.006 & 0.32 & 0.04  \\
10 & 7.03 & 0.45 & 0.4 & 1.24 & 0.6 & 0.838 & 0.006 & 0.31 & 0.02 & 0.84 & 6.2 & 0.590 & 0.004 & 0.44 & 0.03  \\
12 & 6.10 & 0.65 & 0.4 & 1.22 & 0.6 & 0.750 & 0.005 & 0.15 & 0.02 & 0.81 & 6.2 & 0.471 & 0.003 & 0.25 & 0.04  \\
13 & 7.02 & 0.50 & 0.4 & 1.24 & 0.6 & 0.760 & 0.003 & 0.20 & 0.02 & 0.84 & 6.2 & 0.511 & 0.002 & 0.30 & 0.03  \\
14 & 10.52 & 0.98 & 0.4 & 1.30 & 0.6 & 0.706 & 0.003 & 0.18 & 0.02 & 0.92 & 6.2 & 0.481 & 0.002 & 0.27 & 0.03  \\
15 & 3.32 & 0.48 & 0.4 & 1.12 & 0.6 & 0.743 & 0.002 & 0.15 & 0.02 & 0.66 & 6.2 & 0.494 & 0.001 & 0.22 & 0.03  \\
16 & 4.47 & 0.60 & 0.4 & 1.17 & 0.6 & 0.740 & 0.003 & 0.08 & 0.02 & 0.73 & 6.2 & 0.441 & 0.002 & 0.14 & 0.03  \\
17 & 4.94 & 0.52 & 0.4 & 1.19 & 0.6 & 0.601 & 0.003 & 0.14 & 0.03 & 0.76 & 6.2 & 0.413 & 0.002 & 0.21 & 0.04  \\
18 & 3.69 & 0.52 & 0.4 & 1.14 & 0.6 & 1.368 & 0.006 & 0.47 & 0.03 & 0.69 & 6.2 & 0.814 & 0.004 & 0.80 & 0.04  \\
19 & 5.42 & 0.67 & 0.4 & 1.20 & 0.6 & 0.681 & 0.002 & 0.06 & 0.02 & 0.78 & 6.2 & 0.340 & 0.001 & 0.11 & 0.03  \\
20 & 4.57 & 0.57 & 0.4 & 1.17 & 0.6 & 1.442 & 0.013 & 0.63 & 0.04 & 0.74 & 6.2 & 0.934 & 0.008 & 0.98 & 0.05  \\
21 & 4.66 & 0.53 & 0.7 & 1.18 & 1.6 & 1.168 & 0.038 & 0.70 & 0.07 & 0.74 & 16.0 & 0.930 & 0.030 & 0.88 & 0.08  \\
26 & 3.30 & 0.45 & 0.7 & 1.12 & 1.6 & 0.914 & 0.020 & 1.33 & 0.10 & 0.66 & 16.0 & 0.733 & 0.016 & 1.66 & 0.12  \\
27 & 3.88 & 0.18 & 0.7 & 1.15 & 1.6 & 2.087 & 0.141 & 0.33 & 0.08 & 0.70 & 16.0 & 1.446 & 0.098 & 0.47 & 0.10  \\
28 & 4.18 & 0.68 & 0.7 & 1.16 & 1.6 & 1.154 & 0.030 & 0.07 & 0.01 & 0.72 & 16.0 & 0.767 & 0.020 & 0.10 & 0.01  \\
29 & 6.02 & 0.35 & 0.7 & 1.22 & 1.6 & 1.515 & 0.059 & 0.05 & 0.01 & 0.80 & 16.0 & 0.961 & 0.038 & 0.08 & 0.01  \\
30 & 40.02 & 2.51 & 0.7 & 1.47 & 1.6 & 0.927 & 0.009 & 0.10 & 0.01 & 1.15 & 16.0 & 0.594 & 0.006 & 0.16 & 0.02  \\
31 & 36.13 & 1.19 & 0.7 & 1.46 & 1.6 & 1.024 & 0.016 & 0.15 & 0.02 & 1.14 & 16.0 & 0.677 & 0.011 & 0.23 & 0.03  \\
33 & 28.02 & 2.62 & 0.7 & 1.43 & 1.6 & 1.175 & 0.012 & 0.03 & 0.00 & 1.10 & 16.0 & 0.619 & 0.006 & 0.05 & 0.01  \\
34 & 46.19 & 1.11 & 0.7 & 1.49 & 1.6 & 1.425 & 0.032 & 0.13 & 0.01 & 1.18 & 16.0 & 0.930 & 0.021 & 0.20 & 0.02  \\
36 & 6.17 & 0.71 & 0.7 & 1.22 & 1.6 & 1.012 & 0.022 & 0.11 & 0.01 & 0.81 & 16.0 & 0.705 & 0.015 & 0.16 & 0.02  \\
37 & 4.54 & 0.53 & 0.7 & 1.17 & 1.6 & 0.872 & 0.001 & 0.09 & 0.01 & 0.74 & 16.0 & 0.547 & 0.001 & 0.14 & 0.02  \\
38 & 3.48 & 0.39 & 0.7 & 1.13 & 1.6 & 1.623 & 0.059 & 0.14 & 0.02 & 0.67 & 16.0 & 1.012 & 0.037 & 0.23 & 0.03  \\
39 & 3.68 & 0.46 & 0.7 & 1.14 & 1.6 & 1.897 & 0.044 & 0.08 & 0.01 & 0.69 & 16.0 & 1.048 & 0.024 & 0.15 & 0.02  \\
40 & 4.42 & 0.44 & 0.7 & 1.17 & 1.6 & 2.489 & 0.175 & 0.25 & 0.08 & 0.73 & 16.0 & 1.199 & 0.084 & 0.51 & 0.11  \\
42 & 1.78 & 0.32 & 0.7 & 1.02 & 1.6 & 0.708 & 0.004 & 0.12 & 0.02 & 0.46 & 16.0 & 0.478 & 0.003 & 0.17 & 0.03  \\
43 & 4.06 & 0.34 & 0.7 & 1.16 & 1.6 & 1.073 & 0.013 & 0.05 & 0.01 & 0.71 & 16.0 & 0.669 & 0.008 & 0.08 & 0.01  \\
44 & 1.07 & 0.31 & 0.7 & 0.92 & 1.6 & 1.381 & 0.035 & 0.11 & 0.01 & 0.15 & 16.0 & 0.834 & 0.021 & 0.18 & 0.02  \\
45 & 6.49 & 0.68 & 0.7 & 1.23 & 1.6 & 1.304 & 0.029 & 0.05 & 0.01 & 0.82 & 16.0 & 0.763 & 0.017 & 0.08 & 0.01  \\
46 & 3.27 & 0.46 & 0.7 & 1.12 & 1.6 & 1.015 & 0.016 & 0.12 & 0.01 & 0.65 & 16.0 & 0.673 & 0.010 & 0.18 & 0.02  \\
49 & 1.48 & 0.20 & 0.7 & 0.99 & 1.6 & 0.661 & 0.004 & 0.16 & 0.02 & 0.38 & 16.0 & 0.455 & 0.003 & 0.23 & 0.03  \\
50 & 3.15 & 0.37 & 0.7 & 1.12 & 1.6 & 0.680 & 0.001 & 0.11 & 0.01 & 0.64 & 16.0 & 0.451 & 0.001 & 0.16 & 0.02  \\
51 & 5.04 & 0.86 & 0.7 & 1.19 & 1.6 & 0.701 & 0.004 & 0.11 & 0.02 & 0.76 & 16.0 & 0.450 & 0.002 & 0.17 & 0.02  \\
52 & 3.19 & 0.44 & 0.7 & 1.12 & 1.6 & 0.722 & 0.003 & 0.14 & 0.02 & 0.65 & 16.0 & 0.481 & 0.002 & 0.21 & 0.02  \\
53 & 2.76 & 0.28 & 0.7 & 1.09 & 1.6 & 0.727 & 0.005 & 0.11 & 0.02 & 0.61 & 16.0 & 0.457 & 0.003 & 0.18 & 0.02  \\
54 & 2.44 & 0.30 & 0.7 & 1.07 & 1.6 & 0.647 & 0.005 & 0.11 & 0.02 & 0.57 & 16.0 & 0.424 & 0.003 & 0.16 & 0.03  \\
55 & 4.49 & 0.46 & 0.7 & 1.17 & 1.6 & 0.916 & 0.005 & 0.27 & 0.02 & 0.74 & 16.0 & 0.633 & 0.004 & 0.38 & 0.02  \\
58 & 39.98 & 1.08 & 0.8 & 1.47 & 2.3 & 0.875 & 0.007 & 1.33 & 0.04 & 1.15 & 22.5 & 0.697 & 0.005 & 1.67 & 0.05  \\
59 & 27.11 & 2.31 & 0.8 & 1.42 & 2.3 & 0.947 & 0.037 & 0.70 & 0.09 & 1.09 & 22.5 & 0.740 & 0.029 & 0.89 & 0.10  \\
60 & 19.73 & 0.97 & 0.8 & 1.38 & 2.3 & 1.488 & 0.031 & 2.84 & 0.23 & 1.04 & 22.5 & 1.141 & 0.024 & 3.70 & 0.27  \\
62 & 18.98 & 0.87 & 0.8 & 1.38 & 2.3 & 0.799 & 0.008 & 0.69 & 0.04 & 1.03 & 22.5 & 0.611 & 0.006 & 0.90 & 0.05  \\
63 & 9.94 & 0.43 & 0.8 & 1.29 & 2.3 & 1.349 & 0.020 & 1.16 & 0.07 & 0.91 & 22.5 & 0.983 & 0.015 & 1.60 & 0.09  \\
64 & 35.80 & 1.20 & 0.8 & 1.46 & 2.3 & 0.802 & 0.007 & 1.32 & 0.05 & 1.14 & 22.5 & 0.636 & 0.005 & 1.66 & 0.06  \\
65 & 8.00 & 0.57 & 0.8 & 1.26 & 2.3 & 0.872 & 0.011 & 0.38 & 0.03 & 0.87 & 22.5 & 0.622 & 0.008 & 0.53 & 0.04  \\
66 & 16.04 & 0.55 & 0.8 & 1.35 & 2.3 & 0.993 & 0.007 & 0.87 & 0.04 & 1.00 & 22.5 & 0.751 & 0.006 & 1.15 & 0.05  \\
67 & 9.58 & 0.79 & 0.8 & 1.28 & 2.3 & 0.685 & 0.004 & 0.26 & 0.02 & 0.90 & 22.5 & 0.497 & 0.003 & 0.36 & 0.03  \\
68 & 13.57 & 0.48 & 0.8 & 1.33 & 2.3 & 1.044 & 0.008 & 0.78 & 0.03 & 0.97 & 22.5 & 0.780 & 0.006 & 1.04 & 0.04  \\
70 & 6.12 & 0.67 & 0.8 & 1.22 & 2.3 & 0.707 & 0.005 & 0.16 & 0.02 & 0.81 & 22.5 & 0.489 & 0.003 & 0.23 & 0.03  \\
71 & 19.98 & 0.67 & 0.8 & 1.38 & 2.3 & 1.225 & 0.015 & 1.71 & 0.08 & 1.04 & 22.5 & 0.940 & 0.011 & 2.23 & 0.10  \\
72 & 8.18 & 0.56 & 0.8 & 1.26 & 2.3 & 0.835 & 0.010 & 0.38 & 0.03 & 0.87 & 22.5 & 0.597 & 0.007 & 0.53 & 0.04  \\
73 & 3.78 & 0.48 & 0.8 & 1.14 & 2.3 & 0.723 & 0.001 & 0.12 & 0.02 & 0.69 & 22.5 & 0.466 & 0.001 & 0.19 & 0.02  \\
74 & 5.94 & 0.40 & 0.8 & 1.21 & 2.3 & 0.820 & 0.002 & 0.23 & 0.02 & 0.80 & 22.5 & 0.566 & 0.001 & 0.34 & 0.02  \\
75 & 8.03 & 0.53 & 0.8 & 1.26 & 2.3 & 0.756 & 0.003 & 0.31 & 0.02 & 0.87 & 22.5 & 0.540 & 0.002 & 0.43 & 0.03  \\
77 & 3.38 & 0.36 & 0.8 & 1.13 & 2.3 & 0.800 & 0.002 & 0.12 & 0.01 & 0.66 & 22.5 & 0.505 & 0.001 & 0.20 & 0.02  \\
78 & 5.36 & 0.38 & 0.8 & 1.20 & 2.3 & 0.955 & 0.011 & 0.29 & 0.02 & 0.78 & 22.5 & 0.650 & 0.008 & 0.43 & 0.03  \\
\end{tabular}
\end{table*}
\addtocounter{table}{-1}
\begin{table*}
\centering
\caption{-- continued.}
\begin{tabular}{c r@{\,$\pm$\,}l c c c r@{\,$\pm$\,}l r@{\,$\pm$\,}l c c r@{\,$\pm$\,}l r@{\,$\pm$\,}l}
\toprule
 \multicolumn{4}{c}{} & \multicolumn{6}{c}{$n$(H$_{2}$)\,$=10^{3}$\,cm$^{-3}$} & \multicolumn{6}{c}{$n$(H$_{2}$)\,$=10^{4}$\,cm$^{-3}$} \\ \cmidrule(rl){5-10} \cmidrule(rl){11-16}
Source & \multicolumn{2}{c}{$n_{0}$} & $\sigma_{^{13}\textsc{co},gas}$ & \multirow{2}{*}{$\displaystyle\frac{r_{^{13}\textsc{co}}}{FWHM}$} & $P_{\textsc{ext}}/k_{\textsc{b}}$ & \multicolumn{2}{c}{$M_{\textsc{be}}$} & \multicolumn{2}{c}{\multirow{2}{*}{$\displaystyle\frac{M}{M_{\textsc{be}}}$}} & \multirow{2}{*}{$\displaystyle\frac{r_{^{13}\textsc{co}}}{FWHM}$} & $P_{\textsc{ext}}/k_{\textsc{b}}$ & \multicolumn{2}{c}{$M_{\textsc{be}}$} & \multicolumn{2}{c}{\multirow{2}{*}{$\displaystyle\frac{M}{M_{\textsc{be}}}$}} \\
ID & \multicolumn{2}{c}{($\times 10^{4}$\,cm$^{-3}$)} & (kms$^{-1}$) & & (Kcm$^{-3}$) & \multicolumn{2}{c}{(M$_{\odot}$)} & \multicolumn{2}{c}{} & & (Kcm$^{-3}$) & \multicolumn{2}{c}{(M$_{\odot}$)} & \multicolumn{2}{c}{} \\
\midrule
79 & 3.12 & 0.40 & 0.8 & 1.11 & 2.3 & 0.769 & 0.002 & 0.11 & 0.01 & 0.64 & 22.5 & 0.478 & 0.002 & 0.18 & 0.02  \\
80 & 6.48 & 0.45 & 0.8 & 1.23 & 2.3 & 0.846 & 0.003 & 0.22 & 0.02 & 0.82 & 22.5 & 0.590 & 0.002 & 0.31 & 0.02  \\
81 & 4.17 & 0.54 & 0.8 & 1.16 & 2.3 & 0.734 & 0.003 & 0.11 & 0.01 & 0.72 & 22.5 & 0.481 & 0.002 & 0.16 & 0.02  \\
83 & 7.20 & 0.95 & 0.8 & 1.24 & 2.3 & 0.611 & 0.002 & 0.16 & 0.02 & 0.84 & 22.5 & 0.431 & 0.001 & 0.23 & 0.03  \\
85 & 2.75 & 0.46 & 0.8 & 1.09 & 2.3 & 0.705 & 0.005 & 0.11 & 0.02 & 0.60 & 22.5 & 0.425 & 0.003 & 0.19 & 0.03  \\
86 & 6.54 & 0.55 & 0.8 & 1.23 & 2.3 & 0.768 & 0.006 & 0.22 & 0.02 & 0.82 & 22.5 & 0.536 & 0.004 & 0.31 & 0.03  \\
87 & 4.37 & 0.60 & 0.8 & 1.17 & 2.3 & 0.701 & 0.004 & 0.12 & 0.02 & 0.73 & 22.5 & 0.463 & 0.003 & 0.19 & 0.03  \\
88 & 2.34 & 0.26 & 0.8 & 1.07 & 2.3 & 0.926 & 0.004 & 0.16 & 0.02 & 0.55 & 22.5 & 0.536 & 0.003 & 0.27 & 0.03  \\
89 & 3.70 & 0.40 & 0.8 & 1.14 & 2.3 & 0.864 & 0.010 & 0.20 & 0.02 & 0.69 & 22.5 & 0.555 & 0.007 & 0.31 & 0.04  \\
90 & 5.45 & 0.32 & 0.8 & 1.20 & 2.3 & 0.939 & 0.004 & 0.31 & 0.02 & 0.78 & 22.5 & 0.640 & 0.003 & 0.46 & 0.03  \\
92 & 6.75 & 0.64 & 0.8 & 1.23 & 2.3 & 0.720 & 0.003 & 0.18 & 0.02 & 0.83 & 22.5 & 0.504 & 0.002 & 0.25 & 0.02  \\
93 & 9.87 & 0.38 & 0.8 & 1.29 & 2.3 & 1.412 & 0.018 & 1.35 & 0.08 & 0.91 & 22.5 & 1.028 & 0.013 & 1.85 & 0.10  \\
94 & 4.32 & 0.42 & 0.8 & 1.17 & 2.3 & 0.770 & 0.002 & 0.16 & 0.02 & 0.73 & 22.5 & 0.508 & 0.001 & 0.25 & 0.02  \\
95 & 6.72 & 0.21 & 0.8 & 1.23 & 2.3 & 1.638 & 0.015 & 1.19 & 0.05 & 0.83 & 22.5 & 1.146 & 0.011 & 1.71 & 0.07  \\
96 & 5.13 & 0.19 & 0.8 & 1.19 & 2.3 & 1.232 & 0.005 & 0.44 & 0.02 & 0.77 & 22.5 & 0.833 & 0.004 & 0.65 & 0.03  \\
97 & 3.64 & 0.31 & 0.8 & 1.14 & 2.3 & 0.899 & 0.004 & 0.19 & 0.02 & 0.68 & 22.5 & 0.575 & 0.003 & 0.30 & 0.03  \\
98 & 11.28 & 0.48 & 0.7 & 1.31 & 1.6 & 0.722 & 0.005 & 0.30 & 0.01 & 0.93 & 16.0 & 0.532 & 0.003 & 0.41 & 0.02  \\
99 & 17.27 & 1.11 & 0.7 & 1.36 & 1.6 & 0.502 & 0.002 & 0.20 & 0.01 & 1.01 & 16.0 & 0.382 & 0.002 & 0.26 & 0.02  \\
100 & 7.31 & 0.38 & 0.7 & 1.24 & 1.6 & 0.771 & 0.005 & 0.26 & 0.02 & 0.85 & 16.0 & 0.545 & 0.004 & 0.37 & 0.02  \\
101 & 7.20 & 0.33 & 0.7 & 1.24 & 1.6 & 0.781 & 0.003 & 0.22 & 0.01 & 0.84 & 16.0 & 0.551 & 0.002 & 0.31 & 0.01  \\
102 & 9.42 & 1.11 & 0.7 & 1.28 & 1.6 & 0.488 & 0.006 & 0.11 & 0.01 & 0.90 & 16.0 & 0.353 & 0.004 & 0.15 & 0.02  \\
103 & 8.47 & 0.91 & 0.7 & 1.27 & 1.6 & 0.516 & 0.003 & 0.09 & 0.01 & 0.88 & 16.0 & 0.370 & 0.003 & 0.13 & 0.01  \\
104 & 8.25 & 0.74 & 0.7 & 1.26 & 1.6 & 0.538 & 0.004 & 0.13 & 0.01 & 0.87 & 16.0 & 0.385 & 0.003 & 0.19 & 0.02  \\
105 & 3.52 & 1.32 & 0.7 & 1.13 & 1.6 & 0.397 & 0.000 & 0.03 & 0.01 & 0.67 & 16.0 & 0.253 & 0.000 & 0.05 & 0.02  \\
106 & 9.66 & 0.41 & 0.7 & 1.28 & 1.6 & 1.135 & 0.017 & 0.75 & 0.05 & 0.90 & 16.0 & 0.825 & 0.013 & 1.04 & 0.06  \\
107 & 10.38 & 0.67 & 0.7 & 1.29 & 1.6 & 0.599 & 0.005 & 0.20 & 0.01 & 0.92 & 16.0 & 0.438 & 0.003 & 0.27 & 0.02  \\
108 & 2.86 & 0.41 & 0.7 & 1.10 & 1.6 & 0.644 & 0.006 & 0.07 & 0.01 & 0.62 & 16.0 & 0.392 & 0.003 & 0.11 & 0.02  \\
109 & 4.53 & 0.81 & 0.7 & 1.17 & 1.6 & 0.500 & 0.003 & 0.05 & 0.01 & 0.74 & 16.0 & 0.332 & 0.002 & 0.07 & 0.01  \\
110 & 8.96 & 1.35 & 0.7 & 1.27 & 1.6 & 0.451 & 0.003 & 0.07 & 0.01 & 0.89 & 16.0 & 0.325 & 0.002 & 0.10 & 0.01  \\
112 & 5.47 & 1.00 & 0.7 & 1.20 & 1.6 & 0.451 & 0.001 & 0.06 & 0.01 & 0.78 & 16.0 & 0.308 & 0.000 & 0.09 & 0.02  \\
113 & 4.93 & 0.41 & 0.7 & 1.19 & 1.6 & 0.657 & 0.003 & 0.12 & 0.01 & 0.76 & 16.0 & 0.442 & 0.002 & 0.19 & 0.02  \\
114 & 6.29 & 0.98 & 0.7 & 1.22 & 1.6 & 0.471 & 0.001 & 0.06 & 0.01 & 0.81 & 16.0 & 0.327 & 0.001 & 0.09 & 0.01  \\
115 & 7.14 & 0.37 & 0.7 & 1.24 & 1.6 & 0.699 & 0.001 & 0.15 & 0.01 & 0.84 & 16.0 & 0.492 & 0.001 & 0.22 & 0.01  \\
116 & 8.57 & 0.99 & 0.7 & 1.27 & 1.6 & 0.494 & 0.003 & 0.10 & 0.01 & 0.88 & 16.0 & 0.355 & 0.003 & 0.13 & 0.02  \\
117 & 8.11 & 0.60 & 0.7 & 1.26 & 1.6 & 0.575 & 0.001 & 0.15 & 0.01 & 0.87 & 16.0 & 0.411 & 0.001 & 0.21 & 0.02  \\
\bottomrule
\end{tabular}
\end{table*}
\setlength{\tabcolsep}{6pt}

\subsection{Virial analysis of cores in Cepheus}

We performed a virial analysis on our cores, in order to test their stability against collapse, and hence to test the validity of the Bonnor-Ebert analysis above.  We estimate the terms in the virial equation, in order to determine the stability of our cores.  Throughout the following analysis, we assume that our cores obey a Gaussian core density profile, consistent with the core fitting process discussed in Section~\ref{sec:source_extraction}.

We use the virial equation in the form
\begin{equation}
\frac{1}{2}\ddot{\mathcal{I}}=\Omega_{\textsc{g}}+2\Omega_{\textsc{k}}+\Omega_{\textsc{p}},
\end{equation}
where $\ddot{\mathcal{I}}$ is the second derivative of the moment of inertia $\mathcal{I}$, $\Omega_{\textsc{g}}$ is the gravitational potential energy of the core, $\Omega_{\textsc{k}}$ is the internal thermal energy of the core, and $\Omega_{\textsc{p}}$ is the external pressure energy of the core.  We do not include the internal magnetic energy in the following analysis.  A core with $\ddot{\mathcal{I}}>0$ is virially unbound and dispersing, a core with $\ddot{\mathcal{I}}<0$ is virially bound and collapsing, and a core with $\ddot{\mathcal{I}}=0$ is in virial equilibrium with its surroundings.

We determined external pressure energies for our cores using the equation
\begin{equation}
\Omega_{\textsc{p}}=-4\pi P_{\textsc{ext}}r_{e}^{3}.
\end{equation}
We assume that $^{13}$CO traces material adjacent to our cores, and again assume that the cores are confined by external pressure at a density of either $10^{3}$\,cm$^{-3}$ or $10^{4}$\,cm$^{-3}$, in both cases at the radius $r_{e}$ defined in Section~\ref{sec:edge_radius} (see equations~\ref{eq:edge_radius} and \ref{eq:edge_density}).  These bounding densities are chosen as representing a range of densities from a typical gas density traced by $^{13}$CO ($\sim10^{3}$\,cm$^{-3}$) to the gas density at which $^{13}$CO ceases to be an effective tracer ($\sim10^{4}$\,cm$^{-3}$; \citealt{difrancesco2007}), and as being a physically plausible density at which our cores might be bounded (see discussion in Section~\ref{sec:edge_radius}, above).

We determined gravitational potential energies for each of our cores using the equation for the gravitational potential energy of a spherically symmetrical Gaussian density distribution truncated at a radius $r_{e}$:
\begin{multline}
\Omega_{\textsc{g}}=-16\uppi^{2}G\rho_{0}^{2}\alpha^{5}\left[\frac{\sqrt{\uppi}}{4}\erf\left(\frac{r_{e}}{\alpha}\right)\right. \\ -\left.\sqrt{\frac{\uppi}{2}}e^{-\frac{1}{2}\left(\frac{r_{e}}{\alpha}\right)^{2}}\erf\left(\frac{r_{e}}{\alpha\sqrt{2}}\right)+\frac{1}{2}\frac{r_{e}}{\alpha}e^{-\left(\frac{r_{e}}{\alpha}\right)^{2}}\right].
\label{eq:ceph_omega_g}
\end{multline}
where $\rho_{0}$ is the modelled central density of the core and $\alpha$ is the modelled Gaussian width of the core -- see \citet{pattle2016} for a derivation of this result.

\begin{figure*} 
\centering
\includegraphics[width=\textwidth]{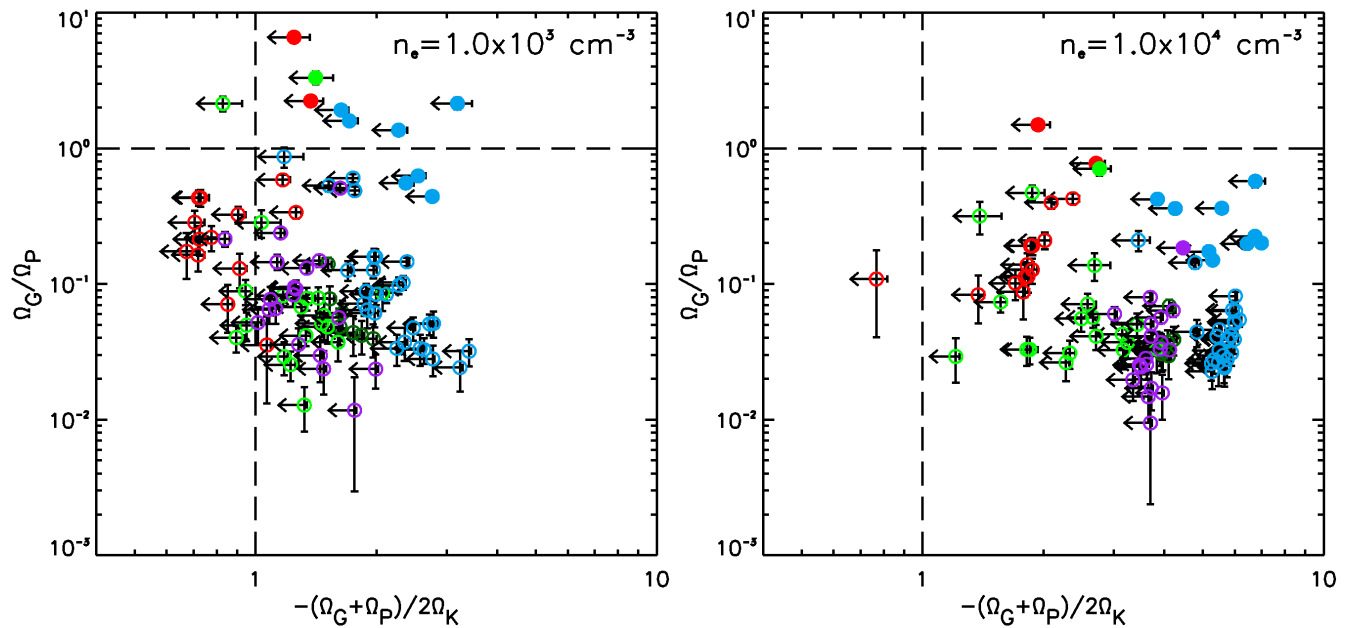}
\caption{The virial plane for our starless cores.  Left panel: bounding density $n_{e}=10^{3}\,{\rm cm}^{-3}$; right panel: bounding density $n_{e}=10^{4}\,{\rm cm}^{-3}$.  Virial stability is plotted on the $x$ axis.  The ratio of gravitational potential energy to external pressure energy is plotted on the $y$ axis.  The vertical dashed line indicates the line of virial stability, with the right-hand side of the plot being bound and the left side being unbound.  The horizontal dashed line marks equipartition between external pressure energy and gravitational potential energy; cores above the line are gravitationally bound, while cores below the line are pressure-confined.  Closed circles indicate cores with a mass greater than their BE-critical mass.  Colour coding as in Figure~\ref{fig:ceph_mass_size} (red -- L1147/58; light green -- L1172; dark green -- L1174; blue -- L1251; purple -- L1228).}
\label{fig:ceph_virial}
\end{figure*}

We were able to put a lower limit on the internal energy of each of our cores by estimating the thermal kinetic energy of the core,
\begin{equation}
\Omega_{\textsc{k},\textsc{t}}=\frac{3}{2}Mc_{s}^{2}=\frac{3}{2}M\frac{k_{\textsc{b}}T}{\mu m_{\textsc{h}}}.
\label{eq:ceph_omega_kt}
\end{equation}
\citet{pattle2015} found that a significant fraction of the internal kinetic energy of starless cores in the Ophiuchus molecular cloud is non-thermal.  Unless starless cores in Cepheus are substantially dissimilar to those in Ophiuchus, there is likely to be a substantial non-thermal component to the internal energy of our cores. Hence the values given by equation~\ref{eq:ceph_omega_kt} are a lower limit on the true value of $\Omega_{\textsc{k}}$.  The values of $\Omega_{\textsc{g}}$, $\Omega_{\textsc{p}}$, $\Omega_{\textsc{k}}$ and the virial parameter $\frac{1}{2}\ddot{\mathcal{I}}$ which we determine are listed in Table~\ref{tab:cepheus_virial}.

\subsection{Virial stability of cores in Cepheus}
\label{sec:ceph_virial}

Our best estimate of the virial plane for Cepheus is shown in Figure~\ref{fig:ceph_virial}.  The virial plane diagram was introduced by \citet{pattle2015} as a means of assessing the virial stability and mode of confinement of starless cores.  The abcissa shows the virial ratio, $-(\Omega_{\textsc{g}}+\Omega_{\textsc{p}})/2\Omega_{\textsc{k}}$.  Cores to the right of the vertical dashed line in the virial plane are virially bound, while cores to the left of this line are virially unbound.  The ordinate shows the ratio of gravitational potential energy to external pressure energy, $\Omega_{\textsc{g}}/\Omega_{\textsc{p}}$.  The dominant mode of confinement of cores below the horizontal dashed line is external pressure, while the dominant mode of confinement of cores above this line is self-gravity.

\setlength{\tabcolsep}{4pt}
\begin{table}
\centering
\caption{Terms in the virial equation for cores in Cepheus: (1) core ID, (2) thermal internal energy (3,4,5) gravitational potential energy, external pressure energy from $^{13}$CO measurements, and the virial parameter, for a bounding density of $n$(H$_{2}$)\,$=10^{3}$\,cm$^{-3}$, (6,7,8) as 3--5 for a bounding density of $n$(H$_{2}$)\,$=10^{4}$\,cm$^{-3}$.}
\label{tab:cepheus_virial}
\begin{tabular}{c c ccc ccc }
\toprule
 & & \multicolumn{3}{c}{$n$(H$_{2}$)$=10^{3}$\,cm$^{-3}$} & \multicolumn{3}{c}{$n$(H$_{2}$)$=10^{4}$\,cm$^{-3}$} \\ \cmidrule(lr){3-5} \cmidrule(lr){6-8}
Source & $\Omega_{\textsc{k}}$ & $\Omega_{\textsc{g}}$ & $\Omega_{\textsc{p}}$ & $\frac{1}{2}\ddot{\mathcal{I}}$ & $\Omega_{\textsc{g}}$ & $\Omega_{\textsc{p}}$ & $\frac{1}{2}\ddot{\mathcal{I}}$ \\ \cmidrule{2-8}
ID & \multicolumn{7}{c}{($\times 10^{41}$\,erg)}  \\
\midrule
3 & 13.0 & -28.04 & -4.3 & -6.4 & -30.08 & -20.2 & -24.4 \\
4 & 15.4 & -29.27 & -13.1 & -11.5 & -36.49 & -47.1 & -52.7 \\
7 & 2.6 & -1.14 & -2.6 & +1.5 & -1.56 & -8.2 & -4.5 \\
8 & 3.2 & -1.42 & -4.4 & +0.6 & -2.23 & -10.7 & -6.5 \\
9 & 1.7 & -0.53 & -1.9 & +1.0 & -0.75 & -5.4 & -2.8 \\
10 & 2.6 & -1.14 & -2.6 & +1.4 & -1.55 & -8.2 & -4.5 \\
12 & 1.2 & -0.24 & -1.9 & +0.2 & -0.42 & -3.7 & -1.8 \\
13 & 1.5 & -0.43 & -1.9 & +0.7 & -0.65 & -5.1 & -2.7 \\
14 & 1.3 & -0.33 & -1.5 & +0.7 & -0.48 & -4.2 & -2.1 \\
15 & 1.2 & -0.24 & -1.4 & +0.7 & -0.37 & -3.6 & -1.6 \\
16 & 0.7 & -0.08 & -1.1 & +0.2 & -0.15 & -1.8 & -0.5 \\
17 & 0.8 & -0.17 & -1.0 & +0.6 & -0.24 & -2.8 & -1.3 \\
18 & 6.3 & -4.00 & -11.9 & -3.3 & -7.51 & -18.9 & -13.8 \\
19 & 0.4 & -0.03 & -0.9 & -0.1 & -0.06 & -0.6 & +0.2 \\
20 & 9.0 & -7.78 & -13.3 & -3.0 & -12.75 & -30.0 & -24.7 \\
21 & 11.7 & -13.24 & -6.2 & +4.0 & -14.10 & -30.0 & -20.6 \\
26 & 12.5 & -27.02 & -8.2 & -10.2 & -28.50 & -40.4 & -43.9 \\
27 & 11.9 & -5.42 & -19.1 & -0.8 & -7.67 & -56.0 & -39.9 \\
28 & 1.4 & -0.13 & -2.6 & +0.1 & -0.20 & -6.4 & -3.8 \\
29 & 1.8 & -0.12 & -3.1 & +0.4 & -0.21 & -6.4 & -3.0 \\
30 & 1.3 & -0.18 & -3.7 & -1.3 & -0.29 & -7.9 & -5.7 \\
31 & 2.0 & -0.42 & -5.4 & -1.8 & -0.67 & -13.2 & -9.8 \\
33 & 0.6 & -0.02 & -1.7 & -0.4 & -0.04 & -1.5 & -0.3 \\
34 & 3.1 & -0.55 & -6.7 & -1.1 & -0.89 & -15.6 & -10.3 \\
36 & 1.7 & -0.27 & -3.0 & +0.2 & -0.37 & -9.0 & -5.9 \\
37 & 1.0 & -0.11 & -3.1 & -1.2 & -0.20 & -6.1 & -4.3 \\
38 & 3.9 & -0.77 & -9.8 & -2.7 & -1.35 & -19.0 & -12.5 \\
39 & 3.2 & -0.34 & -8.1 & -2.1 & -0.68 & -9.3 & -3.6 \\
40 & 10.0 & -3.28 & -38.2 & -21.5 & -6.68 & -21.1 & -7.7 \\
42 & 1.0 & -0.16 & -2.7 & -0.9 & -0.23 & -7.1 & -5.4 \\
43 & 1.0 & -0.06 & -2.2 & -0.3 & -0.11 & -4.3 & -2.5 \\
44 & 2.4 & -0.34 & -6.6 & -2.2 & -0.62 & -11.2 & -7.1 \\
45 & 1.2 & -0.07 & -2.9 & -0.5 & -0.14 & -4.2 & -2.0 \\
46 & 1.7 & -0.28 & -4.2 & -1.0 & -0.44 & -10.3 & -7.3 \\
49 & 1.1 & -0.23 & -3.0 & -1.1 & -0.34 & -8.7 & -6.9 \\
50 & 0.8 & -0.12 & -2.5 & -1.0 & -0.19 & -6.3 & -4.8 \\
51 & 0.8 & -0.12 & -2.9 & -1.4 & -0.20 & -6.4 & -4.9 \\
52 & 1.1 & -0.20 & -3.3 & -1.4 & -0.31 & -8.4 & -6.6 \\
53 & 0.9 & -0.13 & -3.3 & -1.7 & -0.23 & -6.6 & -5.1 \\
54 & 0.7 & -0.11 & -2.5 & -1.1 & -0.17 & -5.9 & -4.6 \\
55 & 2.6 & -0.97 & -6.9 & -2.7 & -1.39 & -20.1 & -16.3 \\
58 & 11.1 & -23.88 & -12.5 & -14.1 & -25.42 & -60.3 & -63.4 \\
59 & 7.5 & -8.19 & -9.4 & -2.7 & -8.97 & -42.7 & -36.7 \\
60 & 37.5 & -162.68 & -76.0 & -163.7 & -184.08 & -322.0 & -431.1 \\
62 & 5.1 & -5.42 & -10.2 & -5.4 & -6.16 & -43.0 & -38.9 \\
63 & 14.9 & -25.21 & -45.5 & -40.8 & -31.79 & -160.5 & -162.4 \\
64 & 9.3 & -19.56 & -12.3 & -13.3 & -20.97 & -58.3 & -60.7 \\
65 & 3.2 & -1.75 & -11.0 & -6.3 & -2.32 & -36.1 & -31.9 \\
66 & 8.3 & -10.89 & -18.1 & -12.4 & -12.66 & -73.0 & -69.1 \\
67 & 1.8 & -0.67 & -5.3 & -2.5 & -0.86 & -18.6 & -15.9 \\
68 & 8.0 & -9.26 & -19.0 & -12.3 & -11.03 & -73.6 & -68.6 \\
70 & 1.2 & -0.28 & -4.6 & -2.4 & -0.40 & -13.3 & -11.3 \\
71 & 19.4 & -50.83 & -37.4 & -49.4 & -57.43 & -158.8 & -177.4 \\
72 & 3.0 & -1.66 & -10.5 & -6.0 & -2.18 & -34.5 & -30.6 \\
73 & 1.0 & -0.16 & -4.7 & -3.0 & -0.26 & -10.5 & -8.9 \\
74 & 2.0 & -0.65 & -7.8 & -4.5 & -0.93 & -22.4 & -19.4 \\
75 & 2.2 & -0.99 & -7.8 & -4.3 & -1.31 & -25.6 & -22.4 \\
77 & 1.1 & -0.18 & -5.6 & -3.6 & -0.32 & -11.4 & -9.6 \\
\end{tabular}
\end{table}
\addtocounter{table}{-1}
\begin{table}
\centering
\caption{-- continued.}
\begin{tabular}{c c ccc ccc }
\toprule
 & & \multicolumn{3}{c}{$n$(H$_{2}$)$=10^{3}$\,cm$^{-3}$} & \multicolumn{3}{c}{$n$(H$_{2}$)$=10^{4}$\,cm$^{-3}$} \\ \cmidrule(lr){3-5} \cmidrule(lr){6-8}
Source & $\Omega_{\textsc{k}}$ & $\Omega_{\textsc{g}}$ & $\Omega_{\textsc{p}}$ & $\frac{1}{2}\ddot{\mathcal{I}}$ & $\Omega_{\textsc{g}}$ & $\Omega_{\textsc{p}}$ & $\frac{1}{2}\ddot{\mathcal{I}}$ \\ \cmidrule{2-8}
ID & \multicolumn{7}{c}{($\times 10^{41}$\,erg)}  \\
\midrule
78 & 2.9 & -1.18 & -12.0 & -7.4 & -1.74 & -33.0 & -28.9 \\
79 & 1.0 & -0.14 & -5.1 & -3.4 & -0.25 & -9.8 & -8.1 \\
80 & 2.0 & -0.62 & -7.1 & -3.6 & -0.87 & -21.2 & -18.0 \\
81 & 0.9 & -0.13 & -3.9 & -2.3 & -0.21 & -9.3 & -7.7 \\
83 & 1.0 & -0.23 & -3.5 & -1.8 & -0.31 & -11.1 & -9.4 \\
85 & 0.8 & -0.12 & -5.1 & -3.6 & -0.23 & -8.7 & -7.3 \\
86 & 1.7 & -0.54 & -6.4 & -3.5 & -0.75 & -19.4 & -16.6 \\
87 & 0.9 & -0.16 & -4.2 & -2.5 & -0.25 & -10.3 & -8.7 \\
88 & 1.6 & -0.33 & -10.2 & -7.5 & -0.63 & -14.3 & -11.9 \\
89 & 1.8 & -0.48 & -9.3 & -6.3 & -0.79 & -20.3 & -17.6 \\
90 & 3.0 & -1.28 & -12.5 & -7.9 & -1.88 & -34.5 & -30.5 \\
92 & 1.4 & -0.34 & -4.7 & -2.4 & -0.47 & -14.5 & -12.2 \\
93 & 17.7 & -34.67 & -55.4 & -54.6 & -43.79 & -195.0 & -203.4 \\
94 & 1.3 & -0.30 & -6.2 & -3.9 & -0.47 & -15.1 & -12.9 \\
95 & 19.2 & -32.43 & -73.5 & -67.5 & -44.88 & -223.8 & -230.3 \\
96 & 5.8 & -3.49 & -24.0 & -15.9 & -5.22 & -64.1 & -57.8 \\
97 & 1.8 & -0.48 & -9.5 & -6.3 & -0.81 & -20.4 & -17.5 \\
98 & 2.2 & -0.98 & -4.1 & -0.7 & -1.20 & -15.1 & -11.9 \\
99 & 1.0 & -0.30 & -1.4 & +0.3 & -0.34 & -5.7 & -4.0 \\
100 & 2.0 & -0.76 & -5.1 & -1.8 & -1.02 & -16.1 & -13.0 \\
101 & 1.8 & -0.57 & -4.3 & -1.2 & -0.77 & -13.6 & -10.7 \\
102 & 0.6 & -0.09 & -1.1 & -0.1 & -0.11 & -3.9 & -2.9 \\
103 & 0.5 & -0.07 & -1.1 & -0.1 & -0.09 & -3.6 & -2.6 \\
104 & 0.7 & -0.14 & -1.7 & -0.3 & -0.18 & -5.5 & -4.2 \\
105 & 0.1 & -0.01 & -0.5 & -0.2 & -0.01 & -1.0 & -0.7 \\
106 & 8.3 & -9.13 & -18.0 & -10.4 & -11.58 & -62.8 & -57.7 \\
107 & 1.2 & -0.34 & -2.3 & -0.3 & -0.43 & -8.4 & -6.4 \\
108 & 0.5 & -0.05 & -2.0 & -1.0 & -0.08 & -3.4 & -2.5 \\
109 & 0.3 & -0.02 & -0.8 & -0.3 & -0.03 & -2.1 & -1.6 \\
110 & 0.4 & -0.04 & -0.7 & -0.0 & -0.05 & -2.3 & -1.7 \\
112 & 0.3 & -0.02 & -0.8 & -0.3 & -0.04 & -2.3 & -1.7 \\
113 & 0.9 & -0.15 & -2.6 & -1.1 & -0.23 & -6.9 & -5.4 \\
114 & 0.3 & -0.03 & -0.8 & -0.2 & -0.04 & -2.4 & -1.8 \\
115 & 1.2 & -0.26 & -2.7 & -0.6 & -0.35 & -8.5 & -6.5 \\
116 & 0.5 & -0.07 & -1.1 & -0.1 & -0.09 & -3.6 & -2.7 \\
117 & 0.9 & -0.19 & -2.0 & -0.5 & -0.24 & -6.6 & -5.1 \\
\bottomrule
\end{tabular}
\end{table}
\setlength{\tabcolsep}{6pt}

It must be stressed that the values shown for the virial ratio $-(\Omega_{\textsc{g}}+\Omega_{\textsc{p}})/2\Omega_{\textsc{k}}$ are upper limits (for the assumed bounding density); Figure~\ref{fig:ceph_virial} shows the greatest extent to which our cores could be virially bound.

Figure~\ref{fig:ceph_virial} suggests that in the high-bounding-density case, our cores are not thermally supported: in the absence of non-thermal internal energy and/or an internal magnetic field, $-(\Omega_{\textsc{g}}+\Omega_{\textsc{p}})>2\Omega_{\textsc{k}}$ in all but one case (core 19, which is marginally unbound).  If the cores' dominant support mechanism were internal \emph{thermal} motions, then all but two of our cores (core 19, and core 3, discussed below) would be simultaneously undergoing pressure-driven collapse.

The physical picture in the low-bounding-density case, is less clear than in the high-density case.  In this case, the cores are less strongly bound and less pressure-confined than in the high-density case.  Figure~\ref{fig:ceph_virial} shows that our cores remain typically virially bound and pressure-confined in the low-bounding-density-case, but that a significant fraction of the cores are in or near virial equlibrium, particularly those in L1147/58, L1174 and L1228.  Many of the cores in L1147/58 are marginally unbound in this analysis, along with four cores in L1174 and one core in L1228.  All of the cores in L1251 and L1172 remain virially bound.  In this case, there are seven gravitationally- and virially-bound cores in the sample, and one core which is gravitationally-dominated but marginally virially unbound.

In either case, these results suggest that a significant fraction of our cores are simultaneously undergoing pressure-driven collapse.  As this scenario is unlikely, we hypothesise three possible alternative scenarios:

(1) The Y97 measurements overestimate the external pressure on our cores.  If the low-resolution, and hence large-scale, Y97 $^{13}$CO measurements do not correspond accurately to the gas immediately surrounding our dense cores, the linewidths measured by Y97 will not accurately represent the pressure confining the core.  As turbulent motions are expected to dissipate on small scales (e.g. \citealt{larson1981}; \citealt{solomon1987}), velocities measured with the Y97 beam size of 2.4 arcmin (corresponding to 12\,pc at a distance of 300\,pc) may not be representative of the velocities of material surrounding the sub-parsec-scale cores we consider here.

(2) The non-thermal motions of the gas surrounding the core do not create the effect of a hydrostatic pressure on the core, or do not do so in such a manner that the measured linewidth accurately represents the pressure on the core caused by non-thermal gas motions (see, e.g., \citealt{maclow2004}).

(3) The dominant mechanism of core support in Cepheus has not been accounted for in our virial analysis.  In this scenario, cores are predominantly supported by some combination of internal non-thermal motions and/or internal magnetic field.  \citet{pattle2015} found that the majority of the starless cores in the highest-column-density regions of the Ophiuchus molecular cloud were typically in approximate virial equilibrium with their surroundings, and marginally pressure-dominated, with the majority of support against collapse provided by non-thermal internal motions.  Figure~\ref{fig:ceph_virial} is consistent with starless cores in Cepheus behaving in a similar way to those in Ophiuchus, presuming that there is sufficient internal support from non-thermal internal motions and/or internal magnetic fields to bring the cores into approximate virial equilibrium with their surroundings.

None of these hypotheses are contradictory, and all may contribute to the apparent over-estimation of the degree to which our cores are pressure-confined.

Values of the virial ratio are marked as upper limits on Figure~\ref{fig:ceph_virial}, as we can identify the information missing from our determination of the cores' virial ratios: an estimate of the cores' non-thermal and magnetic internal energies.  If either hypotheses (1) or (2) above are valid, then the values of the confinement ratio $\Omega_{\textsc{g}}/\Omega_{\textsc{p}}$ shown on Figure~\ref{fig:ceph_virial} are lower limits on the true values.  However, we do not mark them as such on Figure~\ref{fig:ceph_virial}, as we do not know which of our hypotheses best explain the measured values of external pressure.

The high-density analysis suggests that there is one gravitationally-bound prestellar core amongst our sample: core 3 in L1147/58, for which $\Omega_{\textsc{g}}>\Omega_{\textsc{p}}$ and $-(\Omega_{\textsc{g}}+\Omega_{\textsc{p}})>2\Omega_{\textsc{k}}$.  This core is among the 13 cores predicted to be gravitationally unstable by the Bonnor-Ebert criterion.  It is worth noting that the majority of the cores which are only mildly pressure-dominated are unstable according to the Bonnor-Ebert criterion.  If the measured values of the confinement ratio are in fact lower limits on the true values, many of these cores may be prove to be gravitationally bound.

The low-density analysis suggests that there are seven gravitationally-bound prestellar cores amongst our sample, all of which are unstable according to the Bonnor-Ebert criterion.  Three of the cores which are mildly pressure-dominated are unstable according to the Bonnor-Ebert criterion.

The Bonnor-Ebert stability criterion is in better agreement with our virial analysis in the low-density case than in the high-density case.  If the Bonnor-Ebert criterion is in fact an accurate measure of the stability of our cores, it suggests that the low-density model ($n_{e}=10^{3}\,$cm$^{-3}$) is a more accurate description of the energy balance of our cores than the high-density model.

That the low-density model is more likely to be accurate is also supported by the values of the virial ratio shown in Figure~\ref{fig:ceph_virial}: in the low-density case, a significant fraction of the cores have virial ratios consistent with their being in or near virial equilibrium with their surroundings, as might be expected in reasonably long-lived star-forming regions (all of the Cepheus clouds have been forming stars for long enough to form at least some Class II protostars; see K09).

In the high-density case, almost all of the cores appear to be strongly pressure-confined and virially bound, suggesting that they are all effectively imploding under pressure - an unlikely situation for cores in star-forming regions that appear to have been forming stars continuously for a significant length of time, particularly those showing no signs of recent external influence.  Hence, we conclude that of our two models, a density of $10^{3}\,$cm$^{-3}$ is the more likely to be representative of the true density at which our cores are confined by pressure from the surrounding molecular cloud.

\subsection{Resolving the virial balance of cores in Cepheus}

The minimum additional information which is required in order to determine which of our cores are in fact virially bound is a measure of the cores' internal linewidths, i.e. observations of the cores in an optically thin dense gas tracer such as C$^{18}$O or nitrogen-bearing tracers (e.g. NH$_{3}$, N$_{2}$H$^{+}$).  This would allow determination of the level of core support from non-thermal internal motions.

Ideally, a measure of the magnetic field strength in the cores is also required, to determine whether magnetic fields play a significant r\^{o}le in core support in Cepheus.  This might be achieved using a wide-field polarimeter such as POL-2 on the JCMT (\citealt{friberg2016}; Ward-Thompson et al. 2016, in prep.).

Our estimates of the external surface pressure on the cores could be improved with higher-resolution observations of the Cepheus Flare clouds in medium-density tracers such as $^{13}$CO.  Measurements of the linewidth of gas surrounding the cores taken with an instrument with angular resolution comparable to the angular size of the cores (e.g. using HARP-B on the JCMT; \citealt{buckle2010}) would allow us to exclude hypothesis (1), above.

\section{Summary}

In this paper we have extracted sources from the SCUBA-2 data of the L1147/L1158, L1172/L1174, L1251 and L1228 regions of the Cepheus Flare.  We have characterised our sources using their 850-\um\ flux densities and temperatures supplied by the \emph{Herschel} GBS (Di Francesco et al., 2016, in prep.).  We have compared the properties of cores in the different Cepheus Flare regions in order to determine the mode of star formation proceeding in each region.  We have determined the relative importance of gravity and external pressure in confining our cores, and have determined an upper limit on the degree to which our cores are virially bound.

We identified 117 sources across the Cepheus Flare region using the CSAR source extraction algorithm, of which 23 were associated with a protostar in the K09 Spitzer catalogue.  Of our 117 sources, 20 were located in L1147/L1158, 26 in L1174, 9 sources in L1172, 42 in L1251 and 20 in L1228.  We determined the best-fit flux densities of our sources using the multiple-Gaussian fitting algorithm described by \citet{pattle2015}.

We determined masses for each of our sources using our best-fit flux densities and temperatures supplied by the \emph{Herschel} GBS (Di Francesco et al., 2016, in prep.).  We found that our cores typically lie in the `prestellar' part of the mass/size plane.  Our cores typically have temperatures in the range $\sim9-15$\,K, with the exception of cores associated with the L1174 reflection nebula, which have temperatures up to $\sim30$\,K.

We determined source temperatures from the ratio of SCUBA-2 450-\um\ and 850-\um\ flux densities, for those of our sources with a detection $\geq 5\sigma$ at 450\um.  We found that temperatures determined from \emph{Herschel} and SCUBA-2 data were generally in agreement for our sources.  We found a slight tendency for \emph{Herschel}-derived temperatures to be higher than SCUBA-2-derived temperatures, consistent with \emph{Herschel} observations sampling slightly warmer material.  Source masses derived from SCUBA-2 temperatures are correspondingly slightly higher than those derived from \emph{Herschel} temperatures.  We concluded that the SCUBA-2 flux density ratio is a reliable determinant of a source's temperature when neither the 450-\um\ nor the 850-\um\ data point is on the Rayleigh-Jeans tail of a source's spectral energy distribution -- i.e. when $T\lesssim 20\,$K.

We analysed the cumulative distribution functions of core masses for each region in Cepheus, using the maximum likelihood estimator for an infinite power-law distribution, and found that the core mass function in each region shows a sub-Salpeter power-law behaviour, with the exception of L1228, which has a power-law index consistent with the Salpeter IMF. Determining the power-law index over all cores, we found a sub-Salpeter value of $\alpha=1.88\pm 0.09$ over the mass range $M>0.08\,$M$_{\odot}$.  For the highest-mass cores, we found a CMF power-law index $\alpha=2.61\pm0.27$ over the mass range $M>0.5\,$M$_{\odot}$ (again determined over all cores), marginally consistent with the Salpeter IMF.

We compared the number of starless cores detected in each region with the numbers of embedded and Class II sources found by K09.  We found that L1147/L1158 and L1228 have a high ratio of starless cores to Class II sources, while L1251 and L1174 have a low ratio.  This is consistent with L1174 and L1251 being active sites of star formation, while L1147/L1158 and L1228 form stars in a less active mode.

We determined the Bonnor-Ebert critically-stable masses of our cores, and found that the Bonnor-Ebert model predicts that most of our cores have stable BE solutions accessible to them.

We performed a virial analysis on our cores, determining the external pressure on our cores using $^{13}$CO velocity dispersion measurements determined by \citet{yonekura1997}.  We found that, assuming a bounding density for our cores of $10^{4}$\,cm$^{-3}$, all but one of our cores are virially bound and there is only one gravitationally-bound prestellar core among our sample, with the rest of the cores being confined by external pressure.  However, we found that if we assume a bounding density of $10^{3}$\,cm$^{-3}$,  seven of our cores are gravitationally bound, and the majority of the cores are approximately virialised or mildly pressure-confined.  We concluded that, if the \citet{yonekura1997} measurements are representative of the conditions in the gas confining our cores, our cores typically cannot be supported by internal \emph{thermal} energy alone.  In the absence of non-thermal internal motions or an internal magnetic field, a significant fraction of our cores would be significantly out of virial equilibrium and collapsing under pressure.  We hence hypothesise that cores in the Cepheus molecular cloud may not typically be thermally supported.

\section{Acknowledgements}

K.P. wishes to acknowledge STFC postdoctoral support under grant numbers ST/K002023/1 and ST/M000877/1 and studentship support under grant number ST/K501943/1 while this research was carried out.  The James Clerk Maxwell Telescope has historically been operated by the Joint Astronomy Centre on behalf of the Science and Technology Facilities Council of the United Kingdom, the National Research Council of Canada and the Netherlands Organisation for Scientific Research.  Additional funds for the construction of SCUBA-2 were provided by the Canada Foundation for Innovation.  The Starlink software \citep{currie2014} is supported by the East Asian Observatory.  This research used the services of the Canadian Advanced Network for Astronomy Research (CANFAR) which in turn is supported by CANARIE, Compute Canada, University of Victoria, the National Research Council of Canada, and the Canadian Space Agency. This research used the facilities of the Canadian Astronomy Data Centre operated by the National Research Council of Canada with the support of the Canadian Space Agency.  \emph{Herschel} is an ESA space observatory with science instruments provided by European-led Principal Investigator consortia and with important participation from NASA.  This research has made use of the NASA Astrophysics Data System. The authors wish to recognize and acknowledge the very significant cultural role and reverence that the summit of Maunakea has always had within the indigenous Hawaiian community. We are most fortunate to have the opportunity to conduct observations from this mountain.

\bibliographystyle{mn2e_fix}

\vspace{1cm}

\noindent$^{1}$Jeremiah Horrocks Institute, University of Central Lancashire, Preston, Lancashire, PR1 2HE, UK\\
$^{2}$Department of Physics and Astronomy, University of Victoria, Victoria, BC, V8P 1A1, Canada\\
$^{3}$NRC Herzberg Astronomy and Astrophysics, 5071 West Saanich Rd, Victoria, BC, V9E 2E7, Canada\\
$^{4}$Leiden Observatory, Leiden University, PO Box 9513, 2300 RA Leiden, The Netherlands\\
$^{5}$Max Planck Institute for Astronomy, K\"{o}nigstuhl 17, D-69117 Heidelberg, Germany\\
$^{6}$Astrophysics Group, Cavendish Laboratory, J J Thomson Avenue, Cambridge, CB3 0HE\\
$^{7}$Kavli Institute for Cosmology, Institute of Astronomy, University of Cambridge, Madingley Road, Cambridge, CB3 0HA, UK\\
$^{8}$Department of Physics and Astronomy, University of Waterloo, Waterloo, Ontario, N2L 3G1, Canada\\
$^{9}$Joint Astronomy Centre, 660 N. A`oh\={o}k\={u} Place, University Park, Hilo, Hawaii 96720, USA\\
$^{10}$Physics and Astronomy, University of Exeter, Stocker Road, Exeter EX4 4QL, UK\\
$^{11}$LSST Project Office, 933 N. Cherry Ave, Tucson, AZ 85719, USA\\
$^{12}$School of Physics and Astronomy, Cardiff University, The Parade, Cardiff, CF24 3AA, UK\\
$^{13}$European Southern Observatory (ESO), Garching, Germany\\
$^{14}$Jodrell Bank Centre for Astrophysics, Alan Turing Building, School of Physics and Astronomy, University of Manchester, Oxford Road, Manchester, M13 9PL, UK\\
$^{15}$Institute for Astronomy, ETH Zurich, Wolfgang-Pauli-Strasse 27, CH-8093 Zurich, Switzerland\\
$^{16}$Universit\'e de Montr\'eal, Centre de Recherche en Astrophysique du Qu\'ebec et d\'epartement de physique, C.P. 6128, succ. centre-ville, Montr\'eal, QC, H3C 3J7, Canada\\
$^{17}$James Madison University, Harrisonburg, Virginia 22807, USA\\
$^{18}$School of Physics, Astronomy \& Mathematics, University of Hertfordshire, College Lane, Hatfield, HERTS AL10 9AB, UK\\
$^{19}$Astrophysics Research Institute, Liverpool John Moores University, Egerton Warf, Birkenhead, CH41 1LD, UK\\
$^{20}$Imperial College London, Blackett Laboratory, Prince Consort Rd, London SW7 2BB, UK\\
$^{21}$Dept of Physics \& Astronomy, University of Manitoba, Winnipeg, Manitoba, R3T 2N2 Canada\\
$^{22}$Dunlap Institute for Astronomy \& Astrophysics, University of Toronto, 50 St. George St., Toronto ON M5S 3H4 Canada\\
$^{23}$Dept. of Physical Sciences, The Open University, Milton Keynes MK7 6AA, UK\\
$^{24}$The Rutherford Appleton Laboratory, Chilton, Didcot, OX11 0NL, UK.\\
$^{25}$UK Astronomy Technology Centre, Royal Observatory, Blackford Hill, Edinburgh EH9 3HJ, UK\\
$^{26}$Institute for Astronomy, Royal Observatory, University of Edinburgh, Blackford Hill, Edinburgh EH9 3HJ, UK\\
$^{27}$Centre de recherche en astrophysique du Qu\'ebec et D\'epartement de physique, de g\'enie physique et d'optique, Universit\'e Laval, 1045 avenue de la m\'edecine, Qu\'ebec, G1V 0A6, Canada\\
$^{28}$Department of Physics and Astronomy, UCL, Gower St, London, WC1E 6BT, UK\\
$^{29}$Department of Physics and Astronomy, McMaster University, Hamilton, ON, L8S 4M1, Canada\\
$^{30}$Department of Physics, University of Alberta, Edmonton, AB T6G 2E1, Canada\\
$^{31}$University of Western Sydney, Locked Bag 1797, Penrith NSW 2751, Australia\\
$^{32}$National Astronomical Observatory of China, 20A Datun Road, Chaoyang District, Beijing 100012, China

\label{lastpage}

\end{document}